\documentclass{aastex62}

\pdfoutput=1 

\usepackage{graphicx}	% Including figure files
\usepackage{amsmath}	% Advanced maths commands
\usepackage{amssymb}	% Extra maths symbols
\usepackage{physics}
\usepackage{enumitem}
\usepackage{bbm}
\usepackage{color}
\usepackage{natbib}
\usepackage{hyperref}
\usepackage{pifont}
\usepackage{xcolor}
\usepackage{epstopdf}
\newcommand\msun{$M_{\odot}$}
\newcommand\degree{\degr}

\newcommand{\AV}{{$A_{V}$}}

\begin{document}
\title{Dirt-cheap Gas Scaling Relations: Using Dust Absorption, Metallicity and Galaxy Size to Predict Gas Masses for Large Samples of Galaxies}

\shorttitle{Gas scaling relations using {$A_V$}, $Z$, and {$R_{50}$}}
\shortauthors{Yesuf \& Ho}
\author{Hassen M. Yesuf}
\affiliation{Kavli Institute for Astronomy and Astrophysics, Peking University, Beijing 100871, China}
\affiliation{Kavli Institute for the Physics and Mathematics of the Universe, The University of Tokyo, Kashiwa, Japan 277-8583}
\author{Luis C. Ho}
\affiliation{Kavli Institute for Astronomy and Astrophysics, Peking University, Beijing 100871, China}
\affiliation{Department of Astronomy, School of Physics, Peking University, Beijing 100871, China}

\begin{abstract}

We apply novel survival analysis techniques to investigate the relationship between a number of the properties of galaxies and their atomic ($M_\mathrm{HI}$) and molecular ($M_{\mathrm{H_2}}$) gas mass, with the aim of devising efficient, effective empirical estimators of the cold gas content in galaxies that can be applied to large optical galaxy surveys. We find that  dust attenuation, {\AV}, of both the continuum and nebular emission, shows significant partial correlations with $M_{\mathrm{H_2}}$, after controlling for the effect of star formation rate (SFR). The partial correlation between {\AV} and $M_\mathrm{HI}$, however, is weak. This is expected because in poorly dust-shielded regions molecular hydrogen is dissociated by far-ultraviolet photons. We also find that the stellar half-light radius, $R_{50}$, shows significant partial correlations with both $M_{\mathrm{H_2}}$ and $M_\mathrm{HI}$. This hints at the importance of environment (e.g., galactocentric distance) on the gas content of galaxies and the interplay between gas and SFR. We fit multiple regression to summarize the median, mean, and the $0.15/0.85$ quantile multivariate relationships among $M_{\mathrm{H_2}}$, {\AV}, metallicity, and/or $R_{50}$. A linear combination of {\AV} and metallicity (inferred from stellar mass) or {\AV} and $R_{50}$, can estimate molecular gas masses within $\sim 2.5-3$ times the observed masses. If SFR is used in addition, $M_\mathrm{{H_2}}$ can be predicted to within a factor $\lesssim 2$. In this case, {\AV} and $R_{50}$ are the two best secondary parameters that improve the primary relation between $M_\mathrm{{H_2}}$ and SFR. Likewise, $M_\mathrm{HI}$ can be predicted to within a factor $\lesssim 3$ using $R_{50}$ and SFR. %Our new scaling relations can be used to estimate both atomic and molecular gas masses for large samples of galaxies.}
\end{abstract}

\section{INTRODUCTION}

The formation of stars from the cold gas in the interstellar medium (ISM) is a fundamental process of galaxy formation and evolution. The observed tight relationship between star formation rate (SFR) and stellar mass ($M_\star$) in galaxies, known as the star-forming main sequence \citep[][]{Noeske+07}, is interpreted as indicating that galaxies exhibit a self-regulated quasi-equilibrium between external gas accretion, star formation, and gas outflow \citep{Bouche+10,Dekel+13, Lilly+13, Forbes+14, Peng+14,Rodriguez-Puebla+16}. However, the mechanisms that regulate the availability of gas for star formation and the efficiency with which it is converted to stars on galactic scales are not well understood. Energy and momentum inputs from massive stars, supernovae, and accreting supermassive black holes may regulate global star formation by driving galactic outflows or heating the ISM \citep[e.g.,][]{Dekel+09,Hopkins+11,Ostriker+11,Faucher-Giguere+13, Hopkins+14, Harrison+18}.

Although gas observations have strong constraining power for theoretical models of galaxy formation and evolution, directly measuring gas properties for large representative galaxy samples is a difficult and time-consuming endeavor. Neutral atomic hydrogen (\ion{H}{1}) can be directly observed via the 21 cm hyperfine line, but molecular hydrogen (H$_2$) must be inferred indirectly from other tracers, such as the rotational transitions of CO. However, measuring these lines for galaxy sample sizes comparable to those of ultraviolet-optical surveys will remain unattainable for the foreseeable future. Blind \ion{H}{1} surveys such as the Arecibo Legacy Fast ALFA survey do provide global (unresolved) measurements of the \ion{H}{1} content for $\sim 31,500$ galaxies \citep{Haynes+18}, but are only sensitive to the most gas-rich galaxies located at very modest redshifts ($z \lesssim 0.06$). In contrast, the extended GALEX Arecibo SDSS Survey \citep[xGASS; ][]{Catinella+18} provides more sensitive measurements of atomic gas content for 1179 representative galaxies. A follow-up survey, xCOLD GASS \citep{Saintonge+17}, additionally obtained H$_2$ measurements for 477 galaxies using the CO\,(1--0) emission line. Although the GASS samples account for only less than 0.2\% of local galaxies, they are large enough to statistically characterize gas scaling relations and their scatter. \citet{Saintonge+17} and \citet{Catinella+18} showed that typical state-of-the-art hydrodynamic and semi-analytic simulations, which currently implement sub-grid ISM models, do not succeed in reproducing the gas scaling relations based on the GASS data. But encouraging progress is being made \citep[e.g.,][]{Diemer+19}.  In a similar spirit, pushing out to much greater distances, the Plateau de Bure High-z Blue Sequence Survey \citep{Freundlich+19} legacy program collected CO data of $\sim 120$ galaxies at $z = 0.5 - 3$. 
Detecting \ion{H}{1} beyond the local universe remains challenging.  Concerted efforts are being made with current interferometers, but the current samples remain frustratingly small \citep{Verheijen+07,Hess+19}; the most distant \ion{H}{1} detection to date is at $z = 0.316$ \citep{Fernandez+16}. 

An alternative, albeit indirect, method to probe the gas content of galaxies is to estimate the dust mass from the far-infrared emission and convert it to gas mass using a gas-to-dust ratio. Dust is an important constituent of the ISM, and it has a huge impact on the chemistry and thermodynamics of the gas \citep{Krumholz+11,Glover+12, Gong+17}. Dust shields molecules (e.g., CO) from photodissociation by attenuating the interstellar far-ultraviolet radiation field. The atomic-to-molecular hydrogen transition is governed by the balance between H$_2$ formation via dust grain catalysis and destruction by ultraviolet photons from the Lyman-Werner band. With the advent of the {\it Herschel Space Observatory}\ and the Atacama Large Millimeter Array, dust continuum emission has been utilized in numerous studies as a surrogate tool to estimate gas masses \citep[e.g.,][]{Eales+12, Genzel+15, Groves+15, Scoville+16, Janowiecki+18,Shangguan+18, Shangguan+19}. The far-infrared emission, however, is not a simple tracer; there are major systematic uncertainties or modeling challenges to exploiting it as a probe of gas content of galaxies \citep[e.g.,][]{Berta+16}. And rest-frame far-infrared measurements, while more widely available than \ion{H}{1} or CO observations, are still non-trivial to amass for large galaxy samples.

This paper introduces new, efficient, and cost-effective methods to predict gas masses within a factor of $\sim 2-3$ for large samples of galaxies. To that end, we revisit the relationship between the integrated gas mass, SFR, and galaxy morphology of the well-studied local galaxies in the xGASS and xCOLD GASS surveys \citep{Saintonge+11,Saintonge+17,Catinella+18}. We investigate whether parameters such as dust attenuation, galaxy radius, and galaxy morphology are useful in predicting atomic and molecular gas masses.  We conclude that the dust extinction {\AV} of both the continuum and nebular emission significantly correlates with $M_{\mathrm{H_2}}$, after controlling for the effect of SFR. The same does not hold for $M_\mathrm{HI}$. We find that $R_{50}$ also significantly correlates with both $M_{\mathrm{H_2}}$ and $M_\mathrm{HI}$. 

\section{DATA AND METHODOLOGY}\label{sec:data}

\subsection{Stellar and Dust Properties}

We use the publicly available Catalog Archive Server (CAS)\footnote{\url{http://skyserver.sdss.org/casjobs/}} to retrieve measurements of median stellar mass, galaxy size, and optical emission-line fluxes based on the Sloan Digital Sky Survey (SDSS) data release 15 \citep{Aguado+19}. We use the median metallicity estimate given in \verb#galSpecExtra# table for non-AGN galaxies whose principal optical diagnostics emission-line ratios are measured at $> 5\,\sigma$. These data are supplemented with global measurements of $V$-band dust attenuation ({\AV}) and SFR from version 2 of the GALEX-SDSS-WISE Legacy Catalog \citep[GSWLC-2;][]{Salim+16,Salim+18}, along with improved galaxy morphology measurements, derived using machine learning, from \citet{Dominguez+18}. \citet{Salim+16,Salim+18} used spectral energy distribution fitting of ultraviolet, optical, and infrared photometry to calculate the global \AV\ and SFR. To estimate accurate total infrared luminosities ($\sim 0.1$ dex uncertainty), \citet{Salim+18} used luminosity-dependent infrared templates \citep{Chary+01} and calibrations derived from a subset of galaxies that have far-infrared photometry from the {\it Herschel}\ ATLAS survey. They assumed a \citet{Chabrier03} stellar initial mass function. By combining accurate visual classifications from the Galaxy Zoo 2 project \citep{Willett+13} and machine-based classifications from \citet{Nair+10}, \citet{Dominguez+18} provided one of the largest and most reliable ($\gtrsim 97\%$ accuracy) morphological catalogs for the SDSS galaxies. Their classifications apply the convolutional neural networks learning algorithm to reanalyze the SDSS three-color galaxy images to learn the mapping between the images and the measurements in the visual classification catalogs, and also to correct misclassifications in the visual catalogs. The measurements from \citet{Dominguez+18} that we use are the T-types and a set of probabilities that quantify whether a galaxy has a disk/feature or a bar, and whether it is an edge-on system or a merger. The T-types range from $-3$ to 10, whereby 0 corresponds to S0s, $< 0$ corresponds to early-type galaxies, $> 0$ corresponds to spirals (Sa$-$Sm), and 10 represents irregular galaxies. The disk/feature probability quantifies whether a galaxy is smooth or has a disk or features such as spiral arms.

We calculate the fiber $V$-band attenuation using the observed H$\alpha$/H$\beta$ ratio and the dust attenuation curve, as

\citep{Charlot+00,Wild+11a}

\begin{equation}
Q_\lambda = 0.6\,(\lambda/5500)^{-1.3} + 0.4\,(\lambda/5500)^{-0.7}.
\end{equation}

\noindent
Assuming that the intrinsic, dust-free Balmer decrement is H$\alpha$/H$\beta$ = 2.86 for inactive galaxies and  H$\alpha$/H$\beta$ = 3.1 for active galactic nuclei \citep[AGNs; e.g.,][]{Ferland+83,Gaskell+84},

\begin{equation}\label{eq:AV}
 A_{V, {\rm fiber}} = \frac{2.5}{(Q_{4861} - Q_{6563})} \times \log \frac{\mathrm{H}\alpha/\mathrm{H}\beta}{3.1 \,\mathrm{or} \, 2.86}.
\end{equation}

\noindent where $Q_{4861} - Q_{6563} = 0.31$. If the observed ratio of an object is below the intrinsic ratio, we set $A_{V, {\rm fiber}} = 0$.

\subsection{Atomic and Molecular Gas Properties}

We use the publicly available atomic gas data from xGASS\footnote{\url{http://xgass.icrar.org/data.html}} and molecular gas data from xCOLD GASS\footnote{\url{http://www.star.ucl.ac.uk/xCOLDGASS/}}. xGASS observed the \ion{H}{1} properties of 1179 representative galaxies from SDSS DR7, selected based only on stellar mass ($M_\star = 10^{9}-10^{11.5}$ \msun) and redshift ($z = 0.01 - 0.05$). xCOLD GASS is a follow-up survey, which observed the CO\,(1--0) emission of 532 galaxies using the IRAM 30~m radio telescope. The CO emission is converted to molecular hydrogen mass (given in the catalog) using a conversion factor ($X_\mathrm{CO}$) that depends on the gas-phase metallicity and the offset of a galaxy from the star-forming main sequence \citep{Accurso+17}. We note that the uncertainties on $M_\mathrm{H_2}$ are large because of uncertainties of aperture correction of single-beam CO observations and of the $X_\mathrm{CO}$ factor. Our main results do not change if we use the constant Galactic $X_\mathrm{CO}$. The overlap between the xCOLD GASS and the xGASS samples includes 477 galaxies, among which 368 have reliable H$\alpha$ and H$\beta$ ($>3\,\sigma$) measurements and 290 were detected in CO. Whenever {\AV} is used, we require it to be well-measured ($>3\,\sigma$) either from the Balmer decrement or from SED fitting.

\subsection{Statistical Methods}

We apply survival analysis methods \citep{Feigelson+85,Helsel12} to analyze data that include both gas detections and upper limits. Kendall's $\tau$ is a non-parametric (rank-based) correlation coefficient, which can be used to quantify a monotonic association (linear or nonlinear) between two variables for both censored and uncensored data \citep{Helsel12}. The quantity $\tau$ is the probability of concordance minus the probability of discordance for randomly selected pairs of observations. It ranges between $-1$ and 1: $\tau = 1$ is a perfect correlation; $\tau = 0$ is no correlation; and $\tau = -1$ is a perfect inverse correlation. Following \citet{Yesuf+17a}, we compute the Kendall's $\tau$ coefficient using the \verb#cenken# routine in the \verb#NADA R# package\footnote{\url{https://CRAN.R-project.org/package=NADA }} \citep{Helsel12}.

If the Kendall's $\tau$ between a galaxy property and the gas mass is significantly different from 0, we use a partial correlation test to investigate possible additional associations with other parameters. The partial correlation test quantifies the degree of association between two variables after controlling for the effect of a third variable \citep{Akritas+96}\footnote{\url{http://astrostatistics.psu.edu/statcodes/cens\_tau}}. After identifying several parameters that strongly correlate with the molecular or atomic gas masses using the partial correlation test, we fit the association between three or more variables using a censored multiple regression. In particular, we use a censored quantile regression \citep{Portnoy03} to describe the median and the 0.15 and 0.85 quantile variability of the gas mass trends. In contrast to standard linear regression, quantile regression does not assume that the residuals are normally distributed around the mean, and it provides a richer characterization of the variability in the data by quantifying the effects of covariates on the whole distribution of the dependent variable, not just the mean trend. We use the \verb#cqr# routine in the \verb#quantreg R# package\footnote{\url{https://CRAN.R-project.org/package=quantreg}} to fit the censored quantile regression model and estimate the coefficients that describe the quantile relations. The routine also gives the standard errors of the coefficients using bootstrap resampling. For comparison, we also fit for the mean trends using the \verb#cenreg# routine in the \verb#NADA R# package, which computes the regression coefficients using maximum likelihood estimation and tests for their significance. We assume a normal distribution for this calculation but have checked that the results are not sensitive to this assumption using the Buckley-James distribution-free least-squares multiple regression model, as implemented in the \verb#rms R# package\footnote{\url{ https://CRAN.R-project.org/package=rms}}. %We use \verb#bj# with \verb#link="identity"# since the gas masses are already in the $\log$ scale. Since \verb#bj# only models a right-censored response variable (i.e., with lower limits), we multiply the gas masses by $-1$ to transform the data. The regression coefficients are multiplied by $-1$ in the end.

\section{RESULTS}\label{sec:res}
\subsection{Predictors of Molecular Gas Mass}

%######################
\begin{deluxetable*}{lrrrr|rrrr}
\tabletypesize{\footnotesize}
\tablecolumns{9} 
\tablewidth{0pt}
 \tablecaption{Kendall's $\tau$ correlation and partial correlation coefficients between gas mass and galaxy properties. 
 \label{tbl:corr}}
\tablehead{\colhead{Galaxy Property} & \colhead{$\tau_\mathrm{H_2}$} & \colhead{$\tau_\mathrm{H_2}\mid$ \AV}  & \colhead{$\tau_\mathrm{H_2}\mid$ SFR} &  \colhead{$\tau_\mathrm{H_2} \mid R_{50}$} & \colhead{$\tau_\mathrm{HI}$} & \colhead{$\tau_\mathrm{HI}\mid$ SFR} & \colhead{$\tau_\mathrm{HI} \mid R_{50}$} & \colhead{Significance}}
\startdata 
Fiber (nebular) dust attenuation \AV  & \textbf{0.39}  & --- & \textbf{0.24} &  \textbf{0.38} &  0.16 &  0.04 &  0.14 & \checkmark -- \checkmark \checkmark \checkmark \checkmark \checkmark \\
Global (stellar) dust attenuation \AV  & \textbf{0.41} & 0.25 & 0.21 &  \textbf{0.41}  & 0.23 & 0.01 &  0.21 & \checkmark \checkmark \checkmark \checkmark  \checkmark \ding{55} \checkmark \\
Global SFR &  \textbf{0.58} & \textbf{0.54} & --- & \textbf{0.58} & \textbf{0.46} &  ---  & $\textbf{0.41}$ & \checkmark \checkmark -- \checkmark \checkmark --  \checkmark \\ 
Petrosian radius $R_{50}$ & \textbf{0.32}  & \textbf{0.32} &0.19  & --- & \textbf{0.35} & \textbf{0.24} & --- & \checkmark \checkmark \checkmark -- \checkmark \checkmark -- \\
Petrosian radius $R_{90}$ & 0.27 & 0.29 & 0.19  & $0.03$  & 0.29 & 0.21 & 0.01 & \checkmark  \checkmark  \checkmark \ding{55} \checkmark \checkmark  \ding{55} \\
Stellar mass $M_\star$ &  0.18 & 0.24  & 0.19  & 0.08 &  0.09 & 0.10 & $-0.08$ & \checkmark  \checkmark  \checkmark \ding{55}  \checkmark \checkmark  \checkmark \\ 
Disk/feature probability $P_{\mathrm{d/f}}$ & \textbf{0.30} & 0.21 & 0.10 & 0.19 & \textbf{0.37} &  \textbf{0.26} & \textbf{0.28} & \checkmark  \checkmark \checkmark \checkmark \checkmark \checkmark \checkmark \\
T-type  & 0.23 & 0.11  & 0.04 & 0.16 & 0.30 &  0.20 & 0.27 & \checkmark  \checkmark \ding{55} \checkmark \checkmark \checkmark \checkmark\\
Bar probability & 0.19 & 0.14 & 0.03 & 0.12 & 0.22 & 0.12 & 0.16  &  \checkmark \checkmark \checkmark \ding{55} \checkmark \checkmark \checkmark \\
Concentration index $R_{90}/R_{50}$  & $-0.20$ & $-0.11$ & $-0.04 $ &  $-0.17$ & $-0.18$  & $-0.08$ & $-0.18$ & \checkmark \checkmark \ding{55} \checkmark \checkmark  \checkmark \checkmark\\
Mass density $0.5M_\star/(\pi R_{50}^2)$ & $-0.20$ & $-0.14$ & $-0.17$ & $-0.03$ & $-0.10$ & 0.03 & $-0.03$  & \checkmark \checkmark \checkmark \ding{55} \checkmark \ding{55}  \ding{55} \\  
Edge-on probability & $-0.09$ &$ -0.16$ & $\sim 0.00$  & $-0.03$ & $-0.07$ &  $-0.02$ & $-0.02$ & \checkmark \checkmark \ding{55} \ding{55} \ding{55} \ding{55} \ding{55}\\
Merger probability &  0.04 & 0.08 & $\sim 0.01$ & $\sim 0.01$ & 0.01 &  0.01  &  $\sim 0.00$ & \ding{55} \ding{55} \ding{55} \ding{55} \ding{55} \ding{55} \ding{55} \\
\enddata
\tablecomments{$\tau_\mathrm{H_2}$ and $\tau_\mathrm{HI}$ are Kendall's $\tau$ rank-correlation coefficient for the molecular and atomic gas, respectively. $\tau_{y}\mid x : x \in \{ A_V, \mathrm{SFR}, R_{50}\} \, \& \, y \in \{ \mathrm{H_2},\mathrm{HI}\}$ are the partial Kendall's $\tau$ coefficients, after removing the effect of dust attenuation (nebular) or SFR or the half-light radius. The significance of each correlation or partial correlation is indicated in the last column, in the order of the preceding columns: \checkmark denotes that the probability of the null hypothesis $\tau = 0$ (no correlation or partial correlation) being correct is $p < 0.001$; \ding{55} denotes that this probability is higher and the null hypothesis cannot be rejected at $>3\,\sigma$. The T-types range from $-3$ to 10: $\le 0$ corresponds to early-type and S0 galaxies; $> 0$ corresponds to spiral galaxies.} %The $p$-value for the partial correlation $\tau_\mathrm{H_2}/R_{50}$ of the stellar mass is $p = 0.002$ and the $\tau_\mathrm{H_2}$/\AV of merger $p = 0.006$.}
\end{deluxetable*}

Table~\ref{tbl:corr} summarizes the correlation and partial correlation analysis between the molecular or atomic gas and other galaxy properties. The integrated SFR, {\AV}, galaxy size, and disk/feature probability are among the galaxy properties that exhibit high Kendall's $\tau$ correlation coefficients ($\tau \ge 0.3$) with the gas mass. As expected, SFR shows the strongest correlation with gas content ($\tau = 0.58$ for $M_{\mathrm{H_2}}$; $\tau = 0.46$ for $M_{\mathrm{HI}}$). We find a moderately strong ($\tau \approx 0.4$) but highly significant ($>5\,\sigma$) correlation between global or fiber {\AV} and $M_\mathrm{H_2}$. {\AV} correlates better with $M_{\mathrm{H_2}}$ than with $M_\mathrm{HI}$, or the combination of the two. The partial correlation analysis indicates that there is a highly significant correlation between {\AV} and $M_\mathrm{H_2}$ ($\tau \approx 0.2$), {\it after}\ controlling for the effect of the SFR. In contrast, the partial correlation between {\AV} and $M_\mathrm{HI}$ is very weak and hardly statistically significant. The stellar $r$-band half-light radius of the galaxy, $R_{50}$, also has a partial $\tau \approx 0.2$ with both $M_\mathrm{H_2}$ and $M_\mathrm{HI}$ at fixed SFR. However, the correlation between $R_{50}$ and {\AV} is weak ($\tau = 0.1$), which implies that controlling for the effect of the other hardly changes the partial correlation with the gas mass. After controlling for {\AV}, $M_\star$ shows significant partial correction ($\tau = 0.24$). This could be anticipated from the mass-metallicity relation \citep{Tremonti+04} and from the dependence of dust-to-gas ratios on metallicity \citep[e.g.,][]{Janowiecki+18}. It is not so obvious why $R_{50}$ is better correlated ($\tau =0.32$) with $M_\mathrm{H_2}$ after fixing {\AV}. This correlation may reflect the association of $R_{50}$ with metallicity and its deviation from the mass-metallicity relation \citep{Tremonti+04}, and the processes that set the molecular-to-atomic gas ratio or gas mass to SFR ratio (i.e., the star formation efficiency). The disk/feature probability ($P_{\mathrm{d/f}}$) shows a partial correlation of $\tau \approx 0.3$ with $M_{\mathrm{HI}}$ after controlling for SFR or $R_{50}$. After controlling for SFR, no significant correlation exists between molecular/atomic gas mass and the stellar concentration index, stellar surface mass density, or merger or edge-on probability. The atomic gas partially correlates with T-type ($\tau = 0.2$), given the SFR. 

Next, we fit the relationship between gas mass and galaxy properties that show strong statistical correlation. Tables~\ref{tbl:fit_H2}-\ref{tbl:fit_H2_all} (the latter two are in Appendices~\ref{app:xco} and \ref{app:fit_all}) present the result of fitting a censored quantile regression ($q=0.15$, 0.5, and 0.85) or a normal censored linear regression model to the xCOLD GASS molecular mass data \citep{Saintonge+17}. In Table~\ref{tbl:fit_H2}, the coefficients of the equations describing the relationship among molecular gas mass, dust attenuation, gas-phase metallicity ($Z$) and/or galaxy radius are given. We infer $Z$ from the mass-metallicity relation of \citet{Tremonti+04}. We present two sets of fits, using either the global (stellar) attenuation or the fiber (nebular) attenuation. Although the scatter is modeled by the $q=0.15$ and 0.85 regression fits, we additionally quantify the scatter for each relationship using the root mean square deviation ($\operatorname{RMSD}$) and the median absolute deviation ($\operatorname{MAD}$). For normally distributed residuals, $\operatorname{RMSD} \approx 1.48 \operatorname{MAD}$. The calculations of MAD and RMSD include the detections but not the upper limits, and these measures should be regarded as approximate indicators of the true scatter. Table~\ref{tbl:fit_H2_XCO} repeats the analysis for the constant Galactic $X_\mathrm{CO}$. Both cases give consistent fitting formulae.

As shown in Figures~\ref{fig:avf_gas} and \ref{fig:avg_gas}, $M_\mathrm{H_2}$ can be indirectly estimated (within a factor of $\sim 2.5$) for a large sample of galaxies using {\AV} and $Z$, as follows:

\begin{equation}\label{eq:MH2_AV_Z}
\log M_\mathrm{{H_2}} = \alpha + \beta_A A_{V} + \beta_Z \log Z,
\end{equation}

\noindent where $\log Z = 12 + \log {\rm [O/H]} - 8.8$, ($\alpha$, $\beta_A$, $\beta_Z$) = ($8.27 \pm 0.11$, $0.38 \pm 0.13$, $1.44 \pm 0.34$) for fiber (nebular) {\AV}, and ($\alpha$, $\beta_A$, $\beta_Z$) = ($8.23 \pm 0.12$, $0.91 \pm 0.36$, $1.30 \pm 0.31$) for global (stellar) {\AV}. Note that the continuum attenuation is $\sim 2$ times lower than the nebular attenuation \citep{Calzetti+94,Wild+11a}. This may explain the lower $\beta_A$ in the case of fiber {\AV}.
$M_\mathrm{{H_2}}$ can be similarly related to {\AV} and $R_{50}$: 

\begin{equation} \label{eq:MH2_AV_R}
\log M_\mathrm{{H_2}}  = \alpha + \beta_A A_{V} + \beta_R \log R_{50},
\end{equation}

\noindent where ($\alpha$, $\beta_A$, $\beta_R$) = ($7.84 \pm 0.09$, $0.42 \pm 0.06$, $1.27 \pm 0.09$) for fiber {\AV} and ($\alpha$, $\beta_A$, $\beta_R$) = ($7.58\pm 0.11$, $1.41\pm 0.13$, $1.35\pm 0.19$) for global {\AV}.
 
The linear combination of {\AV} and $M_\star$ also has comparable predictive power, but it does not constrain well the scatter from its mean relation. For the mean or median gas relations, however, the coefficients corresponding to $M_\star$, $Z$, and $R_{50}$ are all significant at $\alpha = 0.05$. The Akaike information criterion (AIC) and  the Bayesian information criterion (BIC), which adjust the likelihood/penalize for the extra parameters, indicate that the normal censored regression model that combines {\AV} with either $Z$ or $M_\star$ or $R_{50}$ is preferred over {\AV} by itself. Furthermore, using the inferred mean metallicity from the mass-metallicity relation (e.g., equation 3 of \citet{Tremonti+04}), the gas masses predicted by equation~\ref{eq:MH2_AV_Z} above agree within a factor of $\sim 2.5$ with observed $M_\mathrm{{H_2}}$ of all detections in xCOLD GASS ($\operatorname{MAD=0.22}$ and $\operatorname{RMSD}=0.36$). Thus, this equation may be applied to large samples of galaxies with indirect metallicity estimates. Because the metallicity is measured only for strong emission-line star-forming galaxies, when applying the relation to weak emission-line galaxies, users should verify whether other scaling relations give similar $M_\mathrm{{H_2}}$ estimates. The relation that adds $R_{50}$ to equation~\ref{eq:MH2_AV_Z} (see Table~\ref{tbl:fit_H2}) or just combines {\AV} and $R_{50}$ (equation~\ref{eq:MH2_AV_R}) gives a complementary gas mass estimate in such cases.

It is customary to use SFR as a predictor of gas content, especially for high-redshift studies \citep[e.g.,][]{Boselli+14b,Popping+15,Saintonge+17,Catinella+18}. As shown in Table~\ref{tbl:fit_H2_all} and Figure~\ref{fig:sfr_gas} in Appendix~\ref{app:fit_all}, adding SFR to equation~\ref{eq:MH2_AV_R} improves the fit significantly. This table also gives detailed information for various fits when one or more variables are missing or may not be desirable to include. All these combinations can predict gas mass within a factor of $\sim 3$.  Users should be careful not to use these relations in a circular manner. For example, if SFR is already used by a scaling relation to get $M_\mathrm{{H_2}}$, such a scaling relation should not be employed to study the correlation between $M_\mathrm{{H_2}}$ and SFR. Users should also take into account the quartile relations in the gas mass estimation, as there are systematic trends in the scatters of the scaling relations.

In summary, dust attenuation is significantly correlated with the molecular gas mass. A linear combination of {\AV} and metallicity or {\AV} and $R_{50}$ can indirectly estimate molecular gas masses within $\sim 2.5-3$ times the observed masses. If SFR is further used, $M_\mathrm{{H_2}}$ can be predicted to within a factor $\lesssim 2$. In this case, {\AV} and $R_{50}$ are the two best secondary parameters that improve the primary correlation between $M_\mathrm{{H_2}}$ and SFR.

\subsection{Predictors of Atomic Gas Mass}\label{sec:MHI}

For xGASS data \citep{Catinella+18}, the atomic gas mass can be predicted within a factor of $\sim 2.5$ from the SFR and half-light radius (equation~\ref{eq:MHI_SFR_R}). Table~\ref{tbl:fit_HI} in Appendix~\ref{app:fit_all} presents the result of fitting censored regression models to the xGASS data, describing the relationship among the atomic gas mass, SFR, the half-light radius, T-type, disk/feature probability, and/or dust attenuation. We present two sets of fits, by analyzing the central and satellite galaxies together, or by restricting the sample to central galaxies only. The classification of satellites and centrals is based on the galaxy group catalog of \citet{Yang+07}.  \noindent The median fit for the whole sample is

\begin{equation} \label{eq:MHI_SFR_R}
\log M_\mathrm{{HI}}  = \alpha + \beta_S \log \mathrm{SFR} + \beta_R \log R_{50},
\end{equation}

\noindent where ($\alpha$, $\beta_S$, $\beta_R$) = ($9.07 \pm 0.04$, $0.47 \pm 0.02$, $1.08 \pm 0.11$). The regression coefficients for central galaxies alone are consistent with those for the total sample. In comparison, the corresponding median relation for the molecular gas yields ($\alpha$, $\beta_S$, $\beta_R$) = ($8.77 \pm 0.05$, $0.81 \pm 0.03$, $0.46 \pm 0.11$). The molecular gas has weaker dependence on radius. Hence, the atomic-to-molecular ratio depends on $R_{50}$. Table~\ref{tbl:fit_HI} also gives three other scaling relations that can predict gas mass with accuracy of a factor $\sim 3$ without using SFR but combining $R_{50}$ with either T-type or disk/feature probability or dust attenuation.

For completeness, we give the relations for the total gas mass ($M_\mathrm{HI} + M_\mathrm{H_2}$), $\log M_\mathrm{gas} = \alpha + \beta \log R_{50} + \gamma \log \mathrm{SFR}$, using galaxies contained in both the xGASS and xCOLD GASS surveys.  The median relation yields $\alpha = 9.28 \pm 0.04$, $\beta = 0.87 \pm 0.11$, and $\gamma = 0.70 \pm 0.04$; for $q=0.15$, $\alpha = 8.83 \pm 0.05$, $\beta = 1.05 \pm 0.07$, and $\gamma = 0.88 \pm 0.02$; and for $q = 0.85$, $\alpha = 9.55 \pm 0.05$, $\beta = 0.85 \pm 0.11$, and $\gamma = 0.59 \pm 0.07$.

\begin{deluxetable*}{lcccc|cccc}
\tabletypesize{\footnotesize}
\tablecolumns{10} 
\tablewidth{0pt} 
\tablecaption{Regression model fits for molecular gas, $\log M_\mathrm{{H_2}}  = \alpha + \beta_A A_{V} +\beta_Z \log Z + \beta_R \log R_{50}$. \label{tbl:fit_H2}}
\tablehead{\multicolumn{5}{c}{Nebular $A_{V, \mathrm{Fiber}}$} & \multicolumn{4}{c}{Stellar $A_{V, \mathrm{Global}}$} \\
\colhead{} & \colhead{15\%} &  \colhead{Median} & \colhead{85\%} & \colhead{Mean} & \colhead{15\%} &  \colhead{Median} & \colhead{85\%} & \colhead{Mean}}
\startdata
 $\alpha$ & $ 7.70 \pm 0.07 $ & $ 8.31 \pm 0.04 $ & $8.91 \pm 0.06$ & $8.34 \pm 0.04$ &  $7.63 \pm 0.22$ & $8.04 \pm 0.09$ & $8.71 \pm  0.12$ & $8.12 \pm 0.05$\\
$\beta_A$ & $ 0.47 \pm 0.05$ &$0.47 \pm 0.02$ & $ 0.38 \pm 0.05$ & $0.43 \pm 0.03$ & $1.52 \pm 0.20$ & $1.58 \pm 0.13 $ & $1.39 \pm 0.27$ & $1.49 \pm0.10$\\
 Scatter & & 0.31 & & 0.50 & & 0.35 & & 0.49\\
 \hline
 $\alpha$  & $8.05 \pm 0.08 $ & $ 8.27 \pm 0.11$ & $8.56 \pm 0.07 $ &$8.30 \pm 0.05$ & $ 7.96 \pm 0.03$ & $8.23 \pm 0.12 $ & $ 8.59 \pm 0.11 $ & $ 8.21 \pm 0.06 $ \\
 $\beta_A$  & $0.25 \pm 0.10$ & $0.38 \pm 0.13$ & $0.38 \pm 0.04$ & $0.31 \pm 0.05$ & $ 1.10 \pm 0.06$ & $0.91 \pm 0.36$ & $ 0.91 \pm 0.30 $ & $0.90 \pm 0.13$  \\
 $\beta_Z$ & $ 1.29 \pm 0.36$ & $ 1.44 \pm 0.34 $ & $1.67 \pm 0.19$ & $1.56 \pm 0.20$ & --- & $1.30 \pm 0.31$ & $1.65 \pm 0.26$ & $1.46 \pm 0.20$\\
 Scatter & & 0.22 & & 0.37 & & 0.24 & & 0.35\\
 \hline
 $\alpha$ & $7.41 \pm 0.10 $ & $ 7.84 \pm 0.09$ & $8.49 \pm 0.10$ & $ 7.95 \pm 0.05$ &  $7.48 \pm 0.09 $ & $ 7.58 \pm 0.11$ & $8.16 \pm 0.05 $ &$7.70 \pm 0.06$ \\
 $\beta_A$ & $ 0.35 \pm 0.06 $ &$ 0.42 \pm 0.06 $ & $0.35 \pm 0.04$ & $0.37 \pm 0.03$ & $1.24 \pm 0.29$ & $1.41 \pm 0.13$ & $ 1.34 \pm 0.14$ & $1.34 \pm 0.08$\\
 $\beta_R$ & $1.34 \pm 0.28$ & $ 1.27 \pm 0.09$ & $0.89 \pm 0.12$ & $1.11 \pm 0.09$ & $ 0.97 \pm 0.09$ & $ 1.35 \pm 0.19 $ & $1.12 \pm 0.10$ & $1.19 \pm 0.08$ \\
 Scatter & &0.25 & & 0.41 & & 0.26 & & 0.38\\
  %% : both R50 and Z are important but the coefficients adjust since when one of the variables are missing, likely q=0.15 scatter depends on R and the q=0.85 on Z
\hline
 $\alpha$ & $7.87 \pm 0.07 $ & $ 8.07\pm 0.06$ & $8.56 \pm 0.07 $ &$8.16 \pm 0.06$ & $7.74 \pm 0.13$ &  $7.98 \pm 0.07$ & $8.41 \pm 0.18 $ & $ 8.02 \pm 0.07 $ \\
$\beta_A$  & $0.24 \pm 0.03$ & $0.35 \pm 0.06$ & $0.38 \pm 0.04$ & $0.31 \pm 0.05$ & $1.09 \pm 0.16$ & $0.91 \pm 0.11$ & $0.98 \pm 0.18$ & $ 0.96 \pm 0.12 $  \\
$\beta_Z$ & $0.95 \pm 0.35$ & $1.40 \pm 0.15 $ & $1.67 \pm 0.19$ & $1.29 \pm 0.20$ & --- & $0.98 \pm 0.23$ & $1.13 \pm 0.37$ & $1.02 \pm 0.20$ \\
$\beta_R$ & $0.63 \pm 0.12$ & $ 0.56 \pm 0.15$ & --- & $0.48 \pm 0.12$ & $0.63 \pm 0.18$ & $ 0.74 \pm 0.11$ & $ 0.50 \pm 0.22 $ & $0.62 \pm 0.12$ \\
 Scatter & & 0.21 & & 0.35& & 0.25 & & 0.33\\
 \enddata
\tablecomments{The coefficients of the equations describe the relationship among molecular gas mass ($M_\mathrm{{H_2}}$), dust attenuation ($A_V$), metallicity ($\log Z = 12 + \log {\rm [O/H]} - 8.8$), and/or galaxy radius ($R_\mathrm{{50}}$).  Fits are given for the attenuation in the fiber ($A_{V, \mathrm{fiber}}$) and galaxy-wide attenuation ($A_{V, \mathrm{global}}$). We give two measures of the scatter for the fits: the median absolute deviation (MAD, the left numbers corresponding to the median fits) and the root mean square deviation (RMSD, the right numbers corresponding to the mean fits).  $\operatorname{MAD} =\operatorname{median} (\mid Y_i-\tilde {Y}\mid$), where $Y_i = \log M_\mathrm{H_2}$ is the $i$th observed molecular gas mass and $\tilde{Y}$ is the fitted median relation. Similarly, $\operatorname{RMSD} = \sqrt{\operatorname{mean}( (Y_i - \bar{Y})^2)}$, where $\bar{Y}$ is the fitted mean relation. The fits that include $\log Z$ are based solely on strong emission-line starforming galaxies. They give reasonable prediction when the metallicity is estimated from the stellar mass for all galaxies (see Figure~\ref{fig:avf_gas}).}
\end{deluxetable*}

\section{DISCUSSION}

Dust is often used to estimate indirectly the gas content in galaxies \citep[e.g.,][]{Devereux+90}. We find that dust attenuation, {\AV}, is significantly correlated with the molecular gas mass, $M_\mathrm{H_2}$. A linear combination of {\AV} and metallicity or {\AV} and $R_{50}$ can indirectly estimate $M_\mathrm{H_2}$ within $\sim 2.5-3$ times the observed masses. However, the correlation between {\AV} and $M_\mathrm{HI}$ is weak. A combination of $R_{50}$ and SFR can give an estimate of $M_\mathrm{HI}$ accurate to within a factor of $\sim 3$. Next, we briefly discuss previous studies that aid the interpretation of our results. We surmise that the correlation of $M_\mathrm{{H_2}}$ with metallicity and $R_{50}$ likely comes from physical processes that determine the gas-to-dust ratio and molecular gas fraction.

The ratio of dust mass to total gas mass ($\xi = \mu M_\mathrm{H}/M_\mathrm{dust}$, where $M_\mathrm{H} =M_\mathrm{HI}+M_\mathrm{H_2}$ and the correction factor for helium $\mu \approx 1.4$) varies approximately linearly with metallicity ($\log \xi \propto \beta \log Z \propto \beta \log {\rm [O/H]}$) such that higher metallicity galaxies have lower $\xi$ \citep{Issa+90,Lisenfeld+98,Draine+07,Galametz+11,Leroy+11,Brinchmann+13, Berta+16, DeVis+19}. The amount of dust along a line of sight can be estimated using dust attenuation {\AV}, while the amount of gas is probed by the total hydrogen column density $N_\mathrm{H}$. {\AV}/$N_\mathrm{H}$ can then be used to infer the dust-to-gas ratio. It is known that {\AV}/$N_\mathrm{H}$ positively correlates with metallicity in nearby galaxies \citep[e.g.,][]{Kahre+18}. Therefore, it is natural to anticipate that the molecular gas mass depends on both {\AV} and metallicity. If $f_\mathrm{H_2} \equiv M_\mathrm{H_2}/M_\mathrm{H}$ and $R_\mathrm{H_2} \equiv M_\mathrm{H_2}/M_\mathrm{HI}$, then the molecular mass can be expressed as: 

\begin{equation}\label{eq:MH2_Md}
M_\mathrm{H_2} = M_\mathrm{H}\times M_\mathrm{H_2}/M_\mathrm{H} = \xi \mu^{-1} M_\mathrm{dust} f_\mathrm{H_2} = \xi \mu^{-1} M_\mathrm{dust}\,(1+ R_\mathrm{H_2}^{-1})^{-1}
\end{equation}

\noindent Because previous observations indicate $\xi \propto Z^{\beta_\xi}$ and $R_\mathrm{H_2} \propto Z^{\beta_{R_\mathrm{H2}}}$, $M_\mathrm{H_2}$ depends on both dust mass and metallicity. Depending on the metallicity calibration, $X_\mathrm{CO}$, and sample selection, $\beta_\xi \approx -0.7$ to $-2.5$ \citep{Issa+90, Munoz+09, Leroy+11,Remy-Ruyer+14, Janowiecki+18, DeVis+19} and $\beta_{R_\mathrm{H2}} \approx 2.7-3.4$ \citep{Boselli+14b}. Furthermore, we expect that $f_\mathrm{H_2}$ and $R_\mathrm{H_2}$ also depend on $R_{50}$. We showed in Section~\ref{sec:MHI} that $M_\mathrm{H_2}$ and $M_\mathrm{HI}$ scale differently with $R_{50}$. Based on the empirical correlation between mid-plane gas pressure of galactic disks and $R_\mathrm{H_2}$ \citep[e.g.,][]{Blitz+06}, \citet{Obreschkow+09} presented a phenomenological model that predicts that $R_\mathrm{H_2} \propto M_\mathrm{H}^{-0.24} \propto R_{50}^{-0.39}$. If $f_\mathrm{H_2} \propto R_{50}^{\beta^\prime_R} Z^{\beta^\prime_Z}$, then equation~\ref{eq:MH2_Md} implies that $\log M_\mathrm{H_2} \propto \log M_\mathrm{dust} + \beta_R^\prime \log R_{50}+ \beta_Z \log Z$, where $\beta_Z = \beta_\xi+\beta_Z^\prime$. For galaxies in the Herschel Reference Survey (HRS), \citet{Janowiecki+18} found $\beta_\xi= -0.72 \pm 0.19 $ and $\beta_Z  = 0.67 \pm 0.26$ (without including the $R_{50}$ term; see equations in their Section 2.2). In other words, they found a negative correlation between metallicity and gas-to-dust ratio defined using the total gas phase ($\xi$) or \ion{H}{1} ($\xi^\mathrm{HI} = M_\mathrm{HI}/M_\mathrm{dust}$), but a positive correlation using H$_2$ ($\xi^\mathrm{H_2} = M_\mathrm{H_2}/M_\mathrm{dust}$). Their $\xi^\mathrm{HI}-Z$ relationship has the highest correlation coefficient ($\abs{r} = 0.50 \pm 0.08$) while their relationship with $\xi$ has the smallest scatter ($\sigma=0.23$). The correlation of metallicity with $\xi^\mathrm{H_2}$ is weak ($r = 0.29 \pm 0.09$; see their Figure 1 for all three correlations). Likewise, \citet{Bertemes+18} found $\beta_Z = 0.12$ using $\beta_\xi = -0.85$ \citep{Leroy+11}. Our molecular gas scaling relations (e.g., equation~\ref{eq:MH2_AV_Z}) also imply a positive $\xi^\mathrm{H_2}-Z$ correlation. When {\AV} is estimated from H$\alpha$/H$\beta$ in the fiber and $A_V \ne 0$, we find a median relation $\log M_\mathrm{H_2} \propto \log A_V + \beta_Z \log Z$ with $\beta_Z = 1.11 \pm 0.26$. In equation~\ref{eq:MH2_AV_Z}, we used the term $\beta_AA_V$ in lieu of $\log$ {\AV} and found $\beta_Z = 1.44 \pm 0.34$. But $M_\mathrm{dust}$ likely depends not only on {\AV} (dust column or surface density) but also on galaxy radius. 

%For example, when {\AV} is estimated from H$\alpha$/H$\beta$ in the fiber, the median relation is $\log {\AV}/M_\mathrm{{H_2}} = a + b \times \log Z$, where $a = -8.9 \pm 0.1$, $b = -0.6 \pm 0.3$ and $\log Z = 12 + \log {\rm [O/H]} - 8.8}$. 

The radial profile of dust attenuation or dust surface mass density and the dependence of gas-to-dust ratio on galactocentric distance have been well-studied for local galaxies \citep[e.g.,][]{Issa+90, Boissier+04, Boissier+07, Munoz+09, Pappalardo+12,Sandstrom+13,Smith+16, Giannetti+17, Casasola+17,Chiang+18,Vilchez+19}. Accordingly, in most galaxies the attenuation decreases with the galactocentric distance. The quantity {\AV}/$N_\mathrm{H}$ (or $\xi^{-1}$) also decreases with radius: it is higher in the central, metal-rich regions than in the outer parts of galaxies \citep{Issa+90, Kahre+18}. By combining the signal of 110 spiral galaxies in HRS, \citet{Smith+16} detected dust emission out to about twice the optical radius. They found that the radial distribution of dust is consistent with an exponential, $\Sigma_d(r)  = \Sigma_0 10^{\alpha_r r}$, with a gradient of $\alpha_r = 1.7$ dex/$r_{25}$⁠. Here $r$ is the galactocentric radius and $r_{25}$ is the optical radius at the 25 mag arcsec$^{-2}$ isophote. Moreover, $\Sigma_d$ declines with radius at a similar rate to the stellar mass surface density but more slowly than the surface density of molecular gas or star formation rate \citep[][see Table 1 of the latter work]{Schruba+11,Bigiel+12, Smith+16}. \citet{Schruba+11} showed that the CO radial profiles have exponential distributions with a gradient of $\alpha_r = -2.2$ dex$/r_{25}$. Similarly, using galaxies in the H~I Nearby Galaxy Survey, \citet{Bigiel+12} showed that the total \ion{H}{1} and H$_2$ gas profiles of these galaxies follow exponential distributions beyond the inner 20\% of the optical radius with a gradient of $\alpha_r = -0.7$ dex$/r_{25}$. Because mass is the integral of the surface density profile, if the shape of the profile is ``universal," the gas and dust masses of (unresolved) galaxies should vary in a predictable way with the optical radius, consistent with what we found. The origin of the universal gas profile may be related to the halo angular momentum or the radial distribution of the accretion rate \citep[e.g.,][]{Mo+98,Kravtsov+13,Wang+14}.

For the profile of \citet{Smith+16}, $M_\mathrm{dust} = 2\pi \int^{2r}_0 \Sigma_d(r) r dr \approx 0.4 \Sigma_d(0) r_{25}^2$. Then, for $f_\mathrm{H_2} \propto R_{50}^{\beta^\prime_R} Z^{\beta^\prime_Z}$ and $\xi \propto Z^{\beta_\xi}$, equation~\ref{eq:MH2_Md} leads to $M_\mathrm{H_2} \propto \beta_A A_V + \beta_R \log R_{50}+ \beta_Z \log Z$. Now, $\beta_R$ is at least $\beta^\prime_R+2$ and $\beta_Z = \beta_\xi+\beta_Z^\prime$. The conversion of $\Sigma_d$ to {\AV} is not straightforward. It depends on radiative transfer effects of the relative geometric configuration of dust and ionized gas \citep{Calzetti+94}. \citet{Kreckel+13}, however, showed that empirical fits to {\AV} and $\Sigma_d$ data of local galaxies, with $\Sigma_d$ expressed in both linear and logarithmic form, can predict $\Sigma_d$ given {\AV} to within a factor of $\sim 3$. Because in our analysis (Table~\ref{tbl:fit_H2}) $\beta_R= 0.56 \pm 0.15$ and $\beta_Z =1.40 \pm 0.15$, equation~\ref{eq:MH2_Md} implies $f_\mathrm{H_2} \propto R_{50}^{\beta^\prime_R} Z^{\beta^\prime_Z}$ if the dust profile of our sample is an exponential, whose scale length is a fixed fraction of $R_{50}$. Note that combining equations in \citet{Janowiecki+18} for the $\xi^\mathrm{H_2}-Z$ and $\xi-Z$ relations gives $\xi^\mathrm{H_2}/\xi = f_\mathrm{H_2} \propto Z^{1.39 \pm 0.45}$.

Our molecular gas scaling relation likely reflects physics that governs the amount and spatial distribution of dust ($\xi \propto Z^{\beta_\xi},\, \beta_\xi < 0$) and molecular hydrogen ($f_\mathrm{H_2} \propto R_{50}^{\beta^\prime_R} Z^{\beta^\prime_Z}, \, \beta^\prime_Z > 0$ and $\beta^\prime_R < 0$) in the ISM. We believe that it can usefully constrain theoretical models of galaxy formation and evolution. The observed $\xi-Z$ relation has been useful to constrain dust evolution models \citep[e.g.,][]{Lisenfeld+98, Draine+07,Mattsson+12,Bekki+13,Remy-Ruyer+14,Zhukovska+14, Aoyama+17,Galliano+18, DeVis+19,Li+19,Hou+19}. The models range from analytical models to cosmological hydrodynamic galaxy simulations. They attempt to characterize the complex dust physics by including its production by stellar evolution, growth by accretion of metal in the ISM, destruction by supernova shocks and by thermal sputtering in hot gas, and its dependence on star formation history and gas inflows and outflows. The models are able to reproduce reasonably well the observed $\xi-Z$ relation. They indicate that dust growth in the ISM is crucial, and dust destruction by supernova may greatly influence the chemical evolution of galaxies. Dust-related physical processes, however, have only been self-consistently included in a few simulations. Studying the time evolution of dust and $f_\mathrm{H_2}$ in galaxies using simulations with a self-consistent model for the formation and evolution of dust and H$_2$,  \citet{Bekki+13} found that $f_\mathrm{H_2}$ rapidly increases in the first several Gyr of a star-forming galaxy because of more efficient dust production; it is higher in the inner regions of galaxies than in their outer parts. Regions with low $\xi$ (high dust-to-gas ratio) likely have higher $f_\mathrm{H_2}$. The author also broadly reproduced (with a systematic offset) the observed $M_\mathrm{H_2} \propto M_\mathrm{dust}^{0.77}$ relation, which is based on dust emission of only 35 galaxies in the Virgo cluster \citep{Corbelli+12}. \citet{Bekki+15} suggested that the evolution of dust distributions driven by stellar radiation pressure is important for the evolution of SFR, $f_\mathrm{H_2}$, and $Z$ in galaxies. 

\begin{figure*}
\includegraphics[scale=0.3]{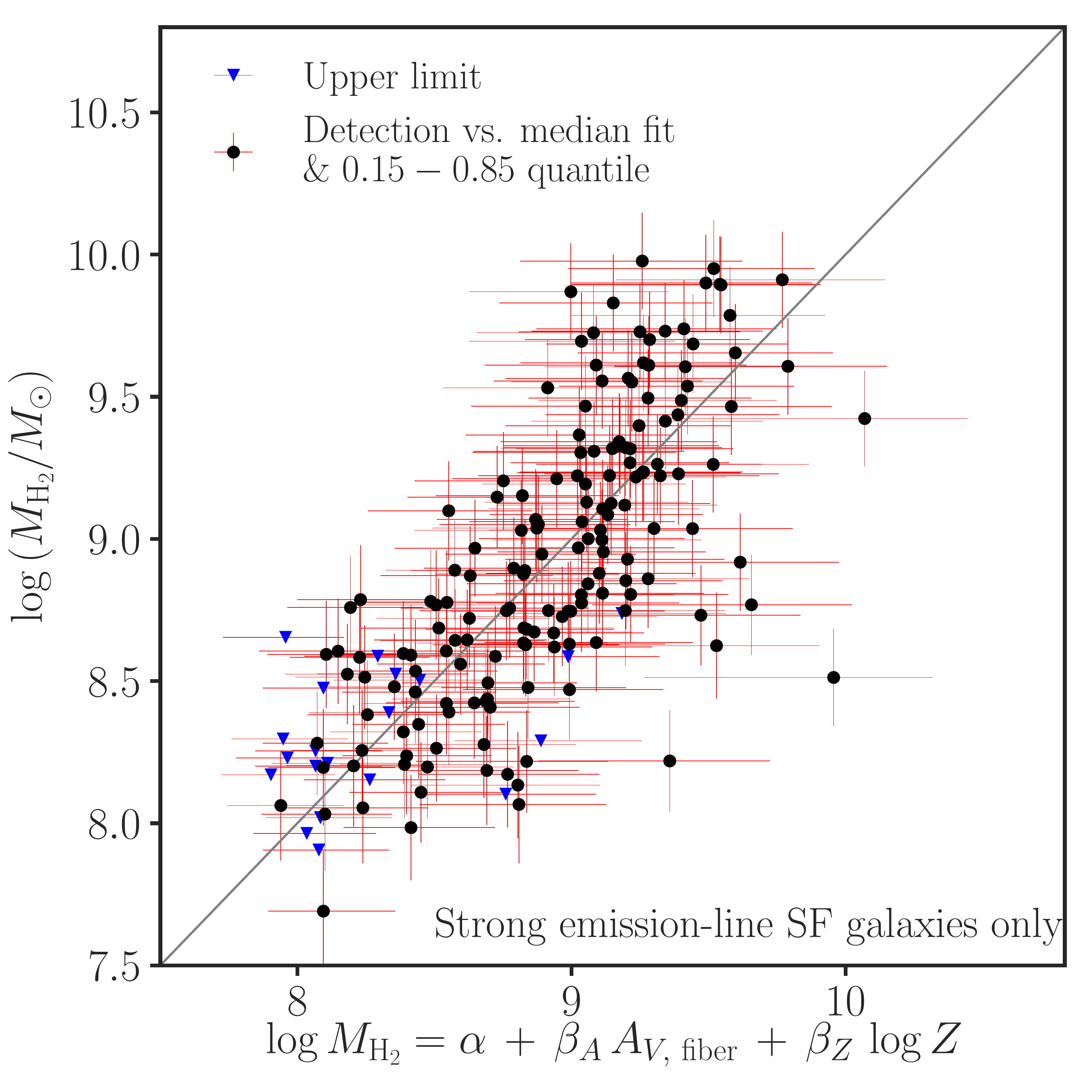}
\includegraphics[scale=0.3]{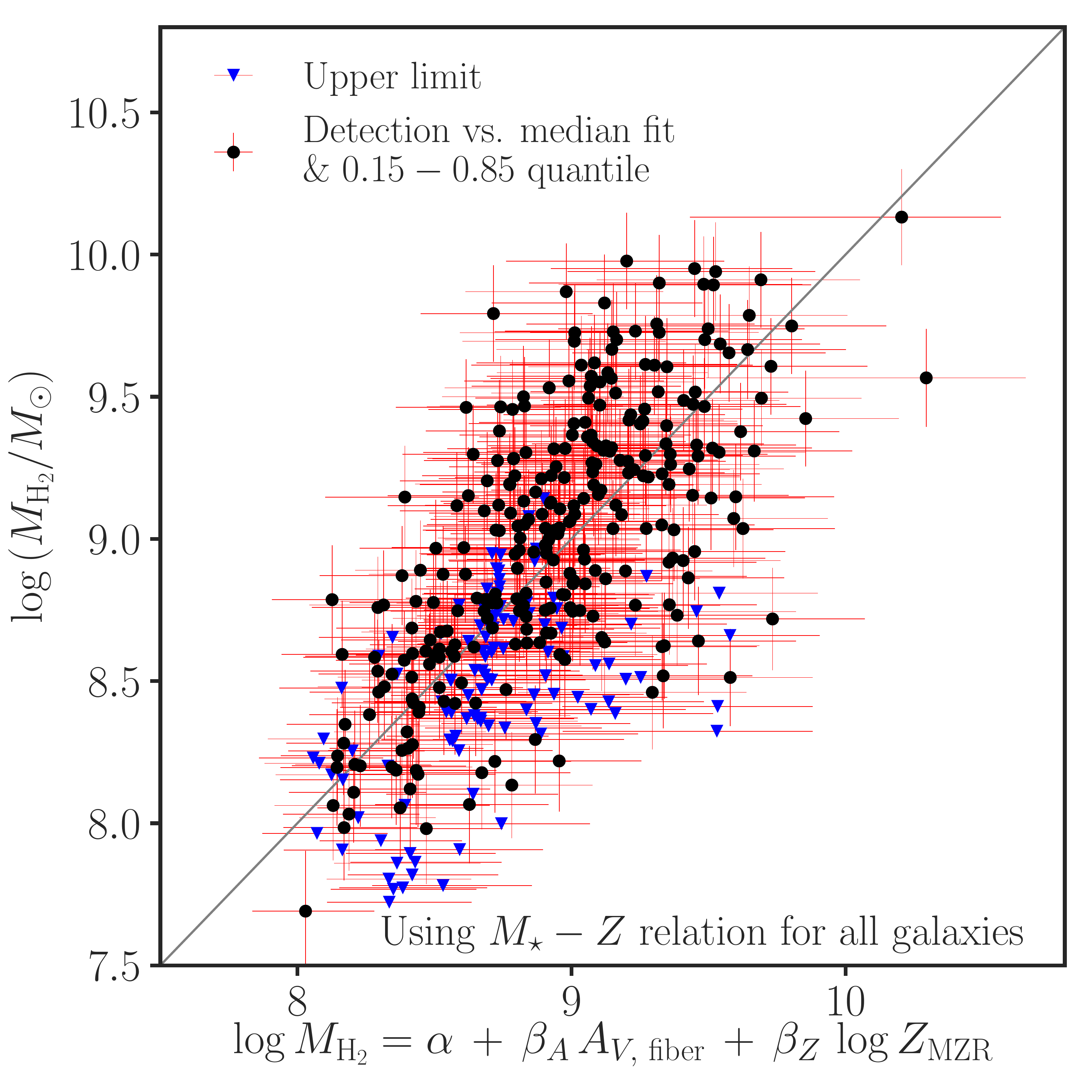}
\includegraphics[scale=0.3]{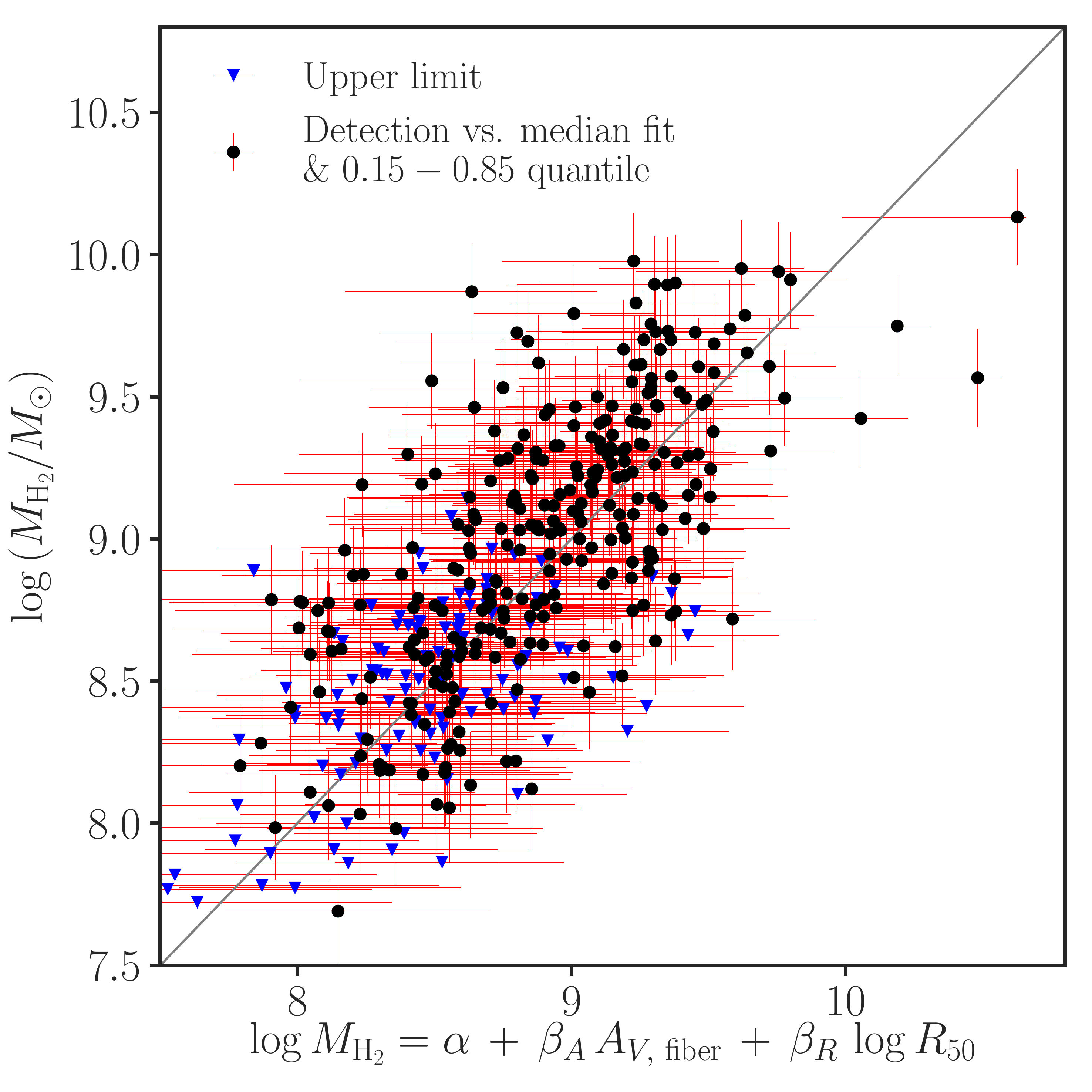}
\hfill
\includegraphics[scale=0.3]{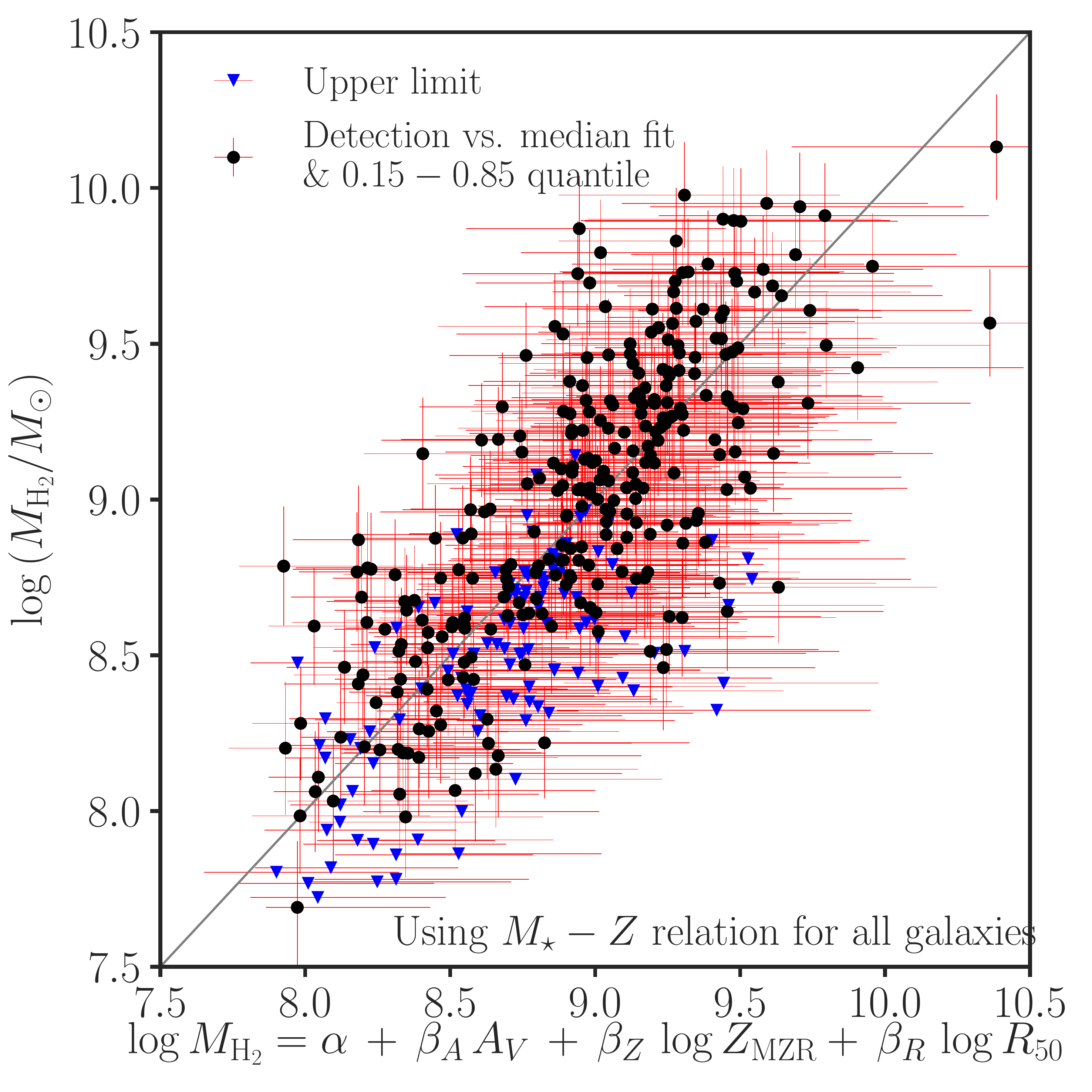}
\caption{Scaling relations among molecular gas mass ($M_\mathrm{H_2}$), fiber {\AV}, gas-phase metallicity ($Z$), and half-light radius ($R_{50}$). The top left panel compares the fitted (predicted) trivariate relation among $M_\mathrm{{H_2}}$, {\AV}, and $Z$ for star-forming galaxies with strong emission lines (see Table~\ref{tbl:fit_H2}). The top right panel shows the same fit applied to the whole sample using metallicity estimates from the mass-metallicity relation (Tremonti et al. 2004). The (black) points denote galaxies with detections, and the (blue) triangles indicate gas mass upper limits, which are included in our analysis. The x-axis positions of the detections and upper limits correspond to the median predictions; the (red) error bars show the predicted $0.15 - 0.85$ quantile ranges in the x-axis and the measurement errors of the gas mass in the y-axis. The diagonal (gray) lines are the 1:1 relation. The bottom left panel shows the fitted trivariate relation of $M_\mathrm{{H_2}}$, {\AV}, and $R_{50}$ for the whole sample. The bottom right panel shows how the prediction from fitting data of the star-forming galaxies generalizes to the whole sample, using metallicity estimates from the mass-metallicity relation. In summary, the figure shows that $M_\mathrm{{H_2}}$ can be estimated with accuracy of a factor of $\sim 2.5$ by combining {\AV} (Balmer decrement) with $Z$ or $R_{50}$ or both. \label{fig:avf_gas}}
\end{figure*}

\section{SUMMARY AND CONCLUSIONS} \label{sec:conc}

We analyze the atomic and molecular gas masses of local representative galaxies ($M_\star = 10^{9}-10^{11.5}$ \msun\, and $z = 0.01 - 0.05$) from the xGASS and xCOLD GASS surveys \citep{Saintonge+17,Catinella+18} in conjunction with a number of galaxy properties derived from SDSS. The central result of this work is that we have discovered remarkable new empirical relations that allow us to predict gas masses easily from galaxy survey data.  Using different statistical approaches (partial correlation and censored quantile regression analyses), we provide useful and convenient formulae in Tables~\ref{tbl:fit_H2}, \ref{tbl:fit_H2_all}, and \ref{tbl:fit_HI} that summarize the median, mean, and $0.15/0.85$ quantile multivariate relationships between gas mass and other galaxy properties that correlate best with it.

Our main conclusions are as follows: 

\begin{itemize}

\item The dust attenuation {\AV}, of both the stellar continuum and gas emission, is significantly correlated with the molecular gas mass (Kendall's rank coefficient $\tau = 0.4$). After controlling for the effect of the SFR, the dust attenuation is still correlated with the molecular gas. Without using SFR, $M_\mathrm{{H_2}}$ can be indirectly estimated (within a factor of $\sim 2.5$) using {\AV} and gas-phase metallicity ($Z$) or {\AV} and $R_{50}$ (Figures~\ref{fig:avf_gas} and \ref{fig:avg_gas} and Tables~\ref{tbl:corr} and \ref{tbl:fit_H2}). In contrast, the correlation between dust attenuation and the atomic gas mass is weak and may be explained by other covariates such as SFR. This result is theoretically expected since in poorly dust-shielded regions molecular hydrogen is dissociated by the far-ultraviolet photons and the atomic gas dominates.

\item The galaxy half-light radius $R_{50}$ is significantly correlated with both molecular and atomic gas (Kendall's $\tau = 0.3$). The correlation is still significant after controlling for the effect of SFR or {\AV}. In practice, $R_{50}$ can be used with {\AV} or SFR to predict the atomic and molecular gas mass within a factor of $\sim 2-3$ for large samples of galaxies and AGNs (detailed equations are given in Tables~\ref{tbl:fit_H2}, \ref{tbl:fit_H2_all}, and \ref{tbl:fit_HI}). 

\item Depending on how it is quantified, galaxy morphology may be correlated with the atomic gas. In particular, the disk/feature probability and the T-type show significant correlation with the atomic gas after accounting for the effect of SFR or $R_{50}$. Combining these morphology indicators with $R_{50}$ (and/or {\AV}) may be useful in predicting atomic gas masses (Table~\ref{tbl:fit_HI}), especially for future AGN studies, as they are easier to measure than the SFR. On the other hand, the merger probability and the edge-on probability have little or no significance in explaining the variability in the gas masses of the xGASS sample. The correlations between the molecular gas and all morphology indicators are insignificant, or weak after accounting for the effects of SFR or $R_{50}$.  

\end{itemize}

As this work was nearing completion, we became aware of a similar effort by \citet{Concas+19}, who proposed an empirical scaling relation between Balmer decrement and molecular gas mass. They reported a significant correlation between the Balmer decrement and $M_\mathrm{{H_2}}$ in a subsample of 222 star-forming galaxies, also from the xCOLD GASS survey. Similar to what we found in Table~\ref{tbl:corr} using the edge-on probability, they showed that star-forming galaxies with high disk inclination angles tend to exhibit high H$\alpha$/H$\beta$ ratios, for the same $M_\mathrm{{H_2}}$, compared to their less inclined counterparts. After correcting for the inclination effect, they did not find residual dependence on galaxy size or star formation rate. This is contrary to what we present in Table~\ref{tbl:corr} for the total xCOLD GASS sample. We show that $M_\mathrm{{H_2}}$ is primarily correlated with SFR and secondarily with {\AV}. Our analysis presented in Appendix~\ref{app:fit_inc} shows that the partial correlation of the inclination angle, $i$, with $M_\mathrm{{H_2}}$ at fixed $\mathrm{H\alpha/H\beta}$ is much weaker than the corresponding partial correlation of SFR. Likewise, fitting together $M_\mathrm{H_2}$, SFR, $i$, and $\log \mathrm{H\alpha/H\beta}$ indicates that the inclination angle does not bring additional information, once SFR and $\log \mathrm{H\alpha/H\beta}$ are used. Using information criteria AIC/BIC and the RMS of the residuals, we conclude that the combination of {\AV} and $Z$ or {\AV} and $R_{50}$ is a better predictor of molecular gas than the combination of {\AV} and $i$. Unlike the combination of H$\alpha$/H$\beta$ and $i$ proposed by \citet{Concas+19}, our gas scaling relations work well for all galaxies, star-forming or not, regardless of whether {\AV} is measured from H$\alpha$/H$\beta$ or stellar continuum absorption. 

We note that our atomic and molecular gas scaling relations predict gas masses consistent with independent gas data of galaxies in the Herschel Reference Survey compiled by \citet[][see Appendix~\ref{app:hrs}]{Boselli+14a}.

\acknowledgements

We thank the anonymous referee for helpful comments and suggestions. This work was supported by the National Science Foundation of China (11721303) and the National Key R\&D Program of China (2016YFA0400702).

\appendix

\section{Checking the Effect of the $\alpha_\mathrm{CO}$ Assumption}\label{app:xco}

%XX should we mention which value of Galactic X_CO you used, and give reference?
% Done
Table~\ref{tbl:fit_H2_XCO} repeats Table~\ref{tbl:fit_H2} for the constant Galactic CO conversion factor $\alpha_\mathrm{CO} = 4.35\, \mathrm{M_\odot \,pc^{-2} \, (K\, km\,s^{-1})^{-1}}$ \citep[e.g.,][]{Bolatto+13}. The results in the two tables are not significantly different. In other words, there is a significant relationship among molecular gas, metallicity, radius, and dust absorption, $\log M_\mathrm{{H_2}} \propto \beta_A A_{V} +\beta_R \log R_{50} + \beta_Z \log Z$, whether the metallicity-dependent or the constant $\alpha_\mathrm{CO}$ is used. Although the difference is not statistically significant, $\beta_Z$ tend to be higher in the latter case. By definition, the absolute amount of molecular gas depends on $\alpha_\mathrm{CO}$. Thus, the intercepts of the scaling relations change with the definition. They are lower for the constant $\alpha_\mathrm{CO}$, as expected.
 
\begin{deluxetable*}{lcccc|cccc}
\tabletypesize{\footnotesize}
\tablecolumns{10} 
\tablewidth{0pt} 
\tablecaption{Regression model fits for molecular gas using constant $X_\mathrm{CO}$, $\log M_\mathrm{{H_2}}  = \alpha + \beta_A A_{V} +\beta_Z \log Z + \beta_R \log R_{50}$. \label{tbl:fit_H2_XCO}}
\tablehead{\multicolumn{5}{c}{Nebular $A_{V, \mathrm{Fiber}}$} & \multicolumn{4}{c}{Stellar $A_{V, \mathrm{Global}}$} \\
\colhead{} & \colhead{15\%} &  \colhead{Median} & \colhead{85\%} & \colhead{Mean} & \colhead{15\%} &  \colhead{Median} & \colhead{85\%} & \colhead{Mean}}
\startdata
$\alpha$ & $7.62 \pm 0.06 $ & $ 8.16 \pm 0.03 $ & $8.99 \pm 0.08$ & $8.26 \pm 0.05$ &  $7.45 \pm 0.04$ & $7.92 \pm 0.10$ & $8.84 \pm  0.08$ & $8.05 \pm 0.06$\\
$\beta_A$ & $ 0.54 \pm 0.06$ &$0.63 \pm 0.08$ & $0.44 \pm 0.06$ & $0.53 \pm 0.04$ & $1.85 \pm 0.10$ & $1.91 \pm 0.16 $ & $1.33 \pm 0.13$ & $1.74 \pm 0.12$\\
 Scatter & & 0.38 & & 0.58 & & 0.45 & & 0.56\\
 \hline
 $\alpha$ & $7.89 \pm 0.14 $ & $ 8.01 \pm 0.11$ & $8.41 \pm 0.02 $ &$8.11 \pm 0.05$ & $7.73 \pm 0.25$ & $7.98 \pm 0.05$ & $8.32 \pm 0.08$ & $7.98 \pm 0.06 $ \\
 $\beta_A$  & $0.20 \pm 0.16$ & $0.45 \pm 0.10$ & $0.35 \pm 0.05$ & $0.32 \pm 0.05$ & $1.05 \pm 0.63$ & $1.01 \pm 0.08$ & $0.99 \pm 0.18$ & $1.00 \pm 0.13$  \\
 $\beta_Z$ & $ 2.46 \pm 0.76$ & $2.43 \pm 0.42 $ & $2.74 \pm 0.14$ & $2.61 \pm 0.20$ & $1.87 \pm 0.47$ & $2.48 \pm 0.16$ & $2.74 \pm 0.14$ & $2.52 \pm 0.20$\\
 Scatter & & 0.23 & & 0.37 & & 0.24 & & 0.34\\
 \hline
 $\alpha$ & $7.15 \pm 0.22$ & $7.70 \pm 0.07$ & $8.30 \pm 0.05$ & $7.76 \pm 0.06$ & $7.07 \pm 0.16 $ & $7.51 \pm 0.06$ & $8.01 \pm 0.14 $ &$7.52 \pm 0.06$ \\
 $\beta_A$ & $ 0.44 \pm 0.10 $ &$0.50 \pm 0.05$ & $0.42 \pm 0.07$ & $0.45 \pm 0.03$ & $1.80 \pm 0.17$ & $1.48 \pm 0.12$ & $1.48 \pm 0.27$ & $1.52 \pm 0.09$\\
 $\beta_R$ & $1.70 \pm 0.55$ & $1.51 \pm 0.13$ & $1.27 \pm 0.08$ & $1.43 \pm 0.10$ & $1.29 \pm 0.23$ & $1.57 \pm 0.09$ & $1.46 \pm 0.18$ & $1.52 \pm 0.09$ \\
 Scatter & &0.27 & & 0.45 & & 0.24 & & 0.41\\
  %%YY : both R50 and Z are important but the coefficients adjust since when one of the variables are missing, likely q=0.15 scatter depends on R and the q=0.85 on Z
\hline
 $\alpha$ & $7.65 \pm 0.16 $ & $7.90\pm 0.09$ & $8.25 \pm 0.07$ &$7.93 \pm 0.06$ & $7.55 \pm 0.07$ &  $7.80 \pm 0.03$ & $8.07 \pm 0.13 $ & $7.74 \pm 0.07$ \\
$\beta_A$  & $0.26 \pm 0.10$ & $0.36 \pm 0.10$ & $0.34 \pm 0.07$ & $0.32 \pm 0.05$ & $1.06 \pm 0.10$ & $0.92 \pm 0.06$ & $0.93 \pm 0.37$ & $1.07 \pm 0.11 $  \\
$\beta_Z$ & $2.02 \pm 0.19$ & $2.01 \pm 0.48$ & $2.23 \pm 0.36$ & $2.23 \pm 0.20$ & $1.45 \pm 0.29$ & $2.14 \pm 0.11$ & $2.44 \pm 0.40$ & $1.94 \pm 0.19$ \\
$\beta_R$ & $0.73 \pm 0.44$ & $0.69 \pm 0.19$ & $0.56 \pm 0.12$ & $0.63 \pm 0.12$ & $0.74 \pm 0.29$ & $0.72 \pm 0.08$ & $0.73 \pm 0.18 $ & $0.77 \pm 0.11$ \\
 Scatter & & 0.21 & & 0.34& & 0.21 & & 0.30\\
 \enddata
\tablecomments{Similar to Table~\ref{tbl:fit_H2} except here we adopt the constant Galactic CO-to-H2 conversion factor $\alpha_\mathrm{CO} = 4.35 \, \mathrm{M_\odot \,pc^{-2} \, (K\, km\,s^{-1})^{-1}}$}
\end{deluxetable*}

\section{Adding SFR, Stellar Mass, and Galaxy Morphology as Gas Mass Predictors} \label{app:fit_all}

\begin{figure*}
\includegraphics[scale=0.3]{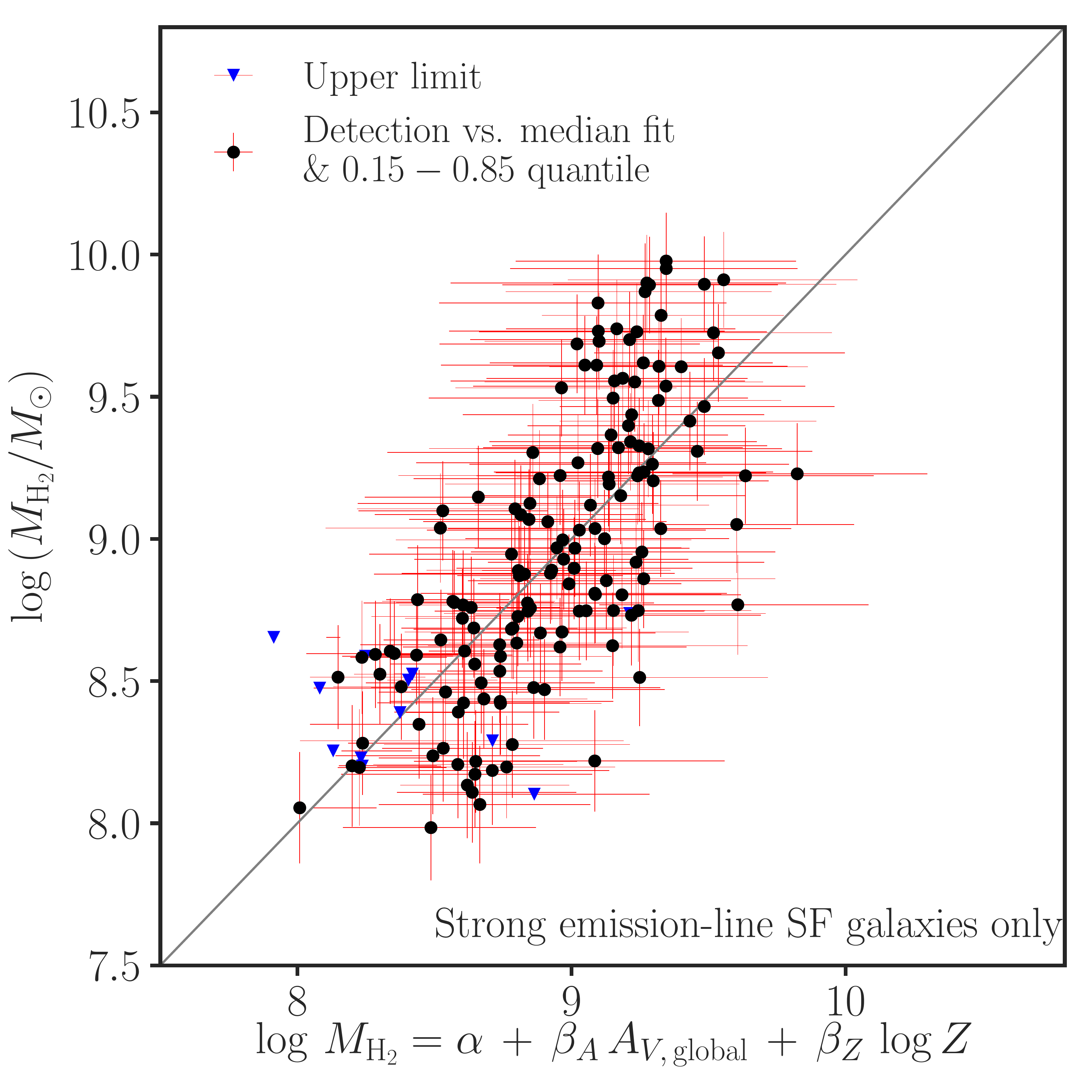}
\includegraphics[scale=0.3]{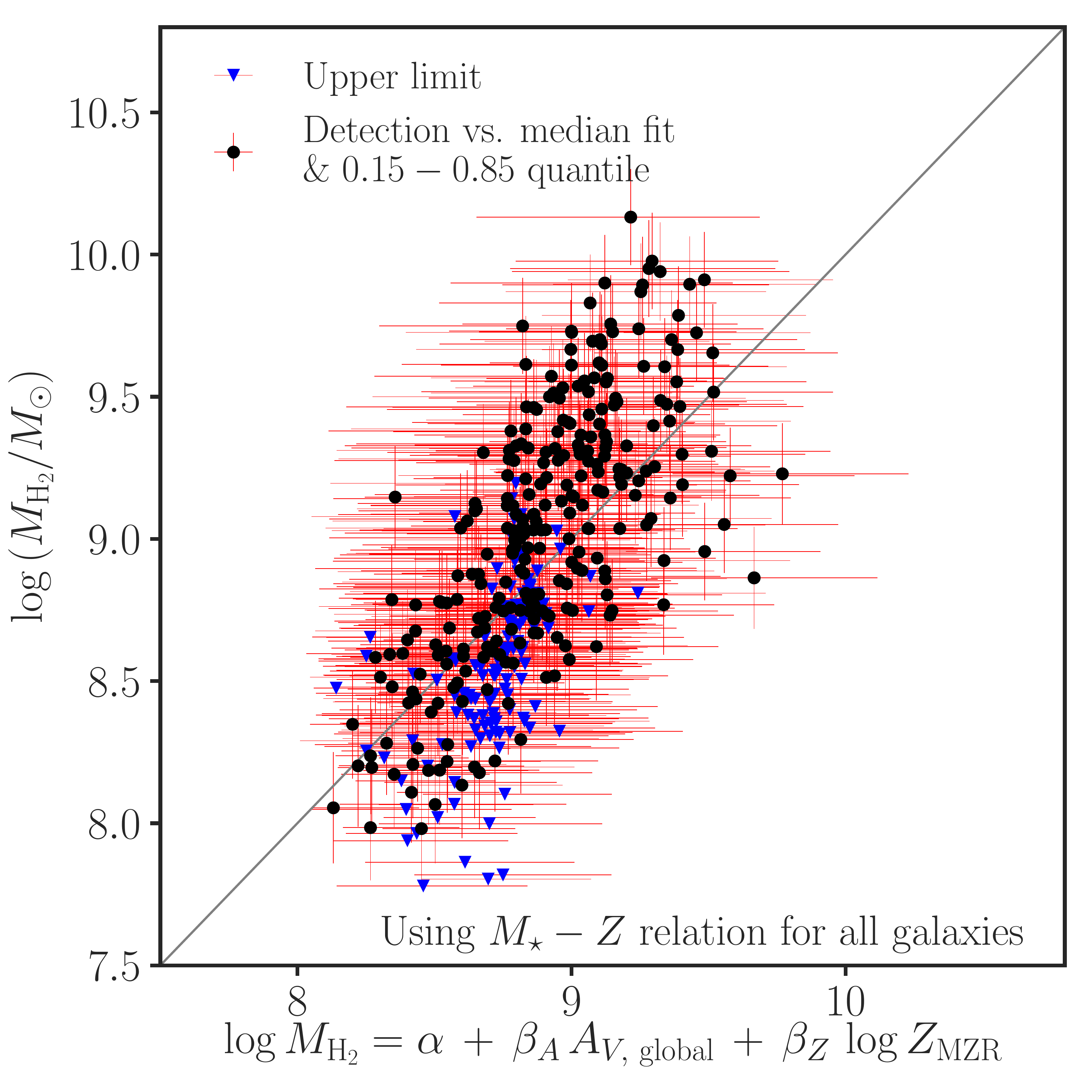}
\includegraphics[scale=0.3]{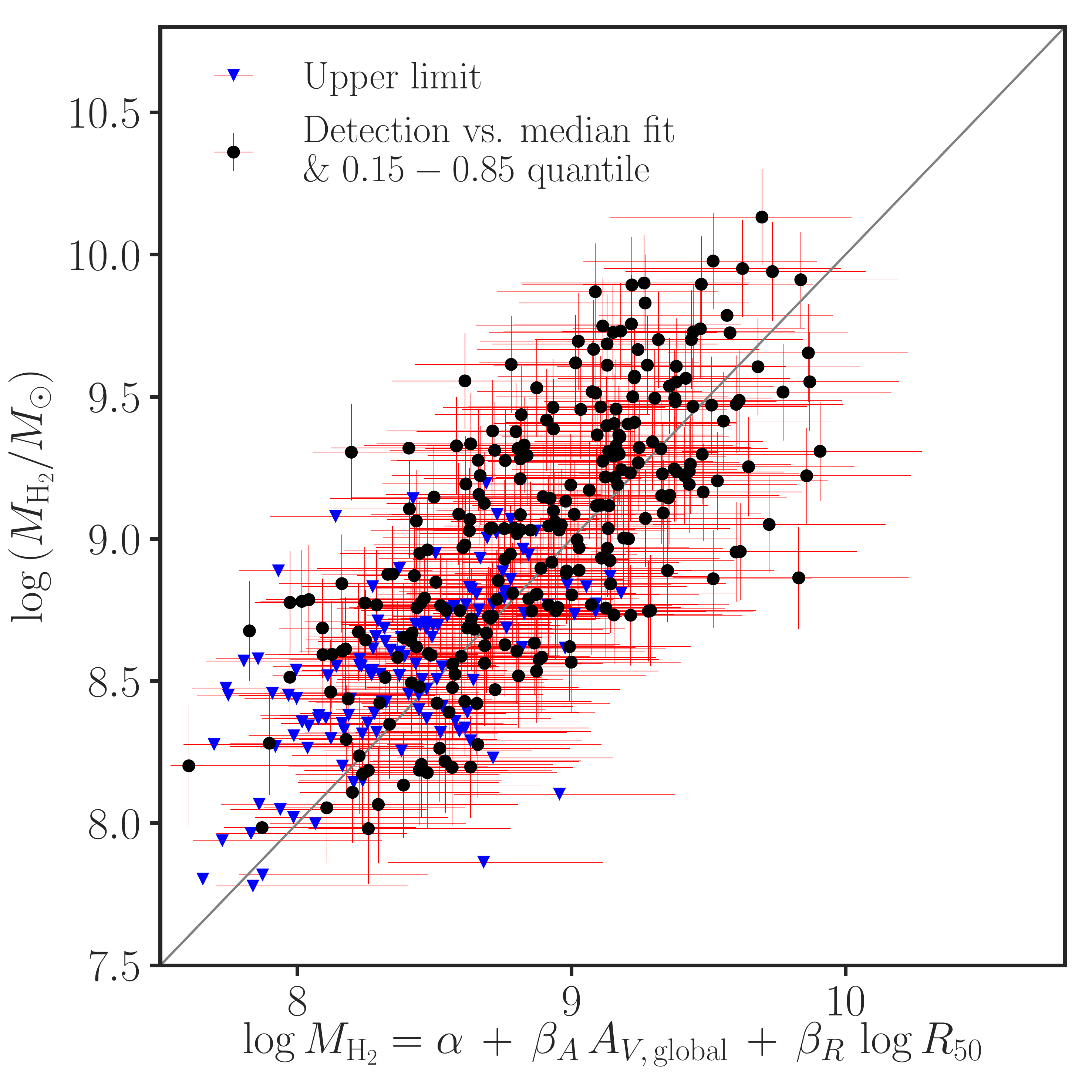}
\hfill
\includegraphics[scale=0.3]{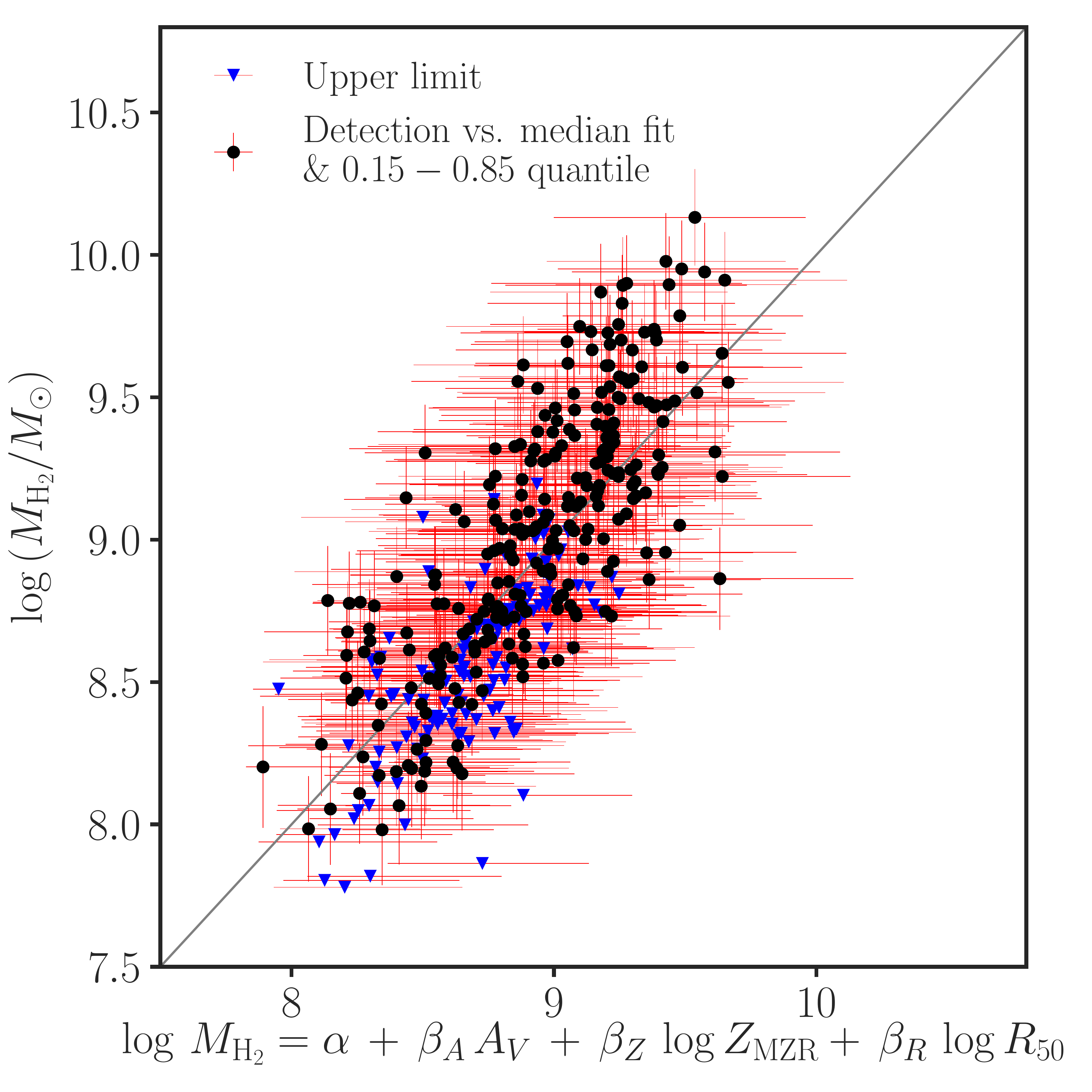}
\caption{Similar to Figure~\ref{fig:avf_gas}, but here galaxy-wide stellar {\AV} is used. \label{fig:avg_gas}}
\end{figure*}

\begin{figure*}
\includegraphics[scale=0.3]{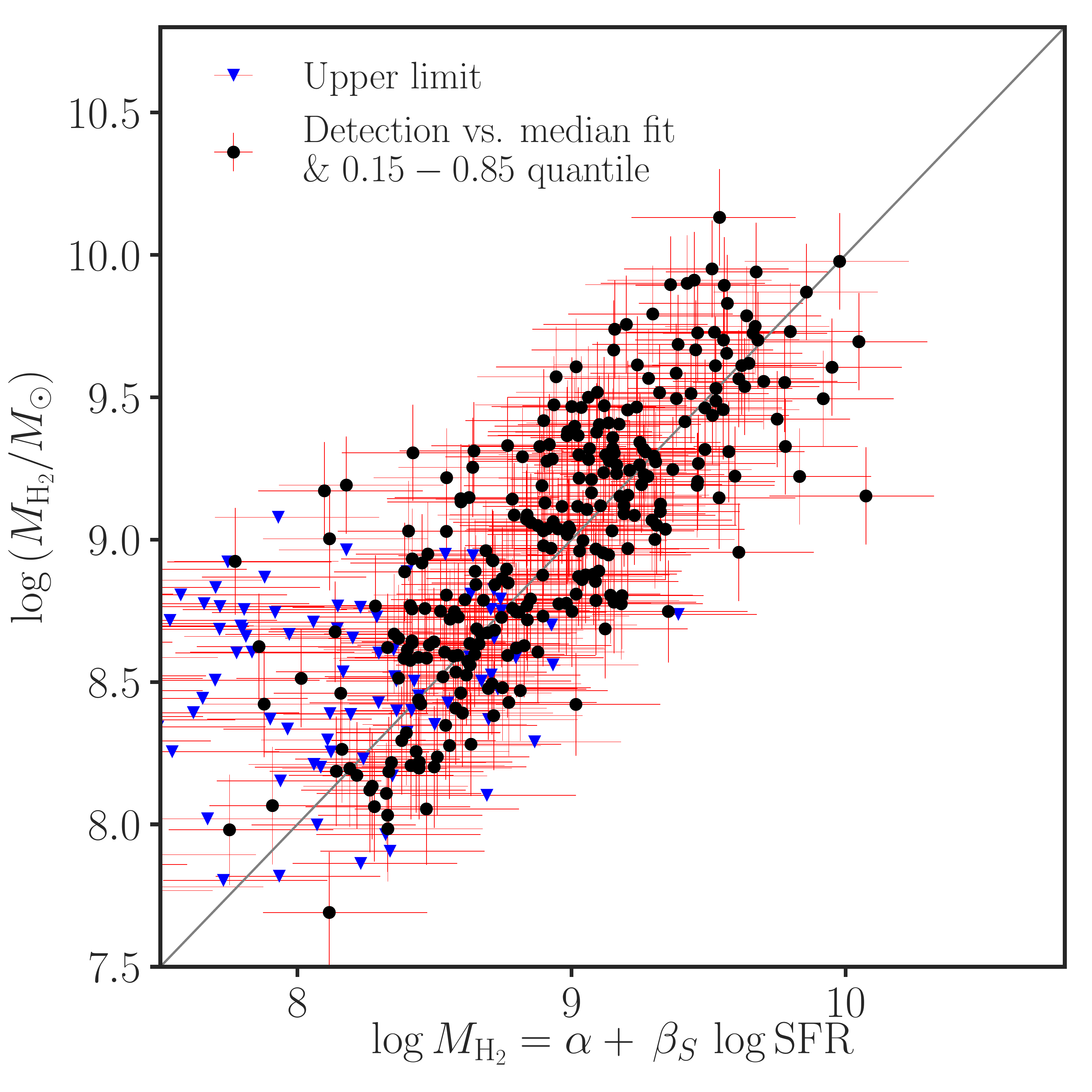}
\includegraphics[scale=0.3]{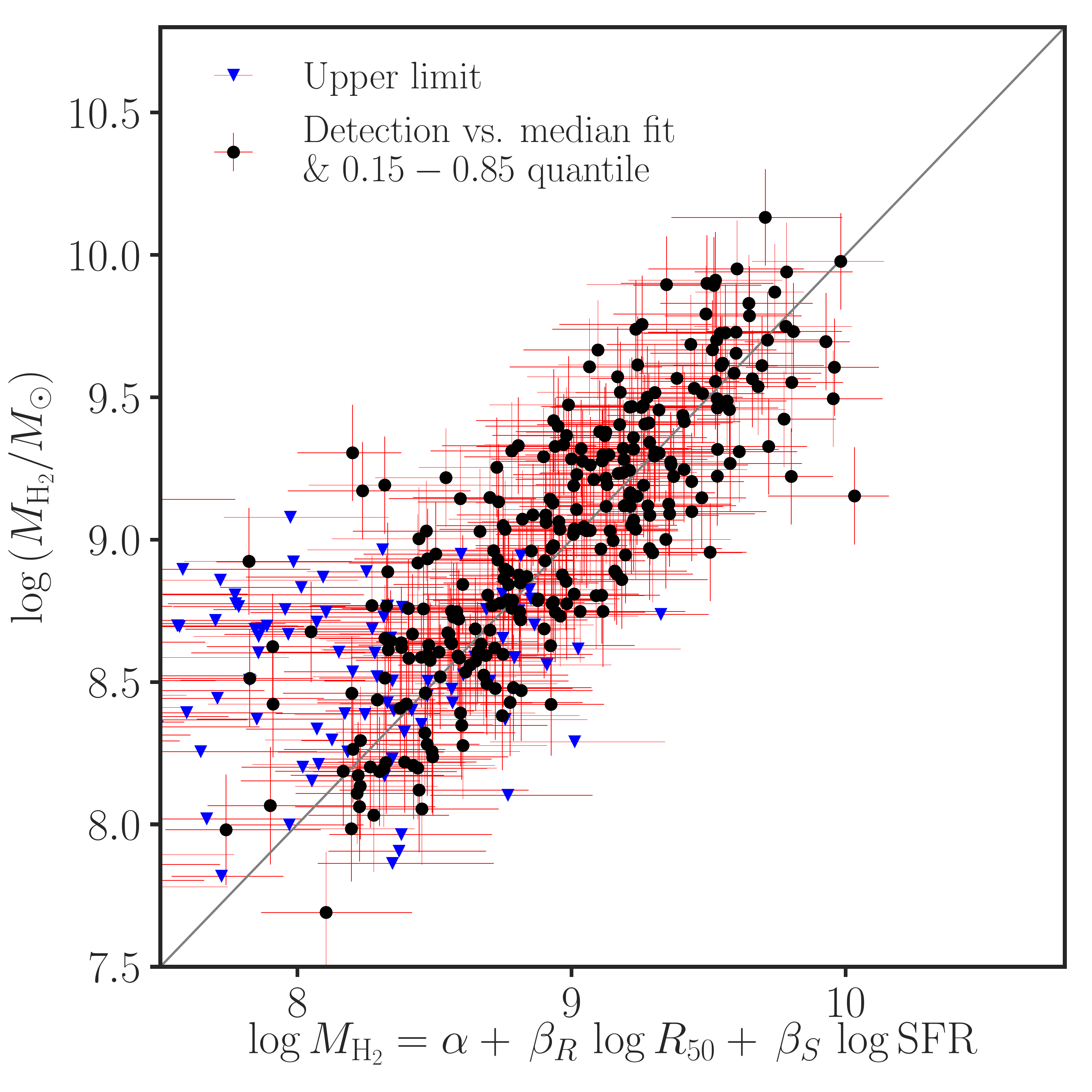}
\includegraphics[scale=0.3]{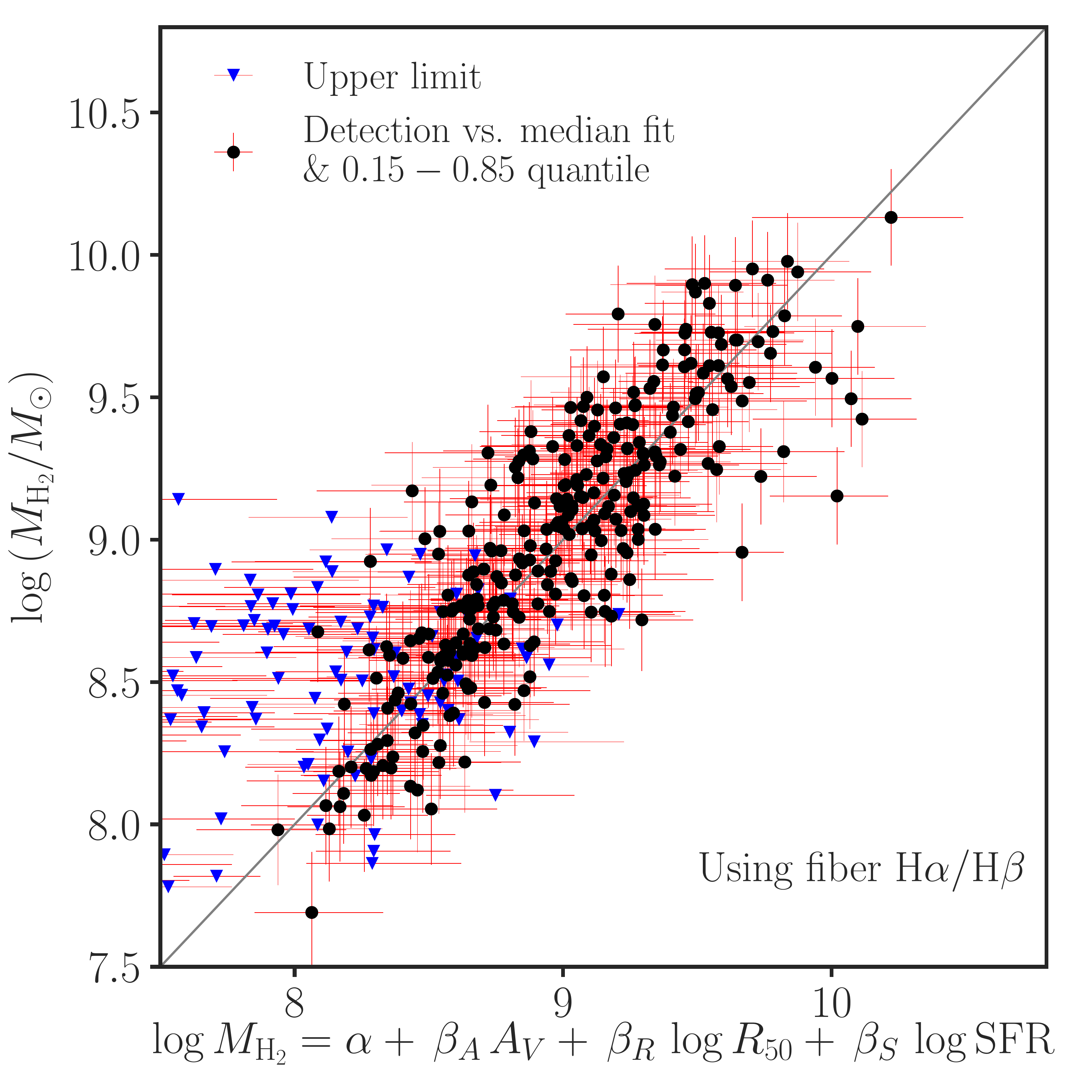}
\includegraphics[scale=0.3]{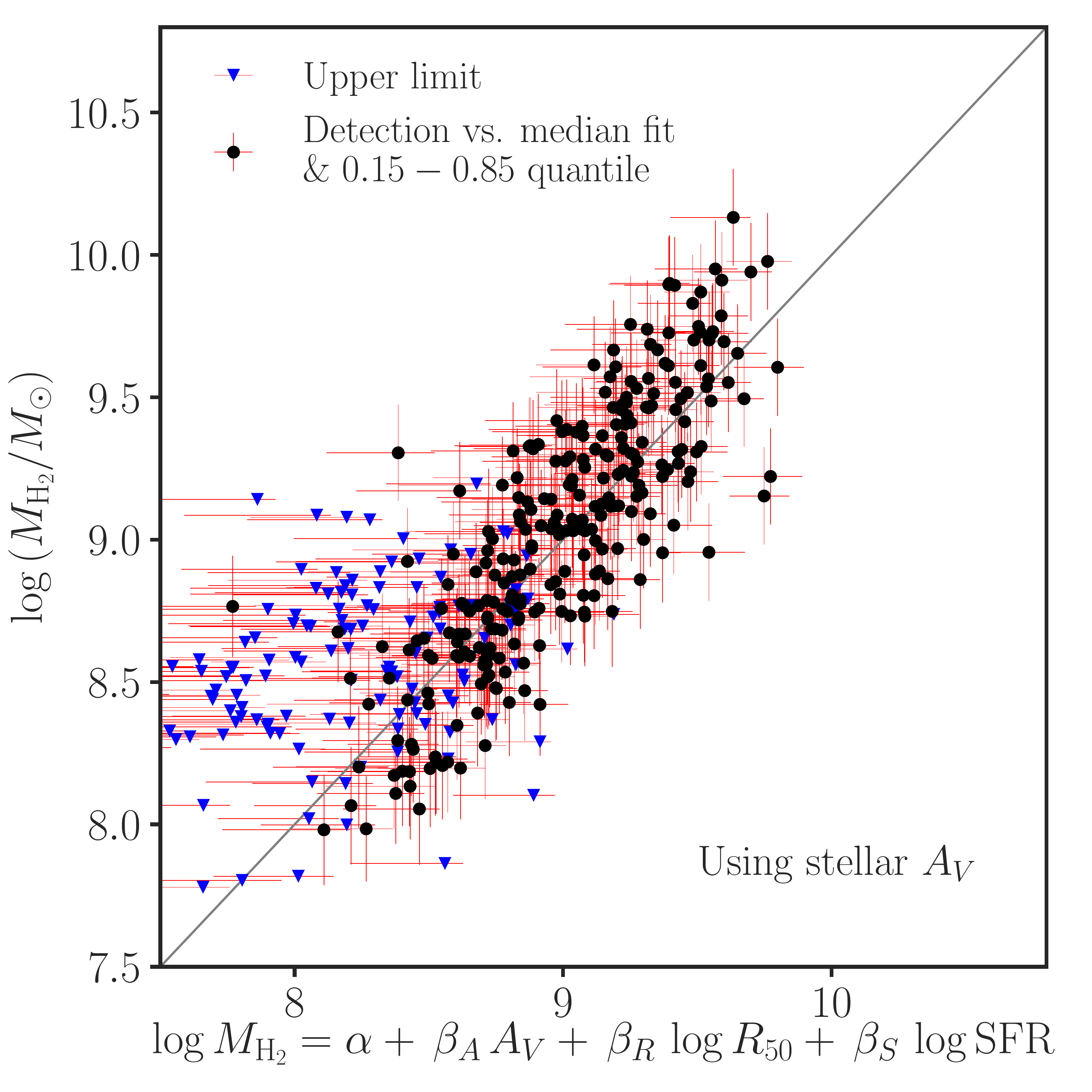}
%\includegraphics[scale=0.29]{lgMH2_AVf_SFR_s1.pdf}
%\hfill
%\includegraphics[scale=0.29]{lgMH2_AVg_SFR_s1.pdf}
\caption{Scaling relations among molecular gas mass ($M_\mathrm{{H_2}}$), {\AV}, $R_{50}$, and SFR. Adding $R_{50}$ or {\AV} significantly improves to the primary correlation between $M_\mathrm{{H_2}}$ and SFR. \label{fig:sfr_gas}}
\end{figure*}

\begin{figure*}
\includegraphics[scale=0.3]{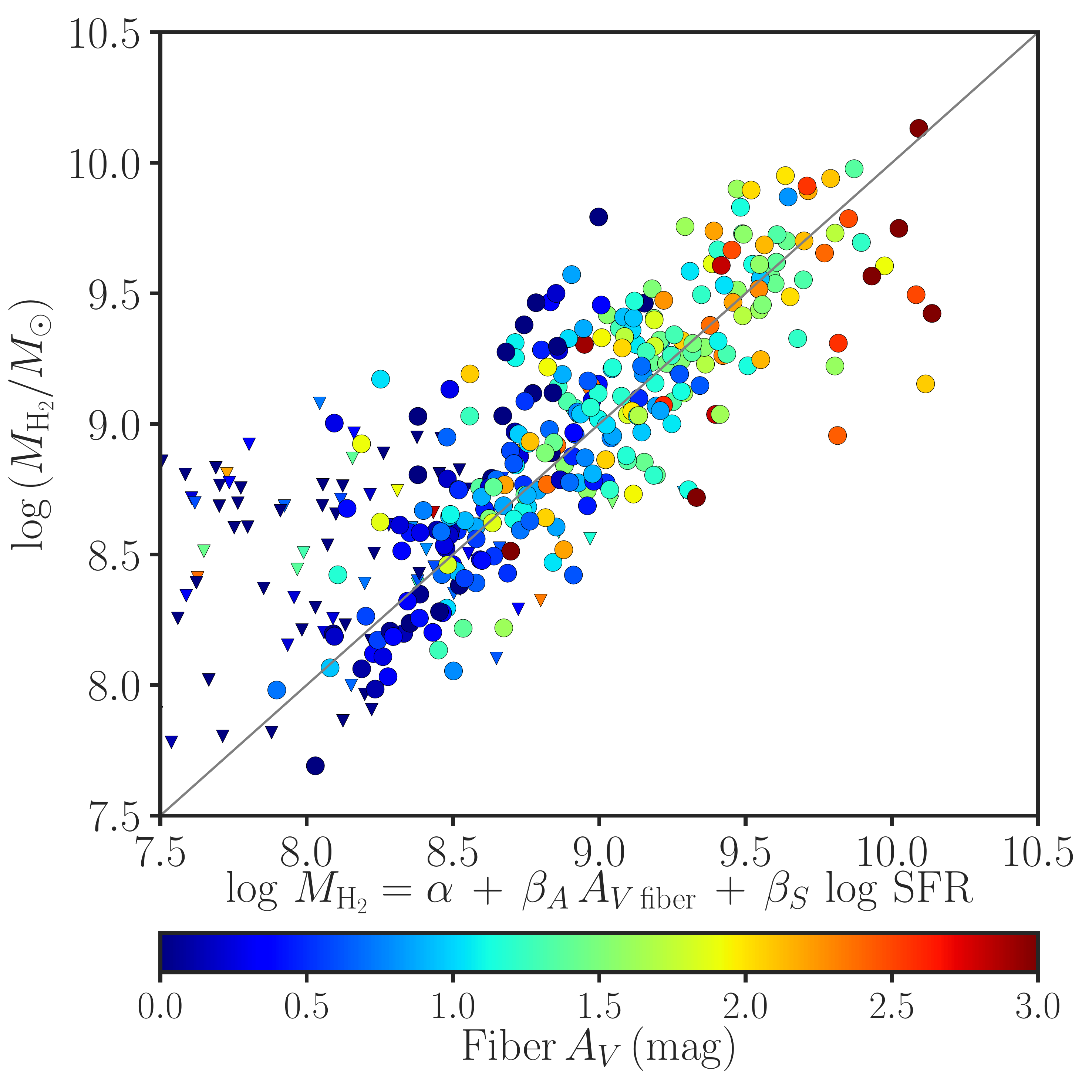}
\includegraphics[scale=0.3]{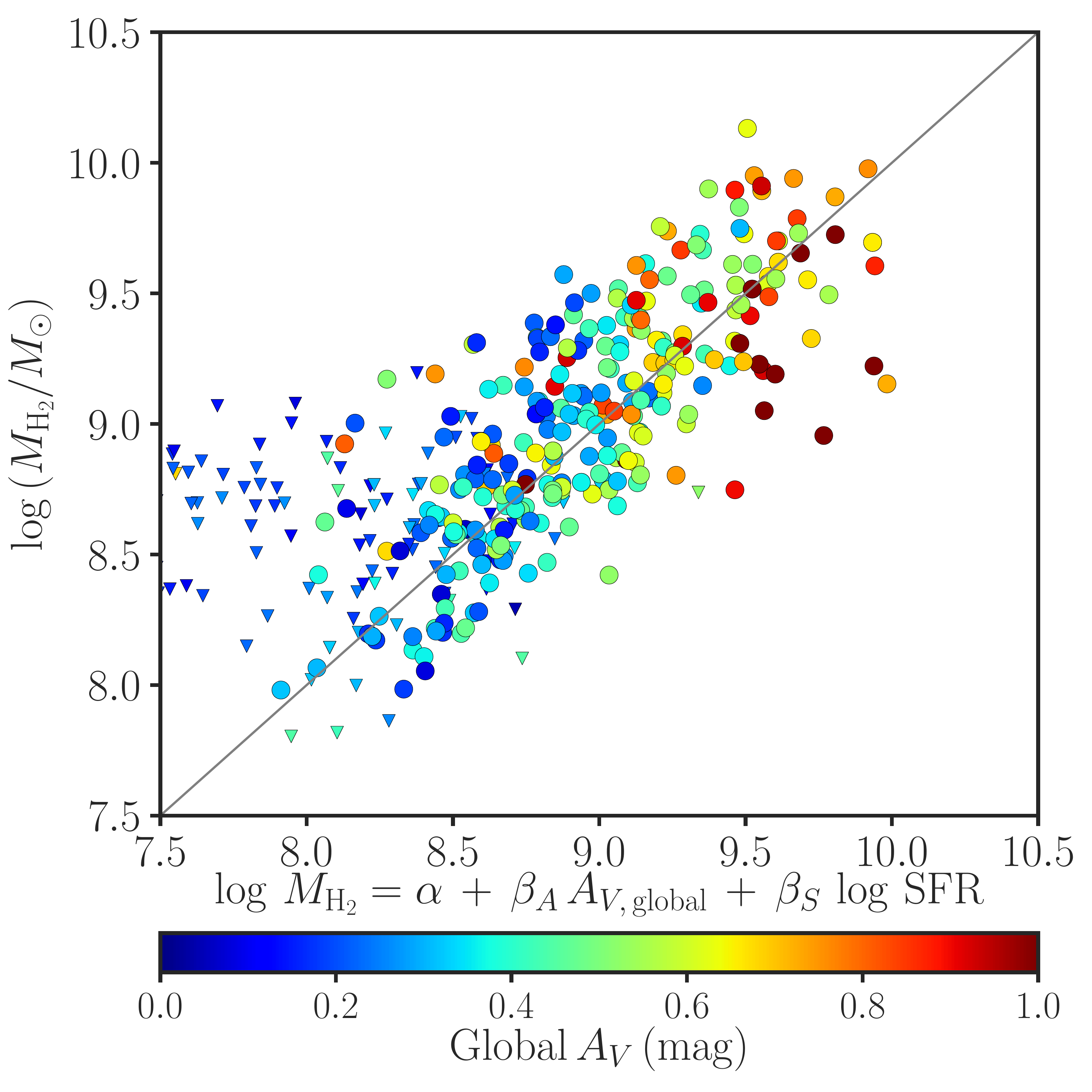}
\includegraphics[scale=0.3]{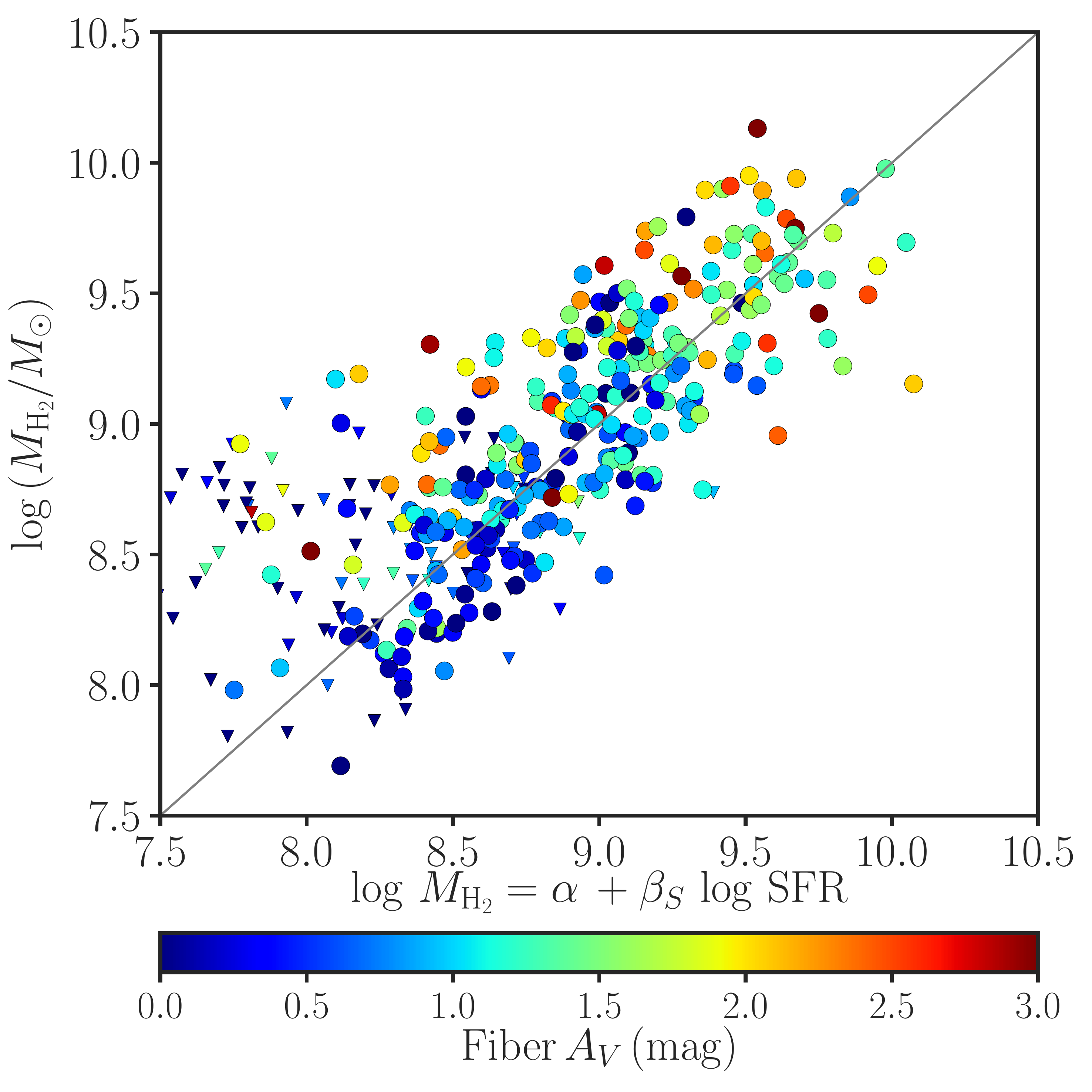}
\hfill
\includegraphics[scale=0.3]{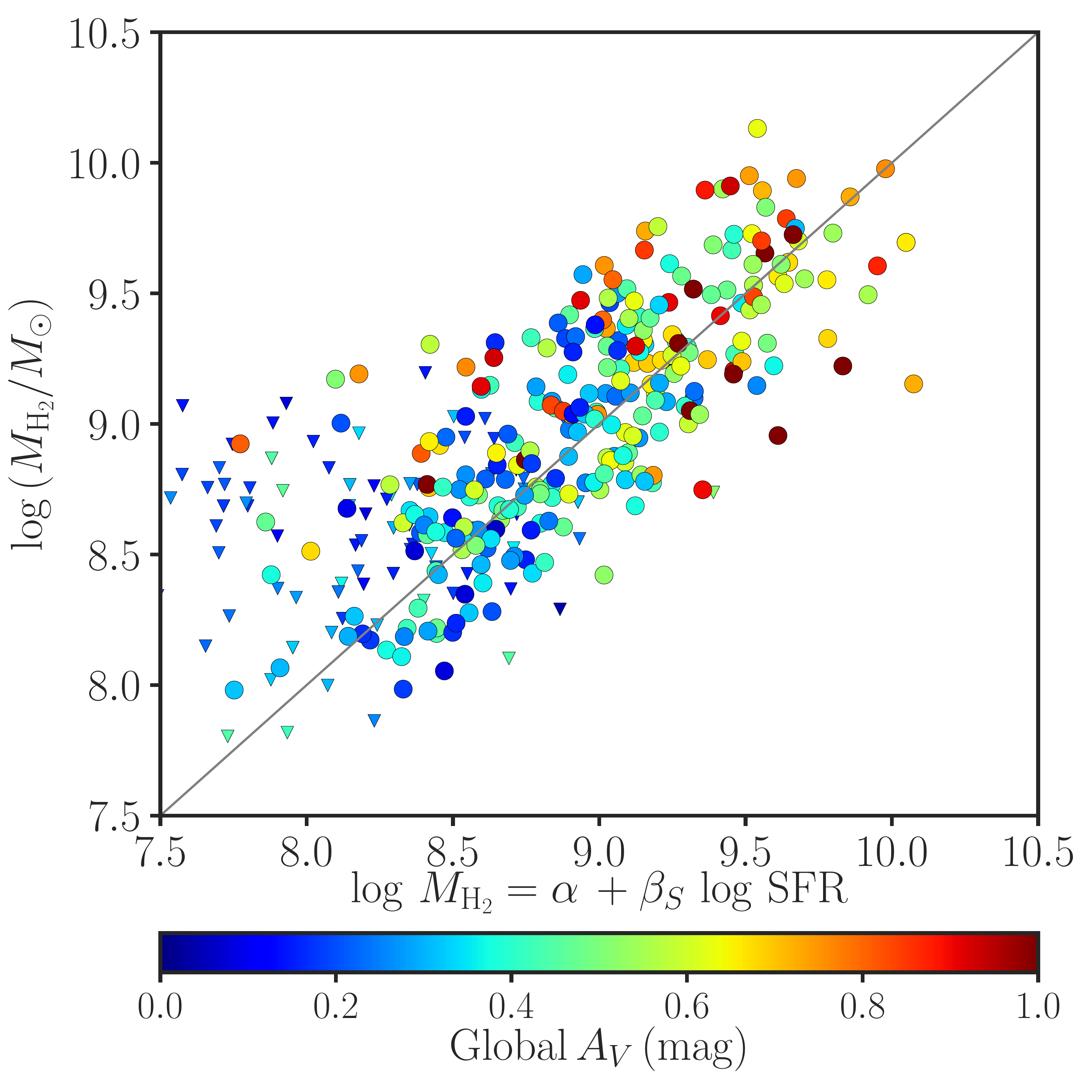}
\caption{The top panels are the median relations with {\AV} and the bottom panels are without {\AV}.\label{fig:sfr_gas2}}
\end{figure*}

\begin{figure*}
\includegraphics[scale=0.3]{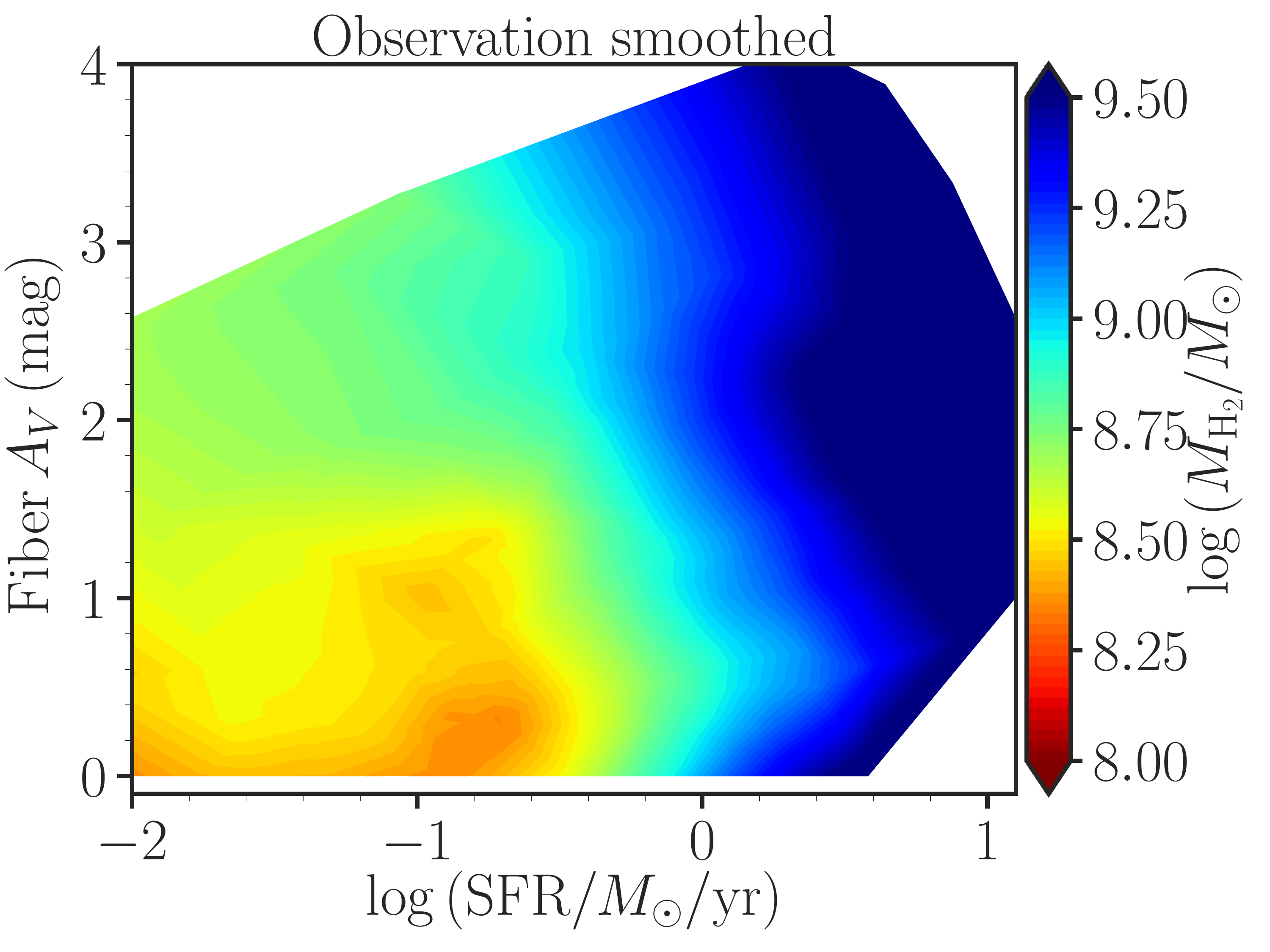}
\includegraphics[scale=0.3]{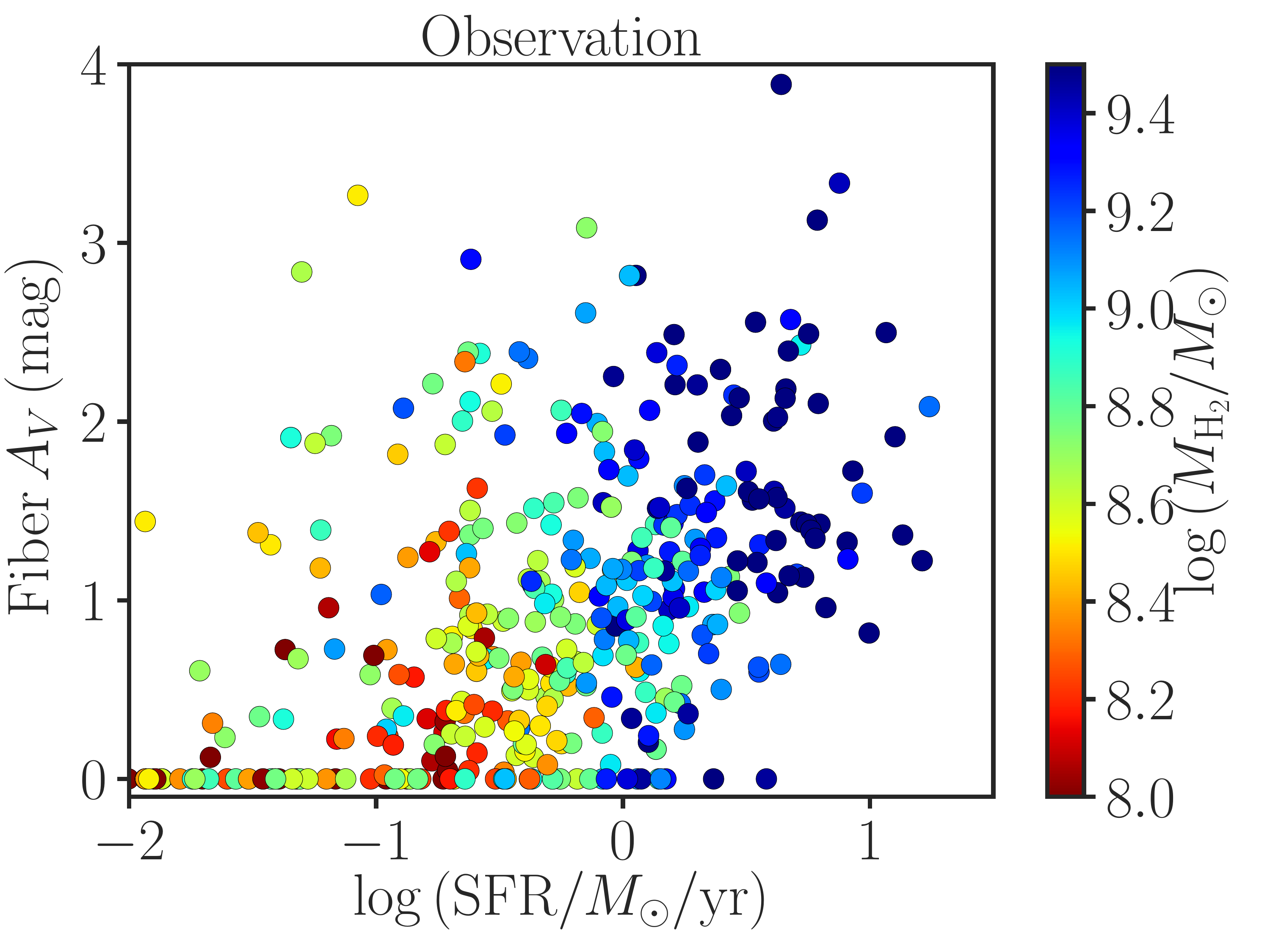}
\includegraphics[scale=0.3]{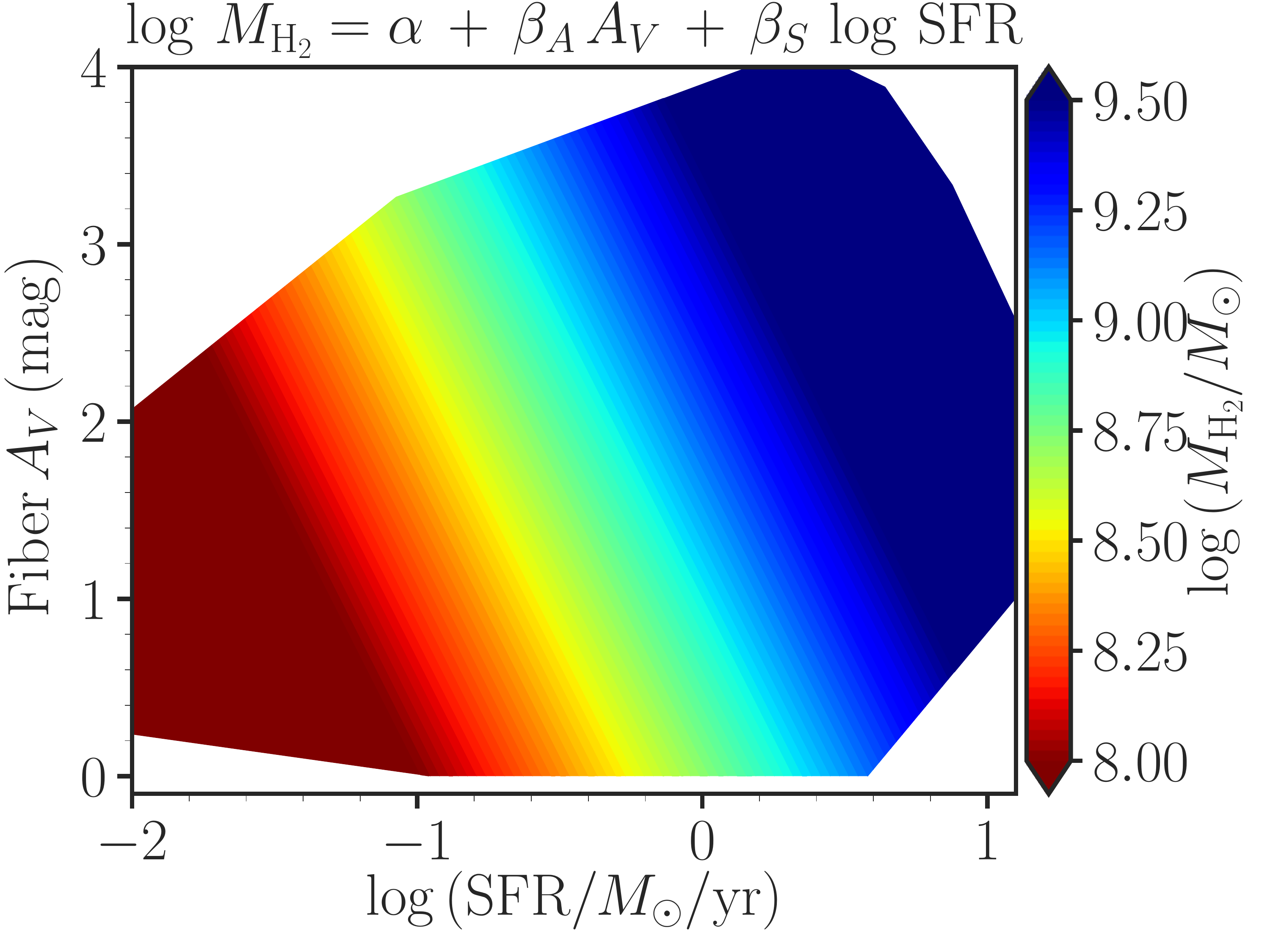}
\includegraphics[scale=0.3]{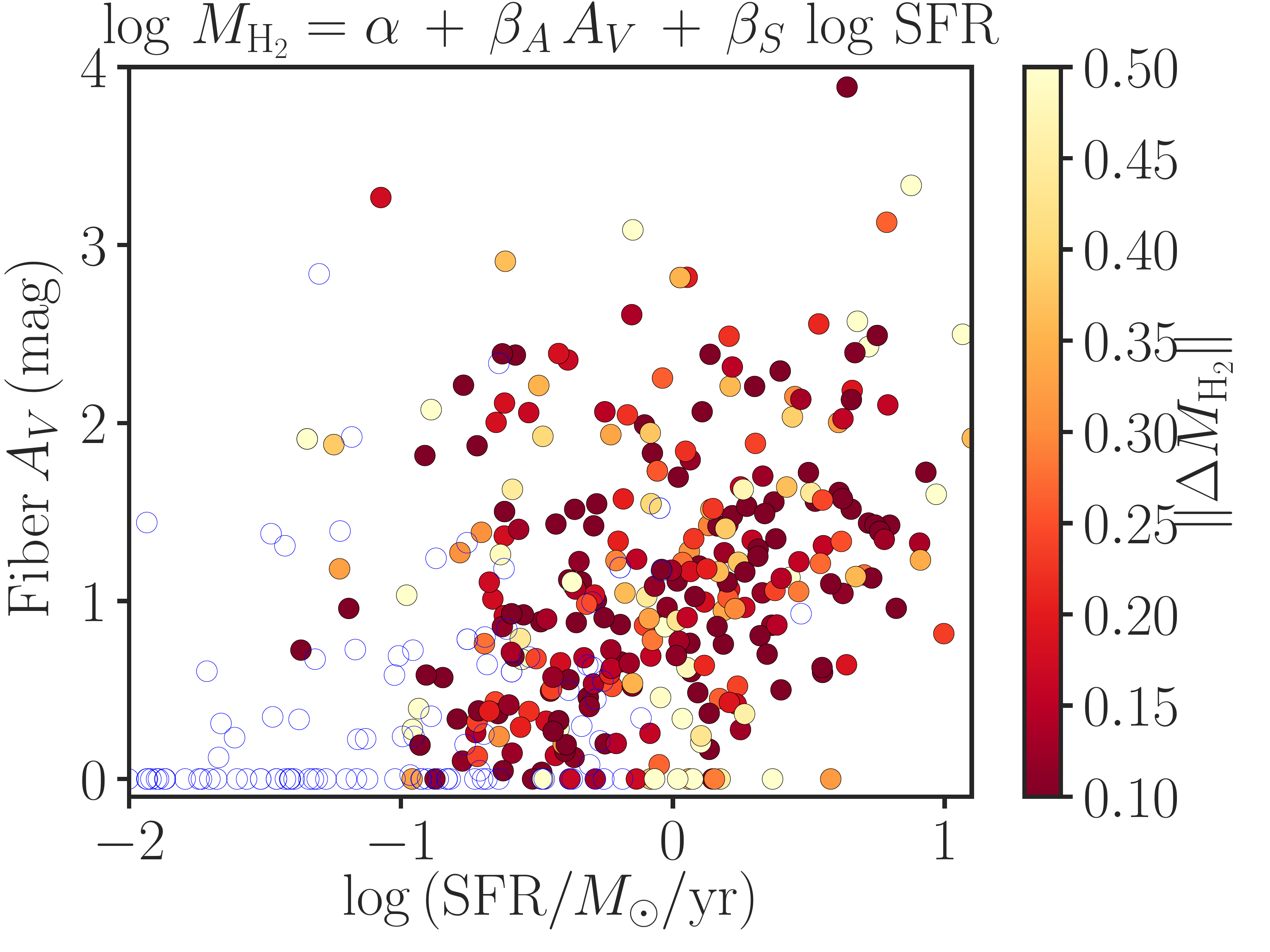}
\includegraphics[scale=0.3]{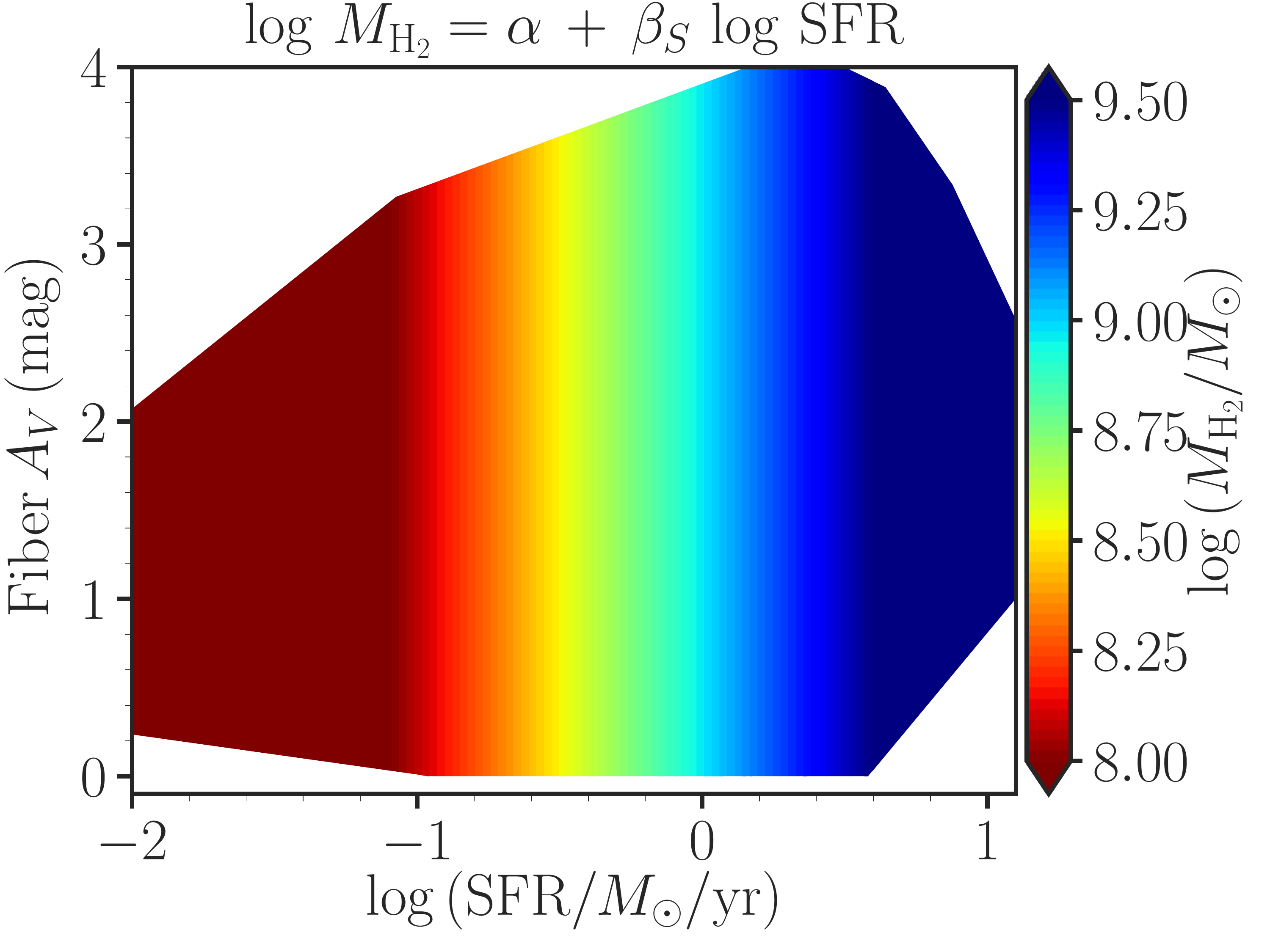}
\includegraphics[scale=0.3]{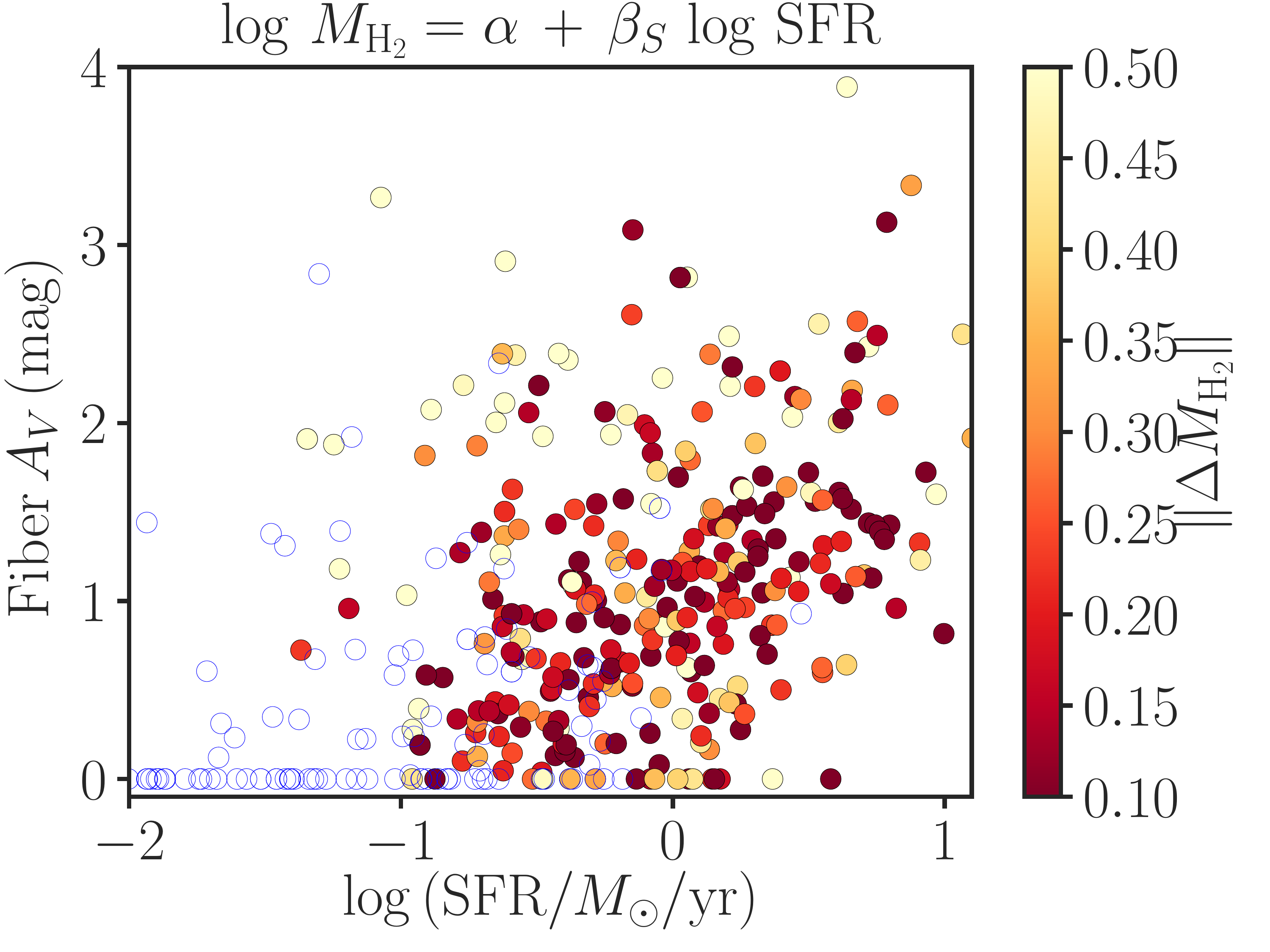}

\caption{The molecular gas mass has a secondary dependence on {\AV}, in addition to SFR. The top panels show trends in the observed data. The top left panel is LOESS-smoothed \citep{Cappellari+13b} using 20\% of the data in the local approximation. Note that almost all data points below 1\,$M_\odot$ yr$^{-1}$ are upper limits, and the smoothing does not treat these points differently. The figure The middle panels show the median fit and its residuals with {\AV}. The bottom panels show the median fit and its residuals without {\AV}. The open blue circles are non-detections. \label{fig:sfr_gas3}}
\end{figure*}

\begin{figure*}
\includegraphics[scale=0.3]{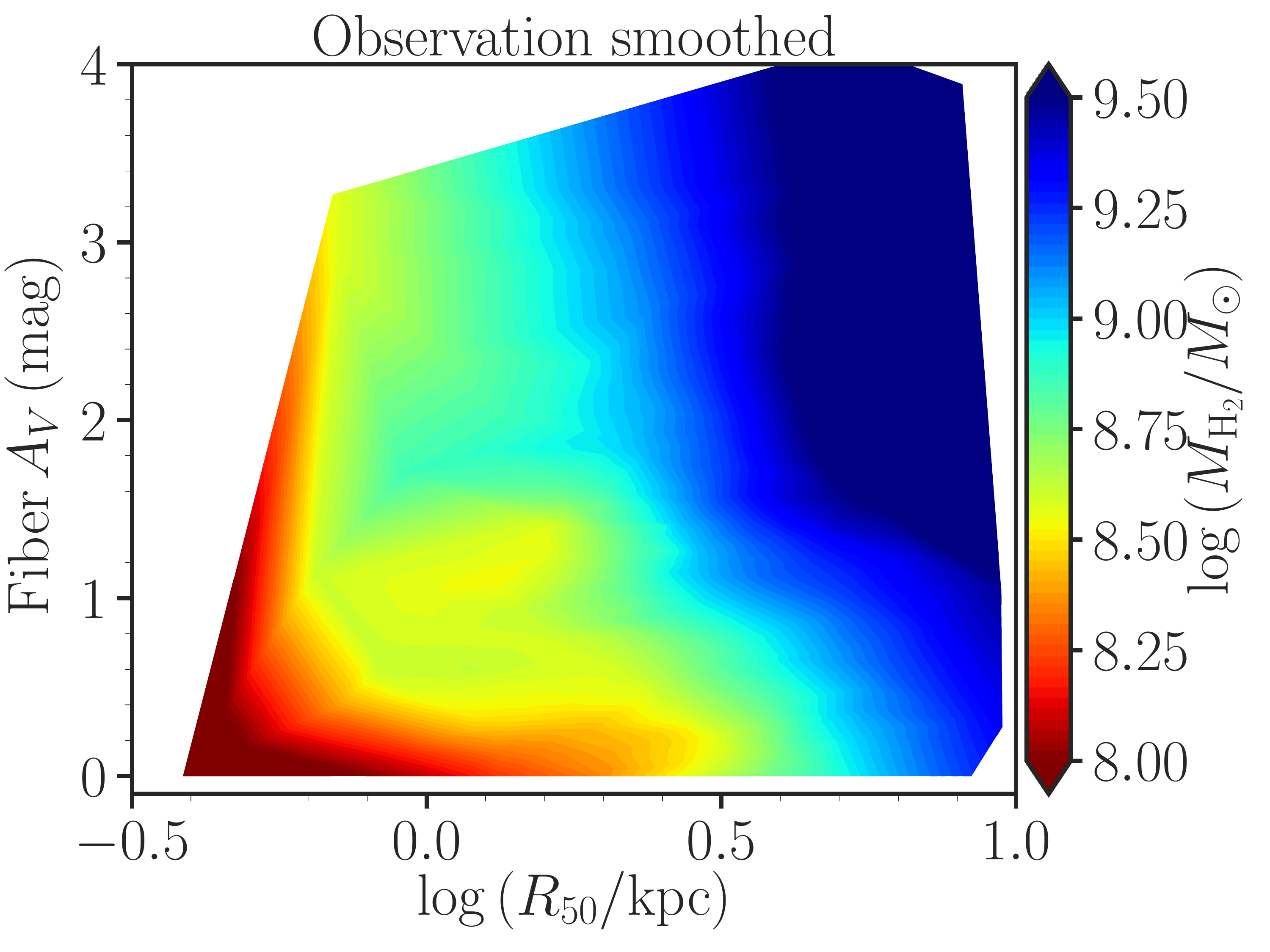}
\includegraphics[scale=0.3]{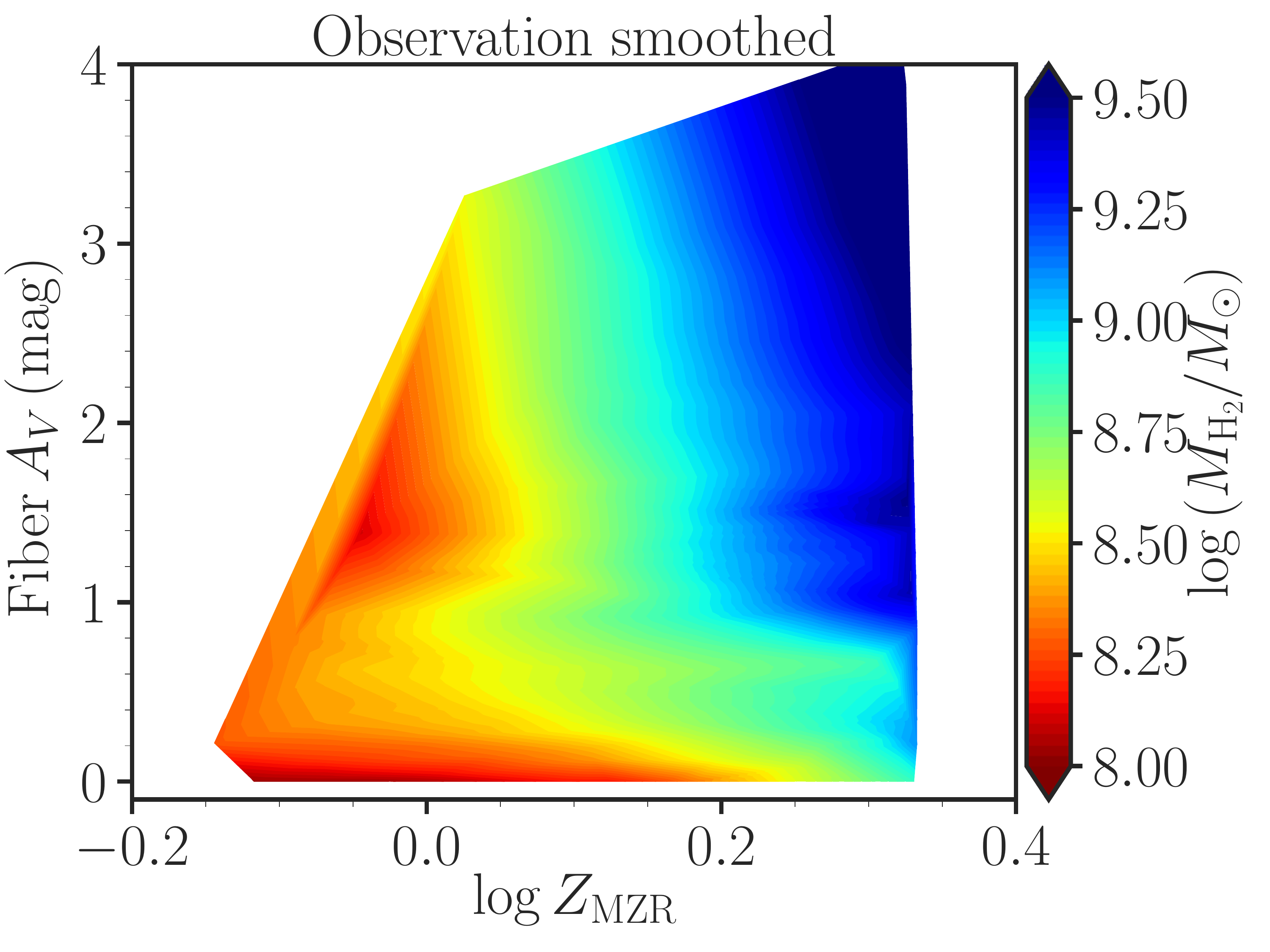}
\includegraphics[scale=0.3]{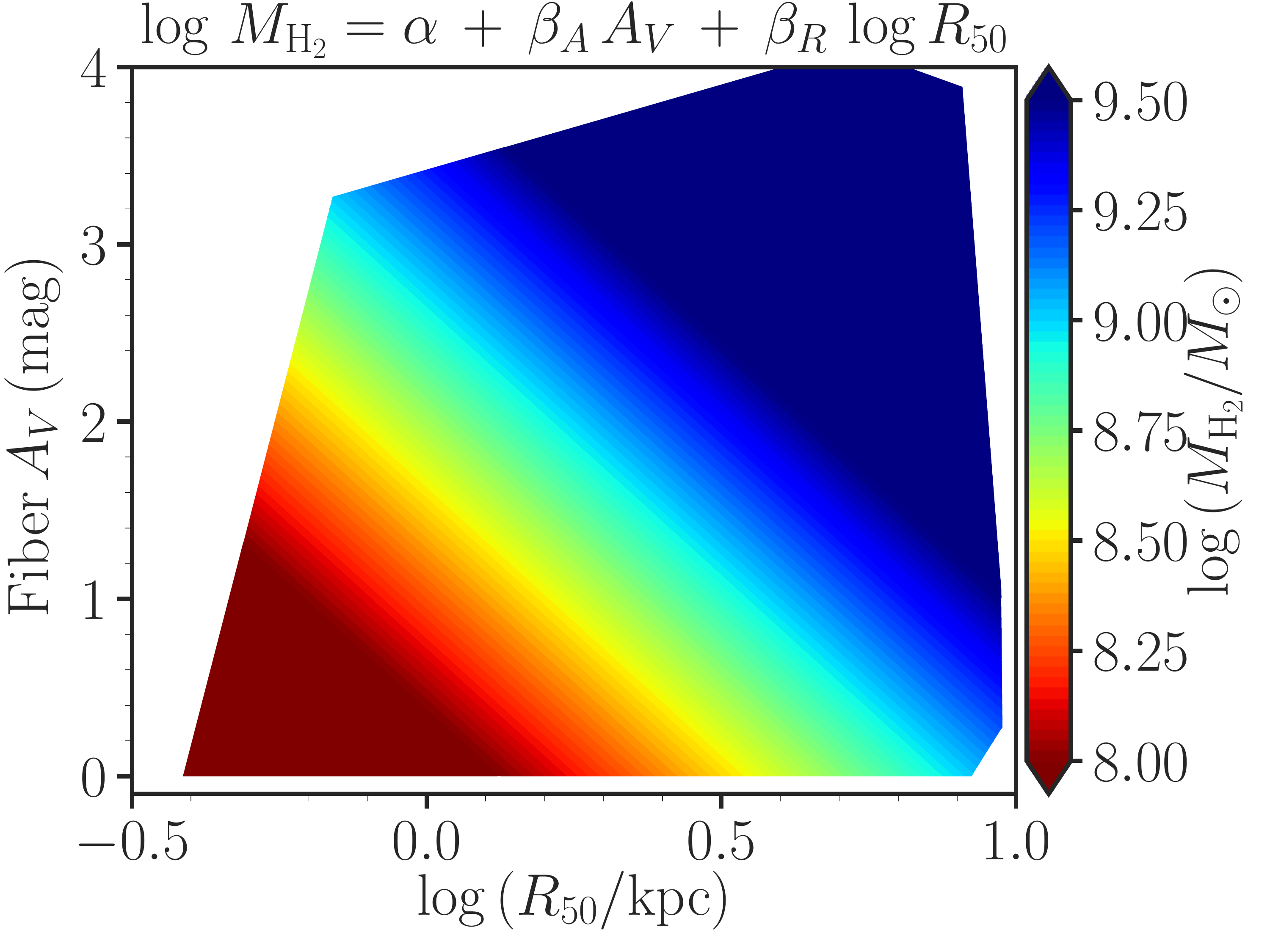}
\includegraphics[scale=0.3]{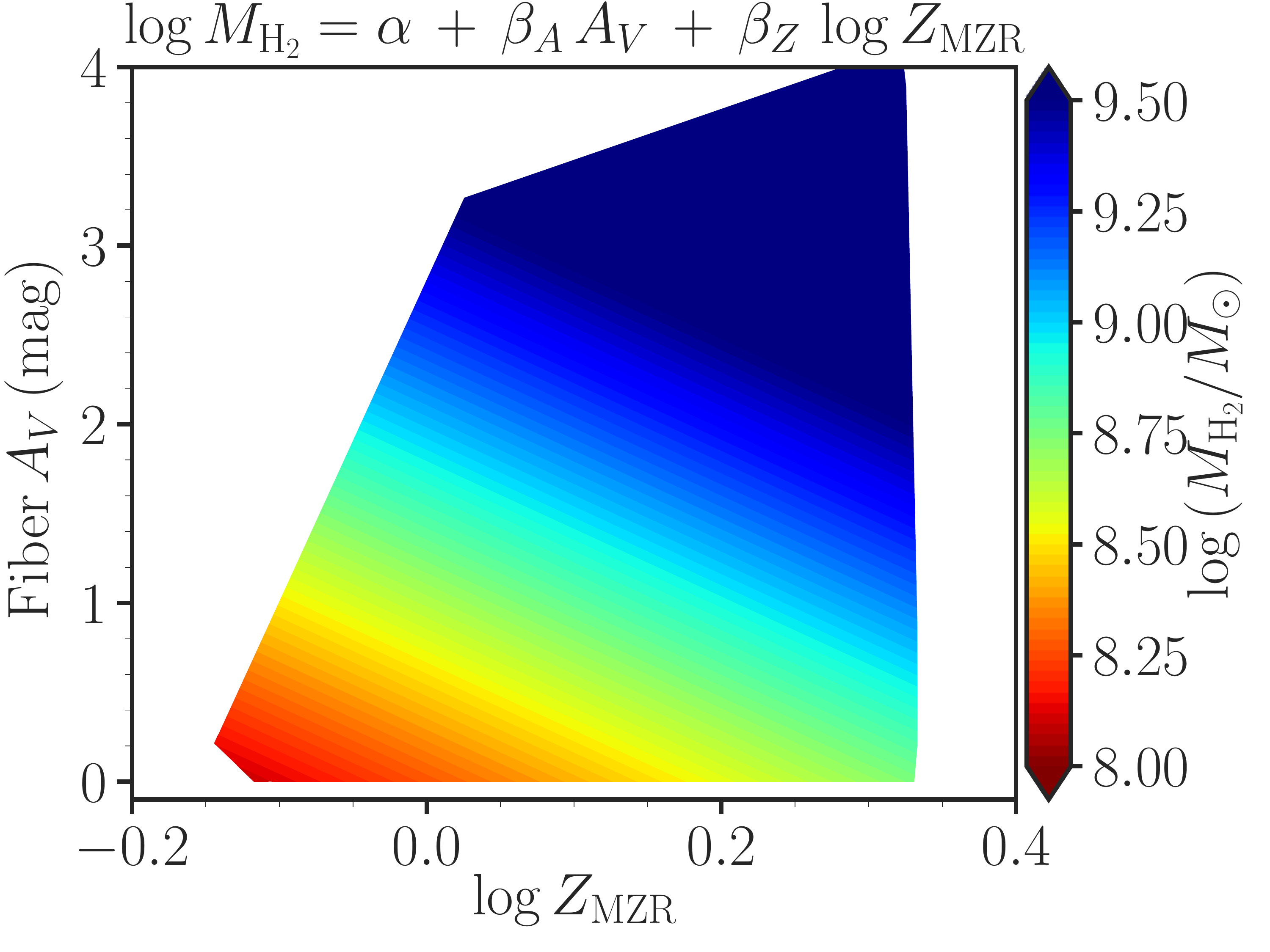}
\includegraphics[scale=0.3]{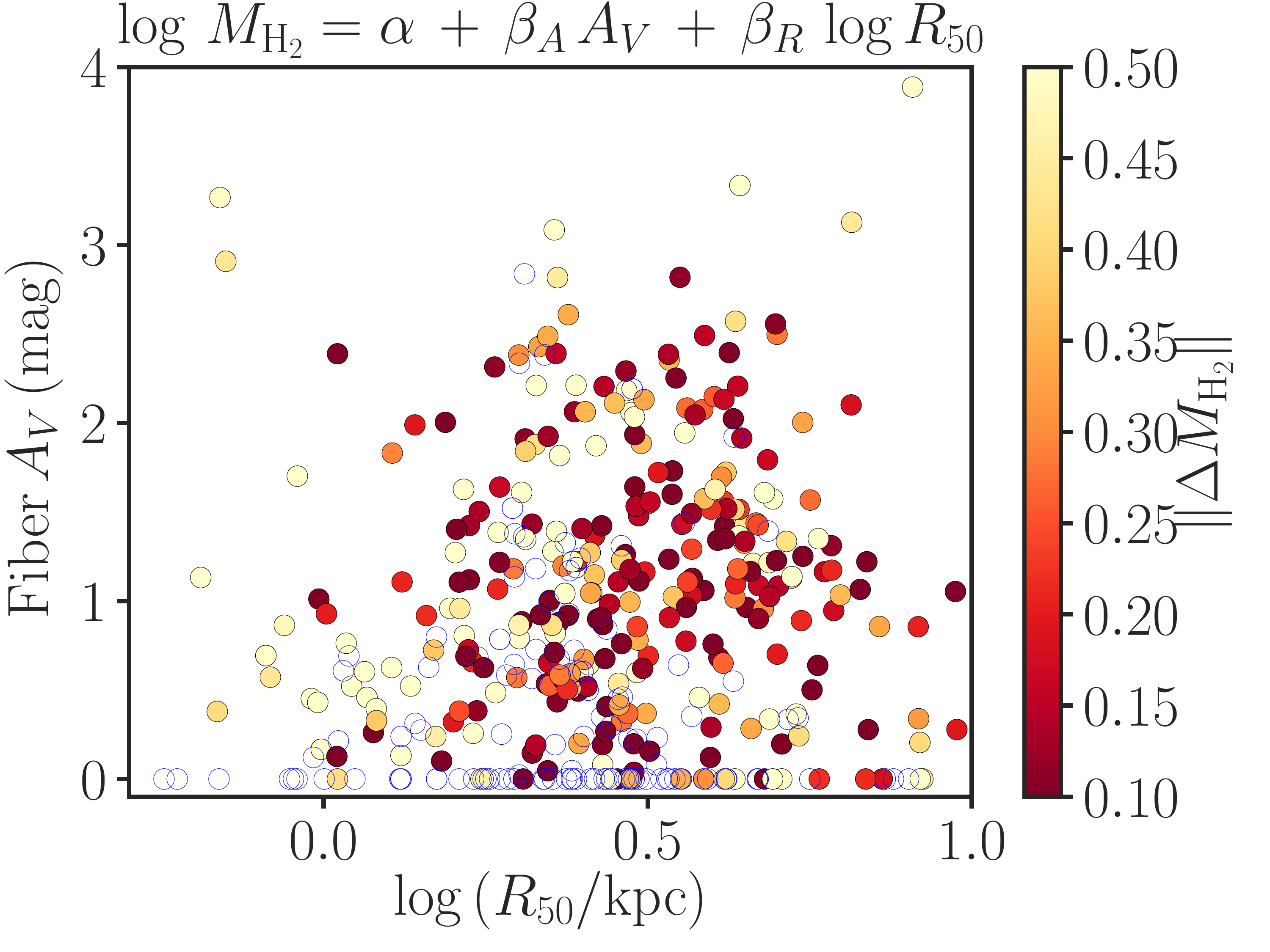}
\includegraphics[scale=0.3]{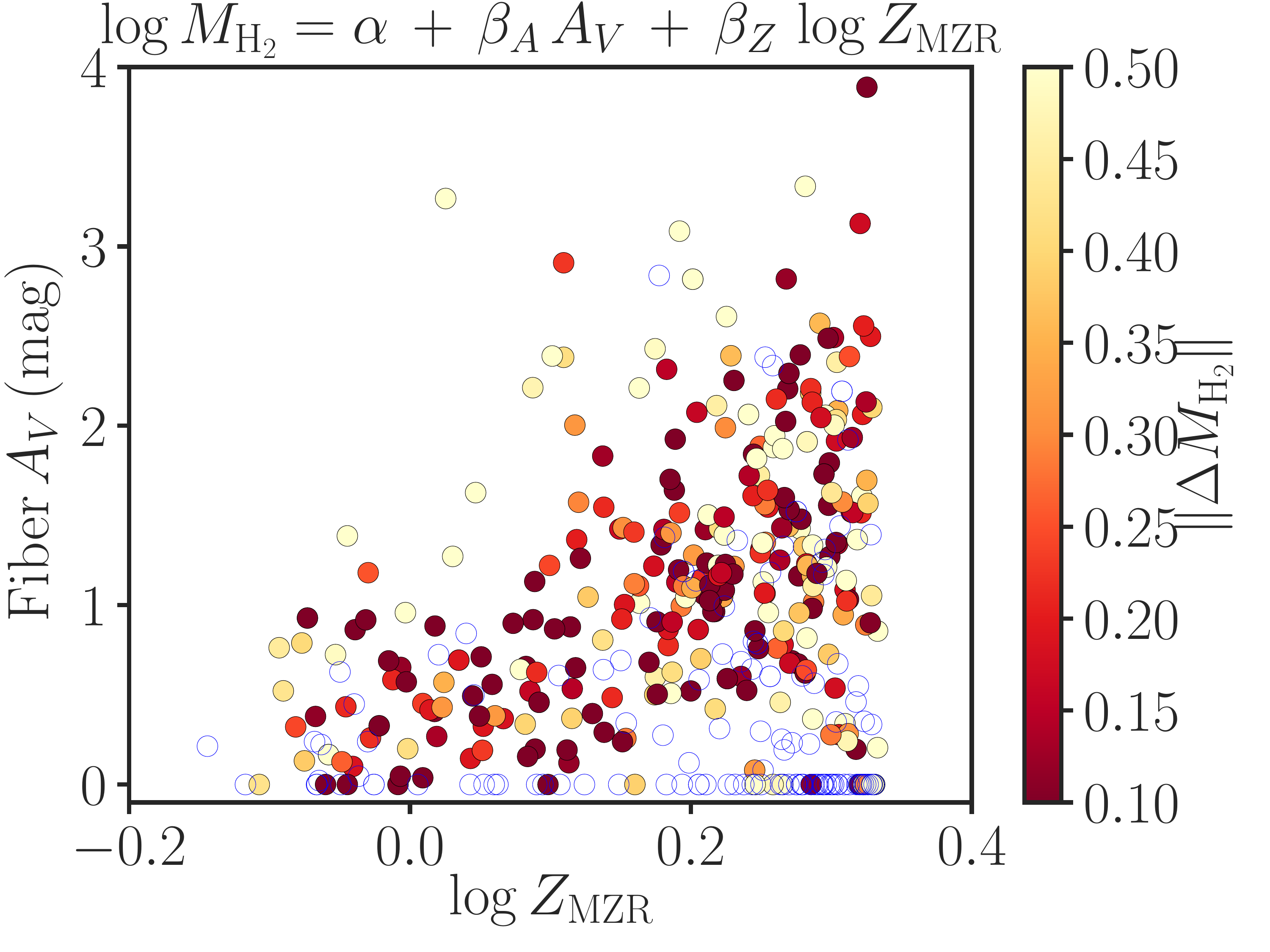}
\caption{Similar to Figure~\ref{fig:sfr_gas3} but here the dependence of $M_\mathrm{H_2}$ on $A_V$, inferred metallicity ($\log Z_\mathrm{MZR} = 12 +\log \mathrm{[O/H]}-8.8)$, and $R_{50}$ is shown. The mean gas-phase metallicity is estimated from the mass-metallicity relation \citep{Tremonti+04}.\label{fig:avzr_2d}}
\end{figure*}

\begin{figure*}
\includegraphics[scale=0.3]{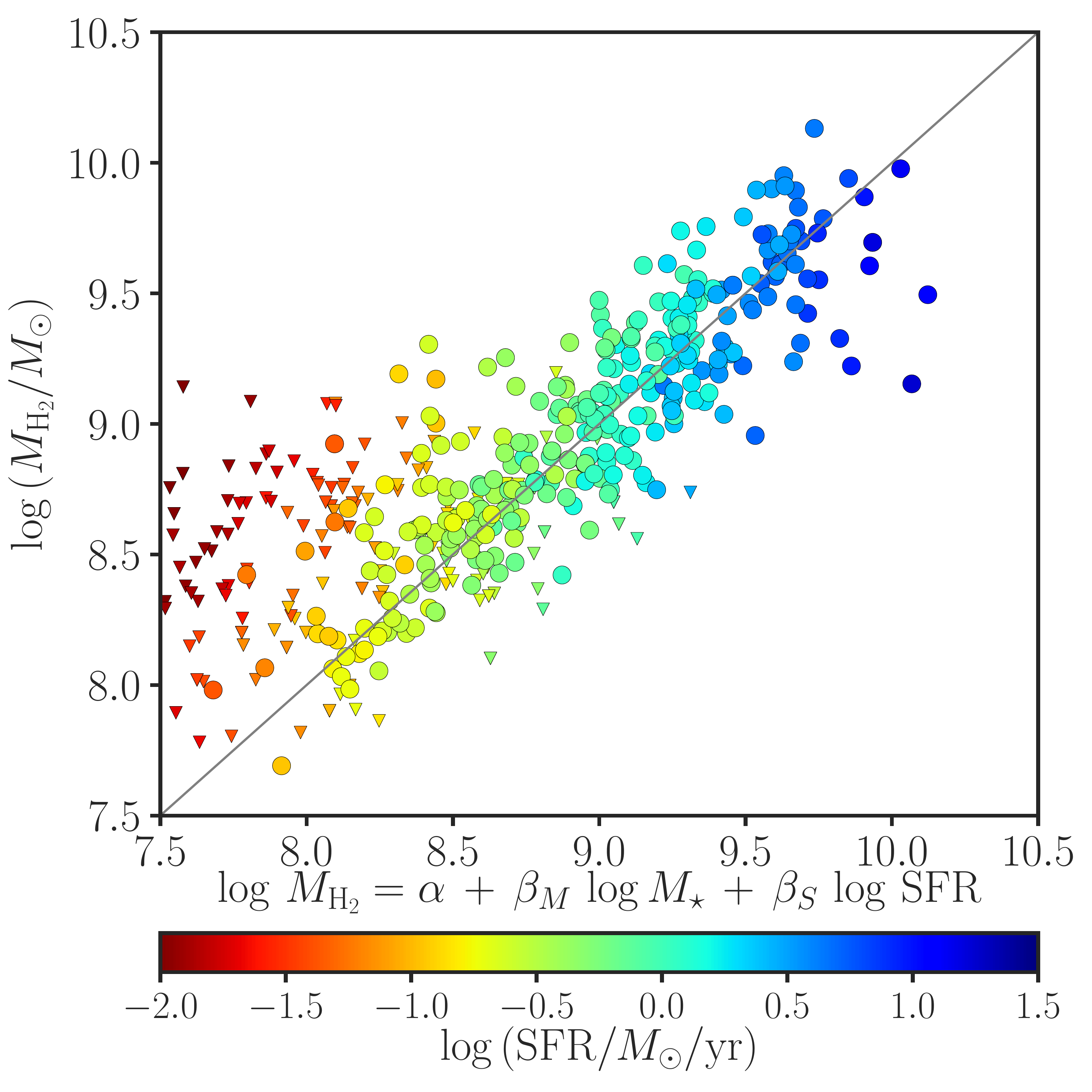}
\includegraphics[scale=0.3]{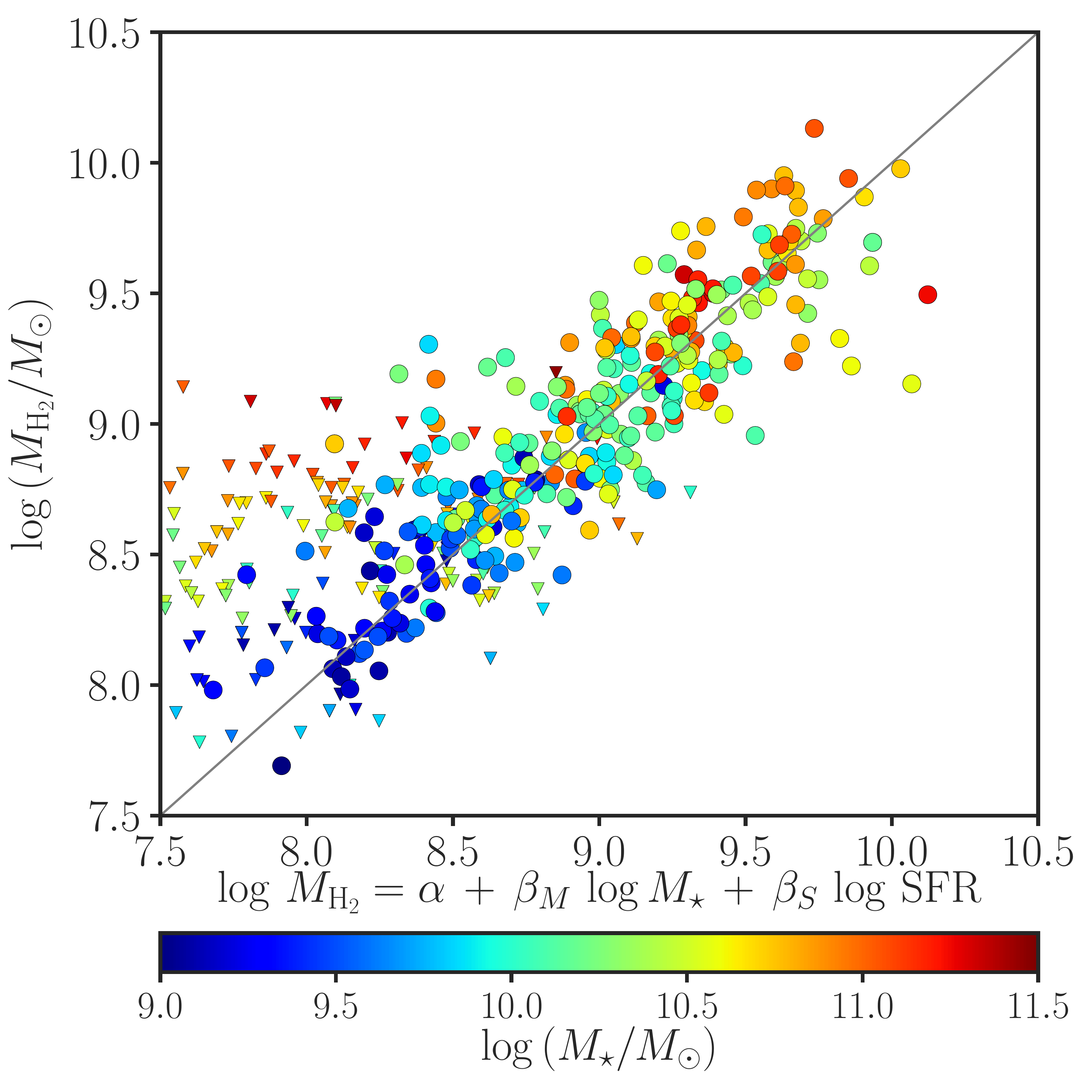}
\caption{The median molecular gas mass predicted by $M_\star$ and SFR. The left panel is color-coded by SFR, while the right panel is color-coded by $M_\star$. The right panel shows that combining $M_\star$ with SFR fails to predict molecular gas masses of gas-poor galaxies (see upper panels of Figure~\ref{fig:sfr_gas2}).\label{fig:MH2_SFR_M}}
\end{figure*}

Figure~\ref{fig:avg_gas} repeats Figure~\ref{fig:avf_gas} using galaxy-wide stellar {\AV} instead of nebular {\AV}. Table~\ref{tbl:fit_H2_all} extends the analysis in the main section by including SFR and stellar mass as predictors of molecular gas mass, in addition to {\AV} or/and $R_{50}$. Figures~\ref{fig:sfr_gas}--\ref{fig:sfr_gas3} show how the fits change when $R_{50}$ and/or {\AV} are added to the primary correlation between $M_\mathrm{{H_2}}$ and SFR. Figures~\ref{fig:avzr_2d} additionally visualizes the dependence of $M_\mathrm{H_2}$ on $A_V$, $R_{50}$, or metallicity. Figure~\ref{fig:MH2_SFR_M} demonstrates that combining $M_\star$ with SFR fails to predict median gas mass for gas-poor galaxies (for xCOLD GASS). When galaxies are on the star formation main sequence, high SFR and high $M_\star$ give high gas mass, but when they are off the sequence the relation predicts higher gas mass for high-mass galaxies and overpredicts the gas upper limits. As Table~\ref{tbl:fit_H2_all} shows, the mass dependence for gas-poor galaxies is weaker (the coefficients change by a factor of 2 and give large changes when multiplied by $M\star$). The true scatter from the median gas relation predicted by the combination of SFR and $M_\star$ is much higher if the upper limits are observed. A priori it is hard to tell which galaxies will be off the median relation. So using this median relation in practice will give less accurate estimates for gas-poor massive galaxies. If SFR is not used, the 0.15 quartile relations (for gas-poor galaxies) seem not to depend on $M\star$. Table~\ref{tbl:fit_HI} presents details of the atomic gas scaling relations, including ones that combine galaxy size with T-type morphology or disk/feature probability.

\begin{deluxetable*}{lcccc|cccc}
\tabletypesize{\footnotesize}
\tablecolumns{10} 
\tablewidth{0pt} 
\tablecaption{Regression model fits for molecular gas, $\log M_\mathrm{{H_2}}  = \alpha + \beta_A A_{V} + \beta_R \log R_{50}+ \beta_M \log M_{\star}+\beta_S \log \mathrm{SFR}$. \label{tbl:fit_H2_all}}
\tablehead{\multicolumn{5}{c}{Without SFR} & \multicolumn{4}{c}{With SFR} \\
\colhead{} & \colhead{15\%} &  \colhead{Median} & \colhead{85\%} & \colhead{Mean} & \colhead{15\%} &  \colhead{Median} & \colhead{85\%} & \colhead{Mean}}
\startdata
 $\alpha$  & $9.30 \pm 0.91$ & $6.62 \pm 0.38$ & $6.46 \pm 0.36$ & $6.91 \pm 0.43$ & $8.44 \pm 0.08$ & $6.67 \pm 0.21$ & $5.92 \pm 0.37$ & $6.34 \pm 0.28$ \\
 $\beta_{A, \mathrm{fiber}}$ & $0.44 \pm 0.07$ &$ 0.39 \pm 0.03$ & $0.29 \pm 0.03$ & $0.35 \pm 0.03$ & $0.13 \pm 0.07$ & $0.16 \pm 0.02$ & $0.15 \pm 0.02$ & $0.15 \pm 0.02$  \\
 $\beta_R$ & $1.61 \pm 0.25$ & $1.13 \pm 0.13$ & $0.62 \pm 0.06$ & $0.96 \pm 0.11$ & $0.34 \pm 0.11$ & $0.30 \pm 0.05$ & $0.22 \pm 0.09$ & $0.29 \pm 0.07$\\
 $\beta_M$ & $-0.22 \pm 0.10$ &$0.13 \pm 0.04$ & $0.22 \pm 0.04$ & $0.11 \pm 0.05$ &  --- & $0.20 \pm 0.02$ & $0.29 \pm 0.04$ & $0.23 \pm 0.03$ \\
 $\beta_S$  & ---& --- & --- & --- & $0.70 \pm 0.09$ & $0.59 \pm 0.03$ & $0.53 \pm 0.05$ & $0.52 \pm 0.02$ \\
 Scatter & &0.23 & & 0.40 & & 0.12 & & 0.24\\
\hline
$\alpha$ & $7.41 \pm 0.10 $ & $ 7.84 \pm 0.09$ & $8.49 \pm 0.10$ & $ 7.95 \pm 0.05$ & $8.44 \pm 0.08$ & $8.53 \pm 0.22$ & $8.69 \pm 0.03$ & $ 8.49 \pm 0.04$ \\
 $\beta_{A, \mathrm{fiber}}$ & $ 0.35 \pm 0.06 $ &$ 0.42 \pm 0.06 $ & $0.35 \pm 0.04$ & $0.37 \pm 0.03$ & $0.13 \pm 0.07$ & $0.22\pm 0.02$ & $0.20 \pm 0.04$ & $0.20 \pm 0.02$  \\
 $\beta_R$ & $1.34 \pm 0.28$ & $ 1.27 \pm 0.09$ & $0.89 \pm 0.12$ & $1.11 \pm 0.09$ & $0.34 \pm 0.11 $ & $0.49 \pm 0.09 $ & $ 0.73 \pm 0.06$ & $0.59 \pm 0.07$\\
 $\beta_S$  & ---& --- & --- & --- & $0.70 \pm 0.09$ & $0.61 \pm 0.06$ & $0.56 \pm 0.05$ & $0.54 \pm 0.03$ \\
 Scatter & &0.25 & & 0.41 & & 0.14 & & 0.27\\
\hline
 $\alpha$  & $ 7.70 \pm 0.07$ & $ 5.21\pm 0.38 $ & $4.44 \pm 0.45$ & $ 5.10 \pm 0.41$ & $8.57 \pm 0.05$ & $6.05 \pm 0.41$ & $5.58 \pm 0.33$ & $5.84 \pm 0.25$ \\
 $\beta_{A, \mathrm{fiber}}$ & $ 0.47 \pm 0.05$ &$0.38 \pm 0.05$ & $ 0.27 \pm 0.05$ & $0.35 \pm 0.03$ & $0.14 \pm 0.06$ & $ 0.14 \pm 0.03 $ & $0.14 \pm 0.03$ & $0.14 \pm 0.02$\\
 $\beta_M$ & --- & $ 0.32 \pm 0.04$ & $ 0.44 \pm 0.04$ & $0.33 \pm 0.04$ &  --- & $ 0.27\pm 0.04 $ & $ 0.34 \pm 0.03$ & $0.29 \pm 0.02 $\\
$\beta_S$ & ---& --- & --- & --- & $0.75 \pm 0.06$ & $0.64 \pm 0.05$ & $0.55 \pm 0.04$ & $0.55 \pm 0.02$ \\
 Scatter & & 0.25 & & 0.44 & & 0.12 & & 0.25\\
\hline
$\alpha$ & $ 7.70 \pm 0.07 $ & $ 8.31 \pm 0.04 $ & $8.91 \pm 0.06$ & $8.34 \pm 0.04$ & $8.57 \pm 0.05$ & $8.72 \pm 0.06$ & $9.04 \pm 0.06$ & $8.76 \pm 0.03$ \\
$\beta_{A, \mathrm{fiber}}$ & $ 0.47 \pm 0.05$ &$0.47 \pm 0.02$ & $ 0.38 \pm 0.05$ & $0.43 \pm 0.03$ & $0.14 \pm 0.06$ & $ 0.23 \pm 0.05 $ & $0.19 \pm 0.03$ & $0.20 \pm 0.02$\\
$\beta_S$  & ---& --- & --- & --- & $0.75 \pm 0.06$ & $0.73 \pm 0.03$ & $ 0.69\pm 0.08$ & $0.63 \pm 0.03$\\
 Scatter & & 0.31 & & 0.50 & & 0.16 & & 0.30\\
 \hline
 $\alpha$ &  --- & $5.20 \pm 0.85 $ & $3.16 \pm 0.53$ & $ 4.67 \pm 0.50$ & $7.47 \pm 0.19$ & $6.14 \pm 0.16$ & $5.74 \pm 0.22$ & $5.83 \pm 0.26$ \\
 $\beta_M$ & ---  &$0.34 \pm 0.09$ & $ 0.60 \pm 0.05$ & $0.39 \pm 0.05$ & $0.12 \pm 0.02$ & $ 0.28 \pm 0.02 $ & $0.34 \pm 0.02$ & $0.31 \pm 0.02$\\
 $\beta_S$  & ---&  --- & --- & --- & $0.83 \pm 0.01$ & $0.78 \pm 0.04$ & $ 0.66 \pm 0.04$ & $0.66 \pm 0.02$\\
 Scatter & & 0.34 & & 0.61 & & 0.14 & & 0.30\\
\hline
 $\alpha$  & --- & $7.83 \pm 0.15 $ & $8.70 \pm 0.05$ & $ 8.02 \pm 0.05$ & $ 8.54\pm 0.04$ & $8.77 \pm 0.05$ & $8.91 \pm 0.08$ & $8.70 \pm 0.04$ \\
 $\beta_R$ & ---  &$1.84 \pm 0.31$ & $ 1.25 \pm 0.07$ & $1.48 \pm 0.11$ & $ 0.35\pm 0.08$ & $0.46 \pm 0.10 $ & $0.70 \pm 0.14$ & $0.59 \pm 0.07$\\
 $\beta_S$  & ---&  --- & --- & --- & $0.79 \pm 0.06$ & $0.81 \pm 0.03$ & $ 0.69 \pm 0.07$ & $0.68 \pm 0.03$\\
 Scatter & & 0.36 & & 0.55 & & 0.17 & & 0.31\\
\hline
 $\alpha$  & $7.48 \pm 0.09 $ & $7.58 \pm 0.11$ & $6.76 \pm 0.65 $ &$7.04 \pm 0.40$ & $8.33 \pm 0.15$ & $ 6.56 \pm 0.37 $ & $ 6.16 \pm 0.47$ & $ 6.32 \pm 0.27 $ \\
 $\beta_{A, \mathrm{global}}$  &$1.24 \pm 0.29$ & $1.41 \pm 0.13$ & $ 1.28 \pm 0.16$ & $1.32 \pm 0.08$ & $ 0.39 \pm 0.12$ & $0.41 \pm 0.09$ & $0.53 \pm 0.09 $ & $0.50 \pm 0.06$  \\
 $\beta_R$ & $0.97 \pm 0.09$ & $ 1.35 \pm 0.19 $ & $0.84 \pm 0.16$ & $1.08 \pm 0.10$ & $0.49 \pm 0.26$ & $0.30 \pm 0.10$ & $ 0.31\pm 0.10$ & $0.37 \pm 0.07$\\
 $\beta_M$  & --- & --- & $ 0.15 \pm 0.07$ & $0.07 \pm 0.04$ & --- & $0.21 \pm 0.04$ & $0.26 \pm 0.05$ & $0.22 \pm 0.03$ \\
 $\beta_S$  & --- & --- & --- &   & $0.59 \pm 0.16$ & $ 0.61 \pm 0.03$ & $0.47 \pm 0.05 $ & $0.48 \pm 0.03$ \\
 Scatter & &0.26 & & 0.38 & & 0.13 & & 0.24\\
 \hline
 $\alpha$  & $7.48 \pm 0.09 $ & $ 7.58 \pm 0.11$ & $8.16 \pm 0.05 $ &$7.70 \pm 0.06$ & $8.33 \pm 0.15$ & $8.51 \pm 0.06 $ & $ 8.60 \pm0.13 $ & $ 8.40 \pm 0.05 $ \\
 $\beta_{A, \mathrm{global}}$  &$1.24 \pm 0.29$ & $1.41 \pm 0.13$ & $ 1.34 \pm 0.14$ & $1.34 \pm 0.08$ & $ 0.39 \pm 0.12$ & $0.44 \pm 0.08 $ & $ 0.58 \pm 0.20 $ & $0.57 \pm 0.07$  \\
 $\beta_R$ & $ 0.97 \pm 0.09$ & $ 1.35 \pm 0.19 $ & $1.12 \pm 0.10$ & $1.19 \pm 0.08$ & $0.49 \pm 0.26$ & $ 0.60 \pm 0.06$ & $0.83 \pm 0.13$ & $0.69 \pm 0.07$\\
 $\beta_S$  & --- & --- & --- &   & $0.59 \pm 0.16$ & $ 0.61 \pm 0.06$ & $0.47 \pm 0.06 $ & $0.49 \pm 0.03$ \\
 Scatter & & 0.26 & & 0.38 & & 0.15 & & 0.27\\
 \hline
 $\alpha$  & $ 7.70 \pm 0.07$ & $ 5.37\pm 0.32 $ & $4.30 \pm 0.20$ & $ 4.96 \pm 0.38$ & $ 8.57 \pm 0.08$ & $6.40 \pm 0.19$ & $5.27 \pm 0.15 $ & $ 5.72\pm 0.25 $ \\
 $\beta_{A, \mathrm{global}}$ & $ 0.47 \pm 0.05$ &$1.49 \pm 0.14$ & $ 1.17 \pm 0.15 $ & $1.38 \pm 0.09$ & $0.31 \pm 0.15 $ & $ 0.39 \pm 0.05$ & $0.50 \pm 0.09$ & $0.43 \pm 0.06$\\
 $\beta_M$ & --- & $ 0.27 \pm 0.04$ & $ 0.43 \pm 0.02$ & $0.32 \pm 0.04$ &  --- & $ 0.24 \pm 0.02 $ & $0.36 \pm 0.02 $ & $0.30 \pm 0.02$\\
 $\beta_S$  & ---& --- & --- & --- & $ 0.75\pm 0.10$ & $0.65 \pm 0.03$ & $0.52 \pm 0.03$ & $0.53 \pm 0.03$ \\
 Scatter & & 0.32 & & 0.43 & & 0.14 & & 0.25\\
 \hline
 $\alpha$  & $ 7.70 \pm 0.07 $ & $ 8.31 \pm 0.04 $ & $8.91 \pm 0.06$ & $8.34 \pm 0.04$ & $8.57 \pm 0.08$ & $8.78 \pm 0.09$ & $9.05 \pm 0.07$ & $8.77 \pm 0.04$ \\
 $\beta_{A, \mathrm{global}}$ & $ 0.47 \pm 0.05$ &$0.47 \pm 0.02$ & $ 0.38 \pm 0.05$ & $0.43 \pm 0.03$ & $0.31 \pm 0.15$ & $ 0.41\pm 0.11 $ & $0.52 \pm 0.11$ & $0.46 \pm 0.08$\\
 $\beta_S$ & ---& --- & --- & --- & $0.75 \pm 0.10$ & $0.73 \pm 0.04$ & $ 0.71\pm 0.06$ & $0.63 \pm 0.03$\\
 Scatter & & 0.31 & & 0.50 & & 0.18 & & 0.31\\
 \enddata
\tablecomments{The coefficients of the equations describe the relationship among molecular gas mass ($M_\mathrm{{H_2}}$), SFR, stellar mass ($M_\star$), dust attenuation ($A_V$), and/or galaxy radius ($R_\mathrm{{50}}$). Dust absorption is estimated from Balmer decrement in the fiber ($A_{V, \mathrm{fiber}}$) or by galaxy-wide SED fitting ($A_{V, \mathrm{global}}$). The scatter for the fits are quantified by the median absolute deviation (MAD, the left numbers corresponding to the median fits) and the root mean square deviation (RMSD, the right numbers corresponding to the mean fits).}
\end{deluxetable*}

\begin{deluxetable*}{lcccc|cccc}
\tabletypesize{\footnotesize}
\tablecolumns{10} 
\tablewidth{0pt} 
\tablecaption{Regression model fits for atomic gas, $\log M_\mathrm{{HI}}  = \alpha + \beta_A A_{V} + \beta_R \log R+ \beta_M \log M_{\star}+\beta_S \log \mathrm{SFR}+ \beta_T \mathrm{T\, types} +\beta_{D} \mathrm{Prob.\, disk}$ \label{tbl:fit_HI}}
\tablehead{\multicolumn{5}{c}{All Galaxies} & \multicolumn{4}{c}{Centrals Only} \\
\colhead{} & \colhead{15\%} &  \colhead{Median} & \colhead{85\%} & \colhead{Mean} & \colhead{15\%} &  \colhead{Median} & \colhead{85\%} & \colhead{Mean}}
\startdata
 $\alpha$  & $8.09 \pm 0.17$ & $8.58 \pm 0.11$ & $9.02 \pm 0.05$ & $8.54 \pm 0.07$ & $8.07 \pm 0.35$ & $8.52 \pm 0.20$ & $9.02 \pm 0.06$ & $8.50 \pm 0.08$ \\
 $\beta_{R_{90}}$ & $0.69 \pm 0.20$ & $0.98 \pm 0.10$ & $0.93 \pm 0.05$ & $0.98 \pm 0.07 $ & $0.75 \pm 0.24$ & $1.05 \pm 0.20$ & $0.96 \pm 0.06$ & $1.04 \pm 0.09$\\
 $\beta_T$ & $0.16 \pm 0.01$ &$0.06 \pm 0.01$ & $0.03 \pm 0.01$ & $0.05 \pm 0.01$ & $0.16 \pm 0.06$ & $0.06 \pm 0.01$ & $0.02 \pm 0.01$ & $0.05 \pm 0.01$ \\
 $\beta_S$  & $0.63 \pm 0.05$ & $0.44 \pm 0.05$ & $0.42 \pm 0.02$ & $0.40 \pm 0.03$ & $0.63 \pm 0.12$ & $0.46 \pm 0.09$ & $0.43 \pm 0.07$ & $0.42 \pm 0.03$ \\
 Scatter & &0.22 & & 0.39 & & 0.22 & & 0.39\\
\hline
$\alpha$  & $8.40 \pm 0.14$ & $8.69 \pm 0.12$ & $9.08 \pm 0.09$ & $8.69 \pm 0.07$ & $8.18 \pm 0.19$ & $8.55 \pm 0.07$ & $9.01 \pm 0.09$ & $8.60 \pm 0.08$ \\
 $\beta_{R_{90}}$ & $0.72 \pm 0.13$ & $1.01 \pm 0.13$ & $0.94 \pm 0.09$ & $0.96 \pm 0.08$ & $1.02 \pm 0.27$ & $1.21 \pm 0.07$ & $1.05 \pm 0.08$ & $1.09 \pm 0.09$\\
 $\beta_S$  & $0.77 \pm 0.10$ & $0.49 \pm 0.06$ & $0.42 \pm 0.09$ & $0.48 \pm 0.03$ & $0.75 \pm 0.14$ & $0.49 \pm 0.06$ & $0.41 \pm 0.06$ & $0.48 \pm 0.03$ \\
 Scatter & &0.23 & & 0.41 & & 0.23 & & 0.41\\
\hline
$\alpha$  & $8.59 \pm 0.04$ & $9.07 \pm 0.04$ & $9.45 \pm 0.04$ & $9.04 \pm 0.04 $ & $8.57 \pm 0.05$ & $9.02 \pm 0.05$ & $9.44 \pm 0.05$ & $9.01 \pm 0.05$ \\
 $\beta_{R_{50}}$ & $1.05 \pm 0.11$ & $1.08 \pm 0.11$ & $0.98 \pm 0.07$ & $1.03 \pm 0.08$ & $1.16 \pm 0.07$ & $1.20 \pm 0.11$ & $1.03 \pm 0.10$ & $1.12\pm 0.09$\\
 $\beta_S$ & $0.69 \pm 0.05$ & $0.47 \pm 0.02$ & $0.43 \pm 0.05$ & $0.44 \pm 0.03$ & $0.69 \pm 0.07$ & $0.47 \pm 0.04$ & $0.42 \pm 0.10$ & $0.45 \pm 0.03$ \\
 Scatter & &0.23 & & 0.41 & & 0.24 & & 0.41\\
\hline
$\alpha$  & $12.32 \pm 0.77$ & $11.16 \pm 0.71$ & $9.18 \pm 0.16$ & $10.30 \pm 0.43$ & $11.34 \pm 0.60$ & $10.22 \pm 0.48$ & $9.16 \pm 0.05$ & $9.82 \pm 0.50$ \\
 $\beta_{A, \mathrm{global}}$ & $0.97 \pm 0.08$ & $0.73 \pm 0.23$ & $0.33 \pm 0.14$ & $0.64 \pm 0.08$ & $0.63 \pm 0.13$ & $0.54 \pm 0.11$ & $0.29 \pm 0.13$ & $0.55\pm 0.10$\\
 $\beta_{R_{50}}$ & $2.42 \pm 0.13$ & $1.99 \pm 0.20$ & $1.09 \pm 0.22$ & $1.68 \pm 0.11$ & $2.47 \pm 0.12$ & $1.97 \pm 0.13$ & $1.18 \pm 0.08$ & $1.77\pm 0.12$\\
 $\beta_{M}$ & $-0.50 \pm 0.09$ & $-0.30 \pm 0.08$ & --- & $-0.20 \pm 0.05$& $-0.38 \pm 0.07$ & $-0.20 \pm 0.05$ & --- & $-0.15 \pm 0.05$ \\
 Scatter & & 0.31 & & 0.49 & & 0.29 & & 0.48\\
\hline
$\alpha$  & $7.85 \pm 0.08$ & $8.41 \pm 0.07$ & $9.14 \pm 0.07$ & $8.49 \pm 0.04$ & $7.52 \pm 0.20$ & $8.42 \pm 0.17$ & $9.13 \pm 0.05$ & $8.47 \pm 0.04$ \\
$\beta_{R_{50}}$ & $1.75 \pm 0.43$ & $1.06 \pm 0.11$ & $1.06 \pm 0.23$ & $1.06 \pm0.09$ & $2.65 \pm 0.44$ & $1.32 \pm 0.22$ & $1.20 \pm 0.10$ & $1.28 \pm 0.11$\\
$\beta_{D}$ & ---& $0.71 \pm 0.05$ & $0.26 \pm 0.09$ & $0.55 \pm 0.05$ & --- & $0.52 \pm 0.23$ & $0.19 \pm 0.06$ & $0.45 \pm 0.06$ \\
Scatter & &0.29 & & 0.47 & & 0.28 & & 0.47\\
\hline
$\alpha$ & $7.48 \pm 0.14$ & $8.54 \pm 0.06$ & $9.18 \pm 0.08$ & $8.56 \pm 0.04$ & $7.44 \pm 0.10$ & $8.47 \pm 0.08$ & $9.13 \pm 0.03$ & $8.52 \pm 0.04$ \\
$\beta_{R_{50}}$ & $1.94 \pm 0.27$ & $1.28 \pm 0.19$ & $1.17 \pm 0.13$ & $1.29 \pm 0.09$ & $2.08 \pm 0.10$ & $1.56 \pm 0.17$ & $1.30 \pm 0.08$ & $1.45 \pm 0.10$\\
$\beta_T$ & $0.16 \pm 0.03$ & $0.10 \pm 0.01$ & $0.03 \pm 0.01$ & $0.08 \pm 0.01$ &  $0.15 \pm 0.02$ & $0.07 \pm 0.02$ & $0.03 \pm 0.01$ & $0.06 \pm 0.01$\\
Scatter & &0.30 & & 0.47 & & 0.28 & & 0.47\\
\hline
$\alpha$  & $8.98 \pm 0.06$ & $ 9.55\pm 0.04$ & $9.96 \pm 0.03$ & $9.50 \pm 0.02$ & $9.00 \pm 0.14$ & $9.55 \pm 0.04$ & $9.99 \pm 0.03$ & $9.51 \pm 0.03$ \\
$\beta_S$ & $0.91 \pm 0.08$ & $0.66 \pm 0.05$ & $0.56 \pm 0.03$ & $0.60 \pm 0.03$ &  $0.88\pm 0.17$ & $0.67 \pm 0.06$ & $0.62 \pm 0.05$ & $0.65 \pm 0.04$\\
Scatter & &0.30 & & 0.49 & & 0.29 & & 0.48\\
\enddata
\tablecomments{Subsets of the general equation are fitted. The notation is similar to Table~\ref{tbl:fit_H2_all}. Morphology as quantified by T-types or disk probability is useful, in combination with radius, to predict atomic gas mass, when SFR is not used.}
\end{deluxetable*}

\section{Galaxy Disk Inclination Angle as a Predictor of Molecular Gas Mass} \label{app:fit_inc}

The aim here is to assess the effects of galaxy inclination on molecular gas mass as proposed by \citet{Concas+19}. To that end, we take inclination angle ($i$) measurements from the bulge-disk decomposition catalog of \citet{Simard+11}, in which the bulge S\'{e}rsic index was assumed to be $n_b=4$. For fair comparison, we also use the best SFR estimates from the xCOLD GASS catalog in this section. Here, we focus on star-forming galaxies, since inclination is proposed to be a useful predictor only for these galaxies. Similar to what we showed in Table~\ref{tbl:corr} using the edge-on probability, the partial Kendall $\tau_{\mathrm{H_2}}$ for the correlation of $i$ with $M_{\mathrm{H_2}}$ at fixed $\log \mathrm{H\alpha/H\beta}$ is $-0.15$ in star-forming galaxies, while the corresponding partial $\tau_\mathrm{H_2}$ of SFR is 0.64. Although the partial correlation of $i$ after taking into account the effect of the Balmer decrement is significant (at $\alpha=0.001$), the strength of its correlation is much smaller than that of SFR. Likewise, given the effect of SFR, the partial $\tau_{\mathrm{H_2}}$ of $i$ is $-0.1$ ($p$ value 0.005). In contrast, given $i$,  the partial $\tau_{\mathrm{H_2}}$ is 0.7 for SFR. Furthermore, fitting together $M_\mathrm{H_2}$, SFR, $i$, and $\log \mathrm{H\alpha/H\beta}$ with the multiple regression models used in the main text indicates that the coefficient for $i$ is not significant (we fail to reject the null hypothesis that its coefficient is zero at $\alpha = 0.05$). In other words, once SFR  and $\log \mathrm{H\alpha/H\beta}$ are used, $i$ does not carry additional information. Using AIC and BIC also indicates that the normal censored regression model in which $M_\mathrm{H_2}$ is fitted with SFR and $\log \mathrm{H\alpha/H\beta}$ describes the data better than the one in which $\log M_{\mathrm{{H_2}}}$ is fitted with $i$ and $\log \mathrm{H\alpha/H\beta}$.

Because it has weak but significant partial correlation, it may be useful to include $i$ when SFR measurements are not available. Following \citet{Concas+19}, fitting the relation $\log M_{\mathrm{{H_2}}} = a + b \,(\log \mathrm{H\alpha/H\beta}-0.62) + c \,(i-56{\degree})$ for star-forming galaxies gives $a = 8.88 \pm 0.03$, $b = 5.70 \pm 0.16$ and $c = -0.007 \pm 0.003$ and $\operatorname{MAD}= 0.22$ dex for the median relation. The mean relation, estimated using the normal censored regression model, has coefficients consistent with those of the median relation and $\operatorname{RMSD} = 0.4$ dex. Our median relation is not inconsistent with the relation of \citet{Concas+19}. They found  $a = 8.93 \pm 0.03$, $b = 5.86 \pm 0.35$ and $c = -0.010 \pm 0.002$. Their fitting does not include star-forming galaxies with $M_{\mathrm{H_2}}$ upper limits. Similarly, fitting only the detections with a simulation-extrapolation (SIMEX) weighted least-square model (which includes measurement errors), we get $a = 8.93 \pm 0.03$, $b = 5.09 \pm 0.48$ and $c = -0.008\pm 0.002$ with RMSD of 0.39. In contrast, the same statistical model gives RMSD of 0.18 when fitting SFR instead of $i$. Namely, $\log M_{\mathrm{{H_2}}} = d + e \,(\log \mathrm{H\alpha/H\beta}-0.62) + f \log \mathrm{SFR}$, where $d = 8.95 \pm 0.14$, $e = 1.92 \pm 0.30$  and $f = 0.80 \pm 0.05$. This gas scaling of star-forming galaxies with SFR and $\mathrm{H\alpha/H\beta}$ is similar to what we found for the total sample in Section~\ref{sec:res}. To convert $e$ to the coefficient of $A_{V, \mathrm{fiber}}$, we divide it by 8.06 (from equation~\ref{eq:AV}).

Having established that we get a consistent result to that of \citet{Concas+19} for star-forming galaxies, we fit the relation among $\log M_\mathrm{{H_2}}$, $i$ and {\AV} estimated from the Balmer decrement or stellar continuum. In other words, we add $i$ to the analysis presented in Table~\ref{tbl:fit_H2_all} and present them in Table~\ref{tbl:fit_Incl}. After {\AV} is combined with $R_{50}$ and $M_\star$, the $q=0.15$ gas relations do not depend on $i$. Similarly, when {\AV}, metallicity ($Z$) and $R_{50}$ are used $i$ does not improve the gas mass prediction.

{\it In conclusion, given other variables, the effect of inclination is marginal. The combination of {\AV} and $Z$ or {\AV} and $R_{50}$ is a better predictor of molecular gas than the combination of {\AV} and $i$. The latter combination, however, does marginally improve the fit of {\AV} by itself, more significantly in star-forming or gas-rich galaxies.} 

\begin{deluxetable*}{lcccc|cccc}
\tabletypesize{\footnotesize}
\tablecolumns{10} 
\tablewidth{0pt} 
\tablecaption{Regression model fits for molecular gas, $\log M_\mathrm{{H_2}}  = \alpha + \beta_A A_{V} + \beta_R \log R_{50}+ \beta_M \log M_{\star}+ \beta_i (i-56\degree)$. \label{tbl:fit_Incl}}
\tablehead{\multicolumn{5}{c}{$A_V,{\mathrm{Fiber}}$} & \multicolumn{4}{c}{$A_V,{\mathrm{Global}}$} \\
\colhead{} & \colhead{15\%} &  \colhead{Median} & \colhead{85\%} & \colhead{Mean} & \colhead{15\%} &  \colhead{Median} & \colhead{85\%} & \colhead{Mean}}
\startdata
 $\alpha$  & $9.30 \pm 0.91$ & $6.28 \pm 0.48$ & $6.33 \pm 0.34$ & $6.47 \pm 0.43$ & $12.32 \pm 0.77$ & $11.16 \pm 0.71$ & $6.48 \pm 0.69$  & $6.95 \pm 0.40$ \\
 $\beta_{A}$ & $0.45 \pm 0.04$ &$ 0.36 \pm 0.03$ & $0.38 \pm 0.03$ & $0.35 \pm 0.03$ & $0.97 \pm 0.08$ & $0.73 \pm 0.23$ &$1.32 \pm 0.07$ & $1.37 \pm 0.08$ \\
 $\beta_R$ & $1.61 \pm 0.25$ & $1.00 \pm 0.10$ & $0.60 \pm 0.12$ & $0.87 \pm 0.11$ & $2.42 \pm 0.13$ & $1.99 \pm 0.20$ & $0.89 \pm 0.28$ & $1.08 \pm 0.10$\\
 $\beta_M$ & $-0.22 \pm 0.10$ &$0.16 \pm 0.05$ & $0.22 \pm 0.03$ & $0.15 \pm 0.05$ & $-0.50 \pm 0.09$ & $-0.30 \pm 0.08$ & $0.17 \pm 0.08$ & $0.07 \pm 0.04$\\
 $\beta_i$  & ---& $-0.004 \pm 0.001$ & $-0.006 \pm 0.001$ & $-0.003 \pm 0.001$ & --- & --- &$-0.005 \pm 0.001$  & $-0.003 \pm 0.001$\\
 Scatter & &0.23 & & 0.40 & &0.24 & & 0.37\\
\hline
\enddata
\tablecomments{Here we add inclination, $i$, to the analysis in Table~\ref{tbl:fit_H2_all}.}
\end{deluxetable*}

\section{Comparison with the Herschel Reference Survey Data}\label{app:hrs}

The Herschel Reference Survey \citep[HRS;][]{Boselli+10} is a volume-limited ($D = 15-25$ Mpc) and $K$-band-selected sample of 322 galaxies. This sample has observations at 250, 350, and 500 $\mu$m, and is a benchmark for studies of dust in the nearby universe. The sample spans a wide range of stellar mass ($M_\star = 10^9 - 10^{11}\, M_\odot$), morphological types, and environments (from the field to the center of the Virgo Cluster).

\begin{figure*}
\includegraphics[scale=0.3]{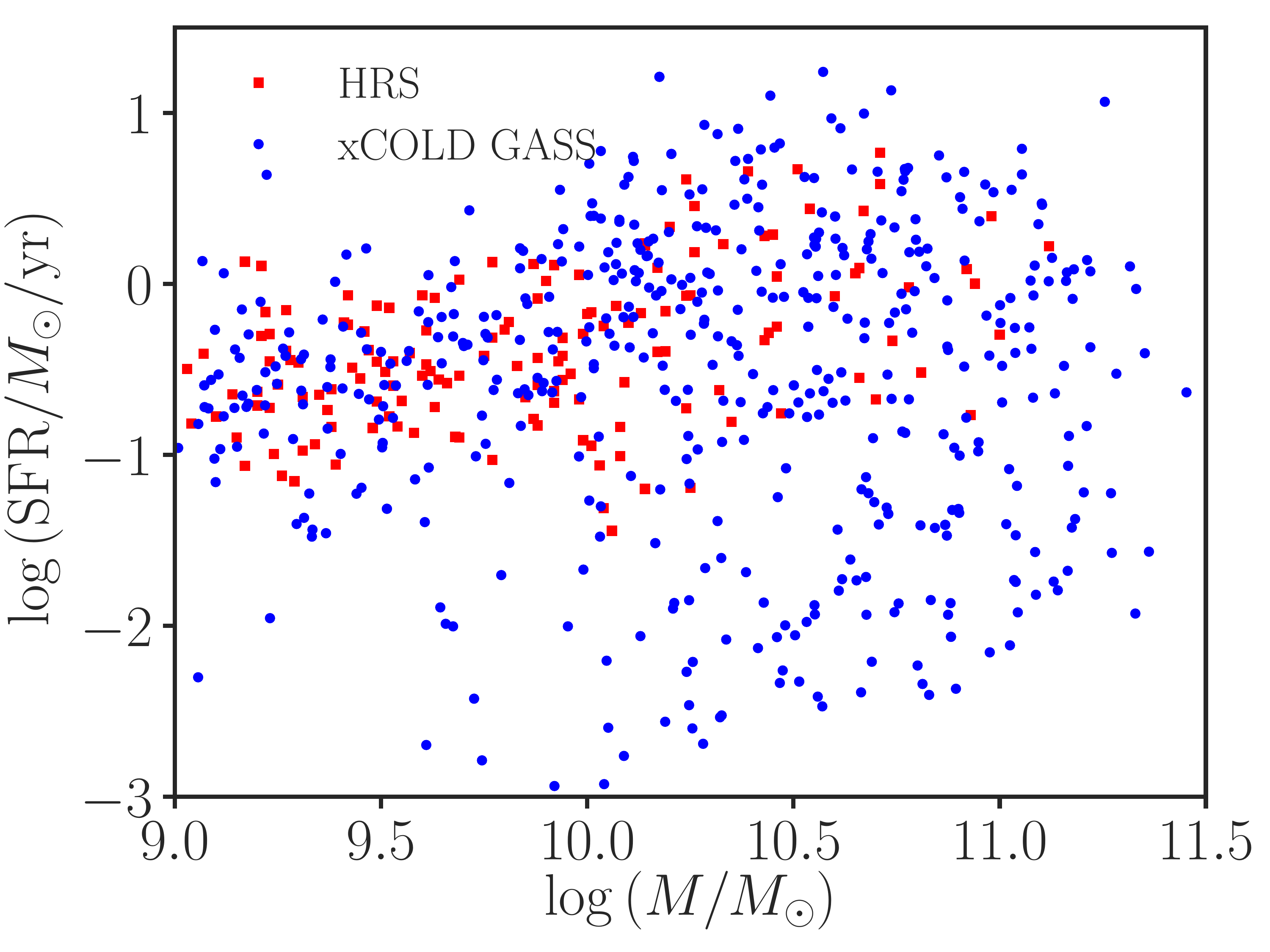}
\includegraphics[scale=0.3]{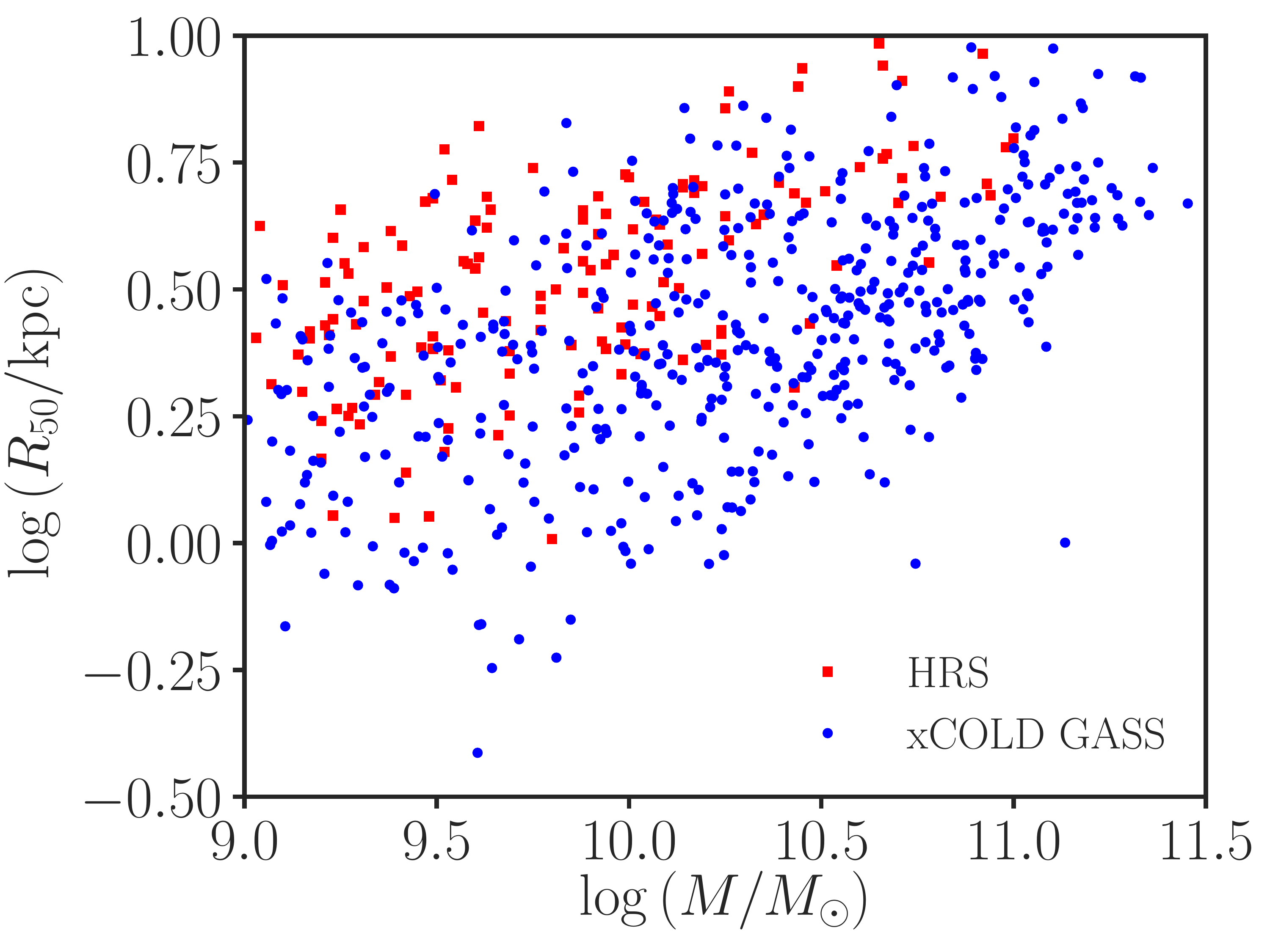}
\caption{Comparison of the Herschel Reference Survey data \citep{Boselli+10} with xCOLD GASS \citep{Saintonge+17}. SFR measurements are lacking for massive quiescent HRS galaxies.\label{fig:HRS}}
\end{figure*}

We use the publicly available Herschel Database in Marseille\footnote{\url{https://hedam.lam.fr/HRS/}} to retrieve measurements of gas masses, SFR, and H$\alpha$/H$\beta$ of the HRS galaxies. \citet{Boselli+14a} obtained new CO\,(1--0) observations for 59 objects, which when combined with literature data produced a molecular gas catalog for 225 galaxies. Among these, only 158 have SFR measurements, and only 115 have both SFR and H$\alpha$/H$\beta$ measurements derived from long-slit spectroscopic observations. As shown in Figure~\ref{fig:HRS}, the massive quiescent HRS galaxies lack SFRs, and the xCOLD GASS sample we analyze in Section~\ref{sec:res} is more representative in terms of coverage of the the $M_\star$ vs. SFR plane.  The SFRs of HRS  were derived from a variety of available data \citep{Boselli+15}. We divide the SFRs by 1.58 to convert them to the \citet{Chabrier03} initial mass function. The aperture-corrected CO fluxes (with 3D exponential disk model) are converted to molecular gas masses using constant (Milky Way) and variable ($H$-band luminosity-dependent) $X_\mathrm{CO}$ conversion factors. \citet{Boselli+14a} also compiled \ion{H}{1} data for 315 HRS galaxies from the literature. For 52 \ion{H}{1} non-detections, 5 $\sigma$ upper limits were estimated assuming a velocity width of 300 km s$^{-1}$. Likewise, for the 82 H$_2$ non-detections, 5 $\sigma$ upper limits were estimated by assuming a CO velocity width equal to the \ion{H}{1} width, when detected, or to 300 km s$^{-1}$ otherwise. Stellar masses and $r$-band half-light radii were taken from \citet{Cortese+12}. The stellar masses were derived from $i$-band luminosities using a relation between $g-i$ color and stellar mass-to-light ratio, assuming a Chabrier initial mass function.

\begin{figure*}
\includegraphics[scale=0.3]{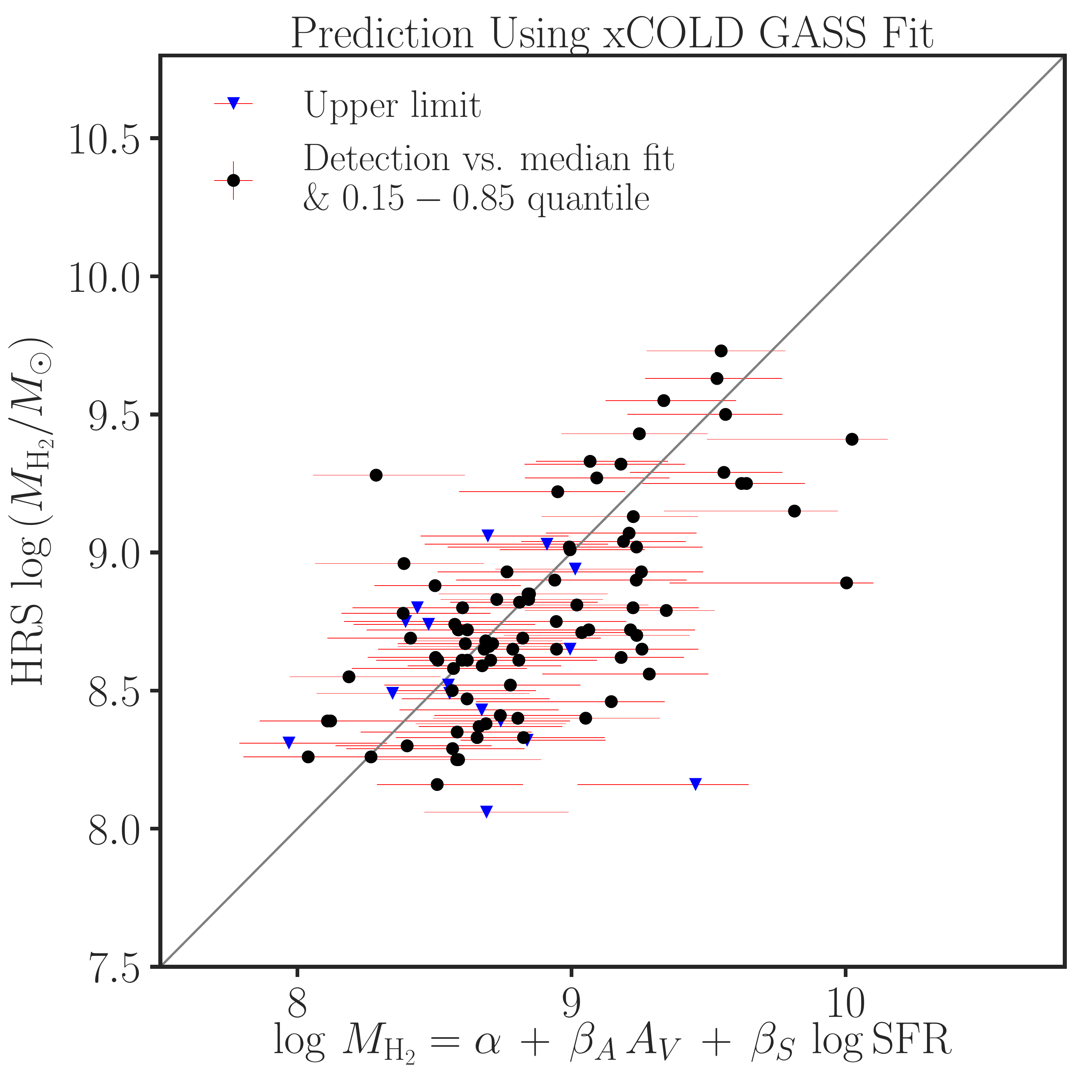}
\includegraphics[scale=0.3]{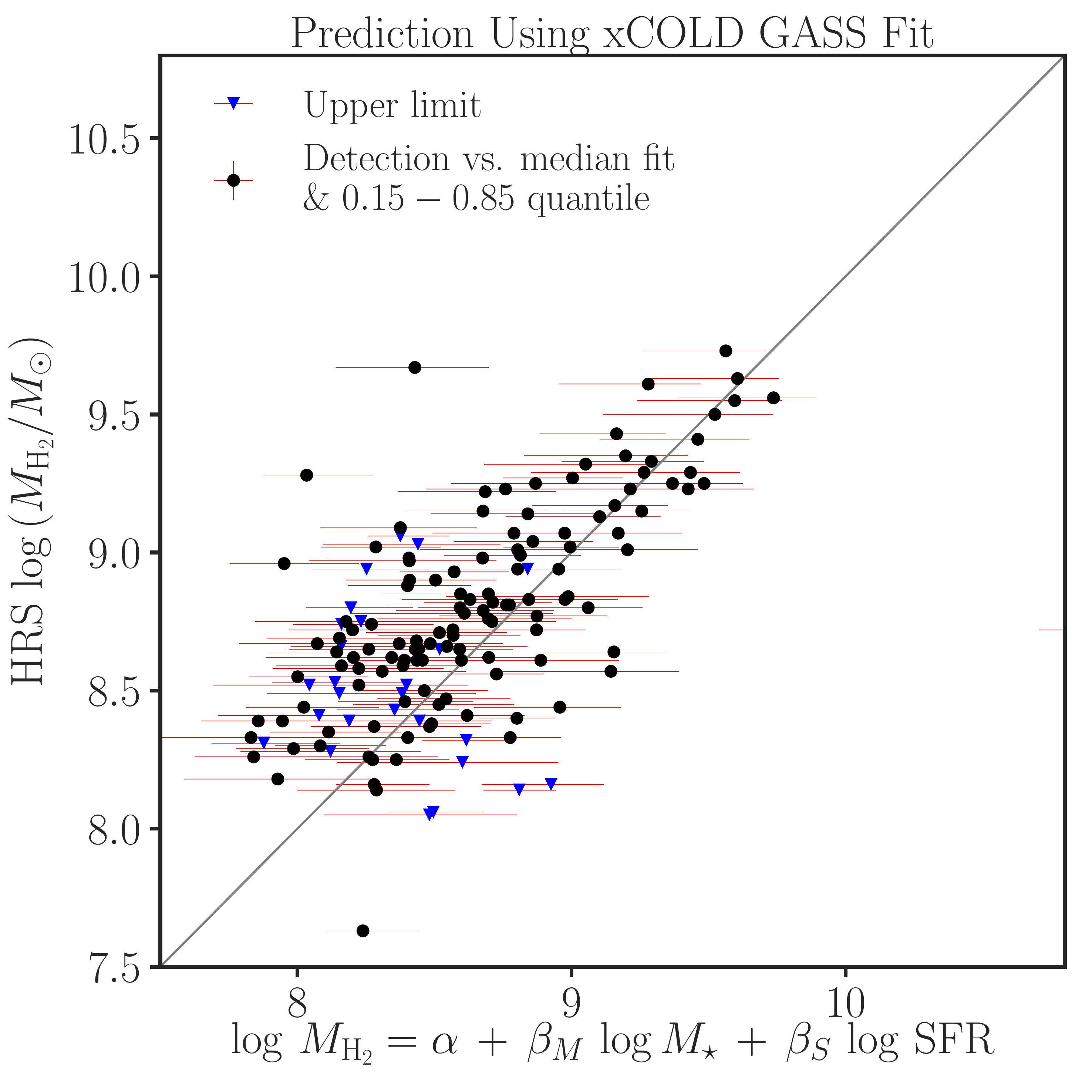}
\caption{Scaling relations of molecular gas derived from xCOLD GASS data are consistent with independent observations of molecular gas of HRS galaxies.\label{fig:HRS_comp1}}
\end{figure*}

\begin{figure*}
\includegraphics[scale=0.3]{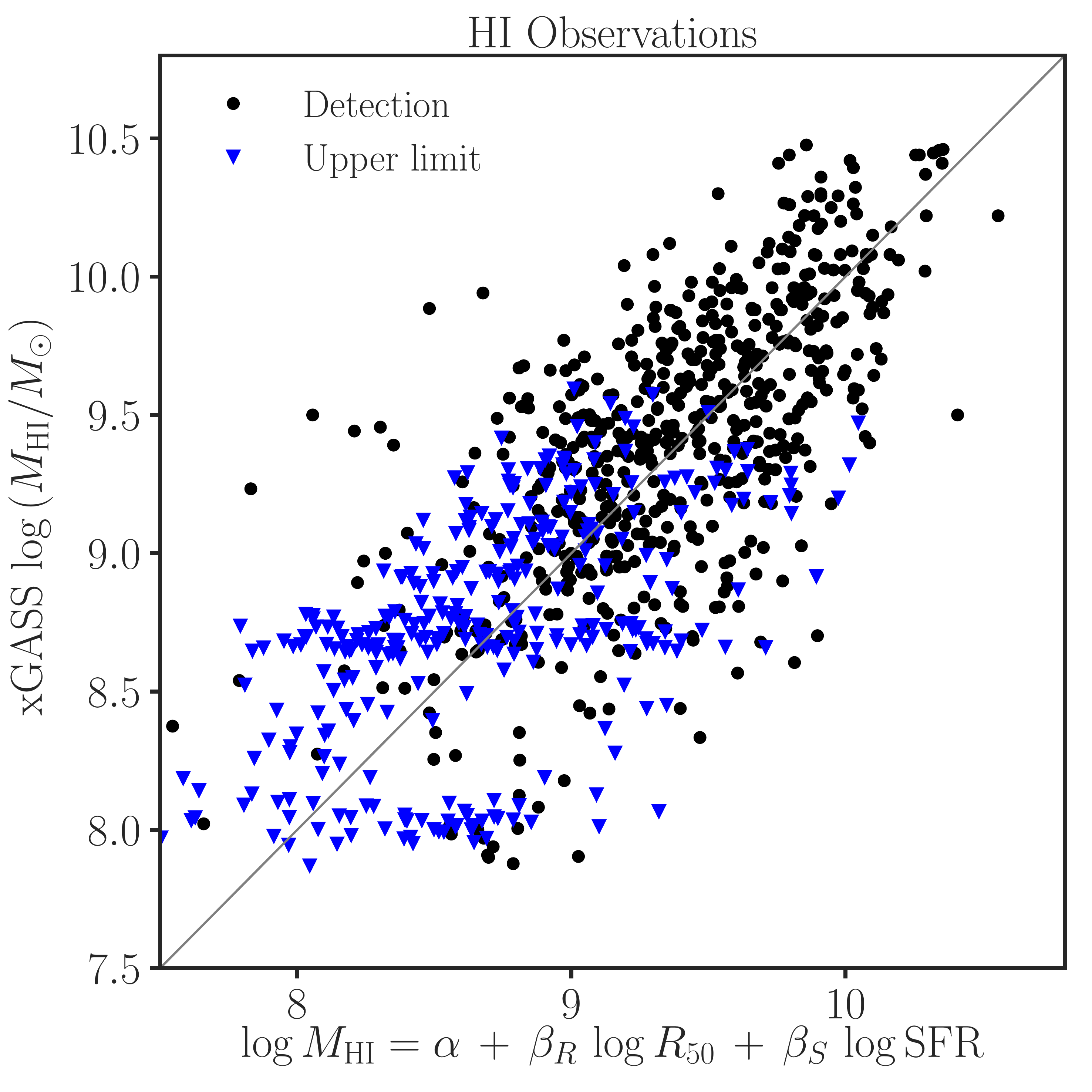}
\includegraphics[scale=0.3]{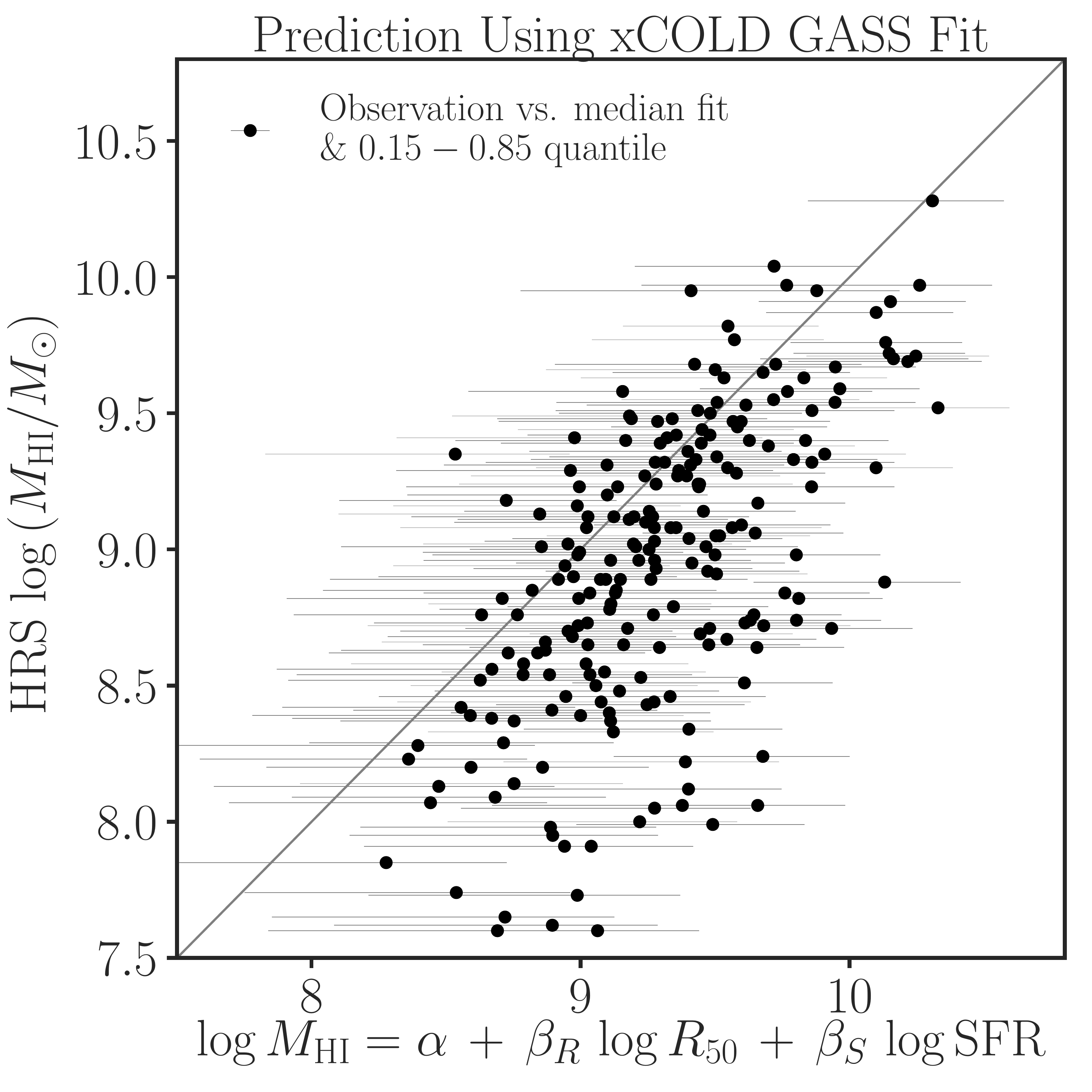}
\caption{The left panel shows our median fit for atomic gas data of the xGASS sample. Our prediction using this fit is consistent with the observed atomic gas masses of galaxies in HRS.\label{fig:HRS_comp2}}
\end{figure*}

\citet{Boselli+14b} used the HRS sample and scaling variables $M_\star$, stellar surface density, specific star formation rate SSFR$\equiv$SFR/$M_\star$, and metallicity to extend the gas scaling relations for massive galaxies in COLD GASS \citep[which is the precursor to xCOLD GASS,][]{Saintonge+11}.
They found a significant correlation between $M_\mathrm{H_2}/M_\star$ and SSFR and between $M_\mathrm{H_2}/ M_\mathrm{HI}$ and metallicity. Here we also do a regression analysis of the HRS sample using $M_\star$ and SFR as independent variables and $M_\mathrm{H_2}$ as dependent variable. As shown in Table~\ref{tbl:fit_hrs}, once the SFR dependence is taken into an account, the $M_\star$ dependence (for the mean relation) is much weaker for this sample compared to xCOLD GASS (Table~\ref{tbl:fit_H2_all}). Furthermore, for late-type galaxies, \citet{Boselli+14b} found $\log M_\mathrm{H_2}/M_\star \propto a \log \mathrm{SSFR}$, where $a=0.94$ for a constant $X_\mathrm{CO}$ and $a=1.01$ for a luminosity-dependent $X_\mathrm{CO}$ (see their Table 3). As $\log M_\mathrm{H_2} \propto (1-a) \log M_\star+ a \log \mathrm{SFR}$, their fit can be interpreted as indicating that the mass dependence is very weak. Thus, for the HRS sample the $M_\mathrm{H_2}/M_\star$ vs. SSFR relation is similar to the $M_\mathrm{H_2}$ vs. SFR relation. Boselli et al.'s method of fitting---using $M_\star$ on both sides--- may induce correlations in the residuals, in addition to forcing SFR and $M_\star$ to share correlated coefficients for the mean gas mass relation. Because the method is implicit, it does not allow physical insights to be gleaned from the coefficients of SFR and $M_\star$ separately. In any event, our fits for the xCOLD GASS/xGASS data are not inconsistent with the HRS sample (Figures~\ref{fig:HRS_comp1} and \ref{fig:HRS_comp2}), despite the difference in the two samples, the $X_\mathrm{CO}$ used, and how SFR and {\AV} are measured.

Similar to \citet{Boselli+14b}, \citet{Saintonge+17} found that $M_\mathrm{H_2}/M_\star$ shows the tightest correlation with SSFR for xCOLD GASS data. Unlike previous work that use these data, we include the upper limits in our analysis and give simple and convenient formulae summarizing the data. We also present scaling relations, for the first time, that do not use SFR.  Despite the emphasis of previous scaling relations on $M_\star$, Tables~\ref{tbl:corr} and \ref{tbl:fit_H2_all} show that the improvement brought by a linear term of $M_\star$ is small if half-light radius is used. Using Random Forest, we confirm that {\AV} and $R_{50}$ have better predictive power than $M_\star$. The scaling relations in the literature using the equivalent (rearranged) relation may give the impression that they have lower dispersions because they did not include upper limits in their analyses. In addition, the $M_\mathrm{H_2}/M_\star$-SSFR relation in \citet{Boselli+14b} looks tighter because the galaxies in their sample with molecular gas measurements are preferentially star-forming, unlike the galaxies in xCOLD GASS, which contains many quiescent galaxies. Nearly all quiescent galaxies in xCOLD GASS are upper limits, and these upper limits have less constraining power when $M_\star$ is combined with SFR (e.g., Figure~\ref{fig:MH2_SFR_M}). Note also that xCOLD GASS/xGASS gives $3\,\sigma$ upper limits, while HRS uses $5\,\sigma$ upper limits, which further reinforces the impression that the scatter is small because the upper limits overlap with the detections. Lastly, adding $M_\star$, in almost all cases, does not statistically significantly improve the prediction of atomic gas masses.

\begin{deluxetable*}{lcccc|cccc}
\tabletypesize{\footnotesize}
\tablecolumns{10} 
\tablewidth{0pt} 
\tablecaption{Regression model fits for the Herschel Reference Survey (HRS) data, $\log M_\mathrm{{H_2}}  = \alpha +  \beta_M \log M_{\star}+\beta_S \log \mathrm{SFR}$.  \label{tbl:fit_hrs}}
\tablehead{\multicolumn{5}{c}{$X_\mathrm{CO,MW}$} & \multicolumn{4}{c}{Variable $X_\mathrm{CO}$} \\
\colhead{} & \colhead{15\%} &  \colhead{Median} & \colhead{85\%} & \colhead{Mean} & \colhead{15\%} &  \colhead{Median} & \colhead{85\%} & \colhead{Mean}}
\startdata
 $\alpha$  & $8.43 \pm 0.3$ & $5.89 \pm 0.07$ & $5.45 \pm 1.40 $ & $8.67 \pm 0.19$ & $8.38 \pm 0.12$ & $8.19 \pm 0.58$ & $8.50 \pm 0.14$ & $8.72 \pm 0.16$ \\
 $\beta_M$ & $0.018 \pm 0.004$ &$0.32 \pm 0.07$ & $0.39 \pm 0.14$ & $0.04 \pm 0.02$ & $0.018 \pm 0.004$ & $0.08 \pm 0.07$ & $0.07 \pm 0.13$ & $0.02 \pm 0.02$ \\
 $\beta_S$  & $0.65 \pm 0.28$ & $0.70 \pm 0.08$ & $0.65 \pm 0.14$ & $0.43 \pm 0.08$ & $0.46 \pm 0.19$ & $0.59 \pm 0.08$ & $0.57 \pm 0.12$ & $0.58 \pm 0.06$ \\
 Scatter & &0.17 & & 0.39 & & 0.17 & & 0.33\\
\enddata
\tablecomments{The median relation of the HRS sample for the case of $X_\mathrm{CO,MW}$ (a constant Milky Way value) is similar to that of the xCOLD GASS sample. In the latter sample, the fits do not change much depending on $X_\mathrm{CO}$.}
\end{deluxetable*}

%\begin{deluxetable*}{lcccc|cccc}
%\tabletypesize{\footnotesize}
%\tablecolumns{10} 
%\tablewidth{0pt} 
%\tablecaption{\textcolor{red}{Regression model fits for the Herschel Reference Survey (HRS) data, $\log M_\mathrm{{H_2}}  = \alpha +  \beta_M \log M_{\star}+\beta_S \log \mathrm{SFR}$}.  \label{tbl:fit_hrs2}}
%\tablehead{\multicolumn{5}{c}{$X_\mathrm{CO,MW}$} & \multicolumn{4}{c}{Variable $X_\mathrm{CO}$} \\
%\colhead{} & \colhead{15\%} &  \colhead{Median} & \colhead{85\%} & \colhead{Mean} & \colhead{15\%} &  \colhead{Median} & \colhead{85\%} & \colhead{Mean}}
%\startdata
% $\alpha$  & $7.84 \pm 1.21$ & $8.69 \pm 0.06$ & $9.04 \pm 0.34$ & $8.48 \pm 0.17$ & $8.04 \pm 0.02$ & $8.59 \pm 0.08$ & $8.84 \pm 0.03$ & $8.53 \pm 0.14$ \\
 %$\beta_M$ & $0.09 \pm 0.13$ &$0.05 \pm 0.01$ & $0.05 \pm 0.03$ & $0.06 \pm 0.02$ & $0.06 \pm 0.02$ & $0.04 \pm 0.01$ & $0.04 \pm 0.01$ & $0.04 \pm 0.02$ \\
% $\beta_S$  & $0.64 \pm 0.18$ & $0.95 \pm 0.14$ & $0.89 \pm 0.09$ & $0.85 \pm 0.07$ & $0.46 \pm 0.19$ & $0.63 \pm 0.06$ & $0.65 \pm 0.05$ & $0.59 \pm 0.06$ \\
 %Scatter & &0.21 & & 0.38 & & 0.20 & & 0.32\\
%\enddata
%\end{deluxetable*}

%\bibliography{references_v3}

\end{document}